\documentclass[twocolumn]{aastex61}
\usepackage{graphicx}

\usepackage{longtable}
\usepackage{rotating}
\usepackage{latexsym}
\usepackage{lineno}
\usepackage{amsmath}

\newcommand{\HI}{H\,{\sc i} }

\newcommand{\wise}{{\it WISE} }

\newcommand{\Ltir}{$L_{\rm TIR}$ }

\received{2017 August 24}
\revised{2017 October 9}
\accepted{2017 October 9}
\published{2017 November 17}
\submitjournal{ApJ}

\shorttitle{Star Formation Relations using \wise}
\shortauthors{Cluver et al.}

\begin{document}


\title{Calibrating Star Formation in \wise using Total Infrared Luminosity}

\correspondingauthor{Michelle Cluver}
\email{michelle.cluver@gmail.com}

\author{M.E. Cluver}
\affiliation{Department of Physics and Astronomy, University of the Western Cape,
Robert Sobukwe Road, Bellville,South Africa}
\affiliation{Inter-University Institute for Data Intensive Astronomy}

\author{T.H. Jarrett}
\affiliation{Department of Astronomy, University of Cape Town, Rondebosch,
 South Africa}

\author{D.A. Dale}
\affiliation{Department of Physics and Astronomy, University of Wyoming, Laramie, WY 82071, USA}

\author{J.-D.T. Smith }
\affiliation{Ritter Astrophysical Observatory, University of Toledo, Toledo, OH 43606, USA}

\author{T. August}
\affiliation{Department of Astronomy, University of Cape Town, Rondebosch,
 South Africa}

\author{M.J.I. Brown }
\affil{School of Physics and Astronomy, Monash University, Clayton 3800, Victoria, Australia}

\begin{abstract}

We present accurate resolved \wise photometry of galaxies in the combined SINGS and KINGFISH sample. The luminosities in the W3 12\micron\ and W4 23\micron\ bands are calibrated to star formation rates (SFRs) derived using the total infrared luminosity, avoiding UV/optical uncertainties due to dust extinction corrections. The W3 relation has a 1-$\sigma$ scatter of 0.15 dex over nearly 5 orders of magnitude in SFR and 12\micron\ luminosity, and a range in host stellar mass from dwarf (10$^7$ M$_\odot$) to $\sim3\times$M$_\star$ (10$^{11.5}$ M$_\odot$) galaxies. In the absence of deep silicate absorption features and powerful active galactic nuclei, we expect this to be a reliable SFR indicator chiefly due to the broad nature of the W3 band. By contrast the W4 SFR relation shows more scatter (1-$\sigma =$ 0.18 dex). Both relations show reasonable agreement with radio continuum-derived SFRs and excellent accordance with so-called ``hybrid" 
H$\alpha + 24 \mu$m and 
FUV$+$24\micron\ indicators.  Moreover, the \wise SFR relations appear to be insensitive to the metallicity range in the sample. We also compare our results with IRAS-selected luminous infrared galaxies, showing that the \wise relations maintain concordance, but systematically deviate for the most extreme galaxies.
Given the all-sky coverage of \wise and the performance of the W3 band as a SFR indicator, the $L_{12\micron}$ SFR relation could be of great use to studies of nearby galaxies and forthcoming large area surveys at optical and radio wavelengths.

\end{abstract}


\section{Introduction} \label{intro}

Mapping the build-up of stellar mass underpins galaxy evolution studies and tests of the $\Lambda$CDM cosmological framework that informs simulations of our Universe. Accurate star formation rates therefore form the cornerstone of extragalactic studies and have been the subject of vigorous study for at least the past three decades \citep[see reviews by][]{Kenn98, Cal13}. However, the determination of a star formation rate (SFR) for any given galaxy is inherently difficult; active star formation is associated with copious dust, which itself obscures the radiation from the hot, young stars one is trying to measure. 
The ultraviolet (UV) directly traces the youngest stellar populations and therefore respresents the ``purest" star formation indicator. However, the UV is heavily extincted by dust, and although numerous methods exist to correct for this dust extinction, variations in dust content and geometry unavoidably produce large scatter \citep[e.g.][]{Cal07}.

The optical nebular recombination lines are less affected by dust compared to the UV, and of these the H$\alpha$ emission line is the least attenuated. However, uncertainties still arise due to the variation of the dust extinction from location to location and the assumption of the underlying stellar absorption \citep[e.g.][]{Cal07}. H$\alpha$ maps, corrected for extinction, have been successfully used to trace star formation \citep[e.g.][]{Cas15} and with increased access to integral field observations, studies of the impact of aperture corrections on particularly single-fiber-derived SFRs advocate caution to avoid systematic bias \citep[e.g.][]{Ig13,Rich16}.

Surveys with multiwavelength coverage can rely on multiband photometry combined with spectral energy distribution (SED) models to determine the physical properties of galaxies using codes such as MAGPHYS \citep{magphys} and CIGALE \citep{Noll09}. However, full radiative transfer solutions \citep[e.g.][]{Nat2014, Groot17} will ultimately be crucial to disentangling star formation processes within galaxies.

Exploiting dust emission itself as a measure of the SFR, i.e. calibrating dust reprocessed starlight, has its own pitfalls \citep[for a detailed review see][]{Cal13}. Using emission from $\sim 5-1000$ \micron, usually termed the total infrared luminosity or $L_{\rm TIR}$, has the advantage of sampling dust heated by young (1 - 100 Myr), UV-luminous O and B stars, as well as intermediate mass (2 - 3 M$_\odot$), UV-faint A and F stellar populations. However, having a well-sampled dust SED is observationally intensive and can only be achieved using space-based observatories. For example, the ongoing DustPedia project \citep{Dav17} leverages {\it Herschel Space Observatory} \citep{Pil10} and {\it Planck} \citep{Pl16} observations with ancillary archival multiwavelength data to produce Bayesian SED fits and photon tracing radiative transfer modeling, providing a legacy dataset for exploring how dust emission is related to its physical properties and origins.

Hybrid SFR indicators compensate for extinction effects by combining a dust-obscured star formation tracer with an unobscured measure of star formation \citep[see, for example,][]{Cal95, Bu99}. Seminal work by \citet{Kenn07} and \citet{Cal07}, combining H$\alpha$ and 24\micron, was expanded by \citet{Kenn09}, and although highly successful, does require some caution when low levels of star formation are measured \citep[e.g.][]{Bos15}. The equivalent FUV$+$ 24\micron\ SFR relation \citep[e.g.][]{Big08, Ler08, Boq16} was made possible by the NUV and FUV bands of the {\it GALEX} satellite \citep{Mar05}. Both these relations have been used extensively to study star formation in nearby galaxies \citep[e.g.][]{Rah11,For13, Mom13, Hees14} and variations include: FUV$+$ 25\micron\ \citep[e.g.][]{Hao11}, UV $+$ 22\micron\ \citep[e.g.][]{Cor12, HK15, Cas17}, and H$\alpha$ $+$ 22\micron\  \citep[e.g.][]{Lee13}.

The {\it Infrared Astronomical Satellite, IRAS} \citep{Neg84}, {\it Infrared Space Observatory, ISO} \citep{Kess03}, {\it Spitzer Space Telescope} \citep{Wer04}, and {\it Herschel Space Observatory} \citep{Pil10} pioneered the use of monochromatic infrared SFR calibrators. In particular, the 24\micron\ band has been characterised by several studies for the determination of global SFRs \citep{Wu05,Zhu08,Riek09,Cal10}. Tracing the warm dust continuum arising from small grains, as well as non-thermal emission from stochastically heated grains, the MIPS 24\micron\ band is relatively free of contamination from the stellar continuum, polycyclic aromatic hydrocarbons (PAHs), and nebular line emission, while also providing adequate spatial resolution for optically-derived surveys. In addition, warm dust emission is more closely associated with recent star formation than emission at longer wavelengths, where heating from old stellar populations become an important consideration \citep[e.g.][]{Pop00}.

Indeed, with the launch of the {\it Wide-Field Infrared Explorer, \wise}\citep{Wr10}, and its all-sky survey, the W4 band \citep[nominally centered on 22\micron, but more recent work by][places the center of the band closer to 23\micron]{Br14b}
 was expected to be the primary SFR indicator. As with the {\it Spitzer} 24\micron\ band, the \wise\ W4 band is not contaminated by emission lines (at $z=0$), measuring the warm dust continuum which, in the absence of active galactic nucleaus (AGN) activity, provides a reliable measure of star formation, comparable to that of Balmer-corrected H$\alpha$ measures \citep{Br17}. However, the widespread use of this band is severely hampered due to a lack of sensitivity \citep{Jar13}.

In contrast to the \wise 23\micron\ band, which benefits from numerous {\it Spitzer} MIPS-24 studies, the 12\micron\ window was last used extensively as part of the {\it IRAS (Infrared Astronomical Satellite)} mission, but was limited by its 4\arcmin\ resolution. Nevertheless, \citet{Tak05} showed that the 12\micron\ luminosity could be used as a reliable measure of $L_{\rm TIR}$. 

Compared to the {\it Spitzer} mid-infrared bands, the \wise W3 band is somewhat unusually broad, covering 7.5 -- 16.5\micron\ at half-power \citep{Jar11}, which at $z=0$ samples part of the broad 7.7\micron\ PAH feature, the 8.5\micron\ PAH, the 10\micron\ silicate absorption feature, the 11.3\micron\ PAH feature, the S(2) line of pure rotational H$_2$ at 12.3\micron, and the 12.81\micron\ $[$Ne {\sc ii}$]$ and 13.7\micron\ $[$Ne {\sc iii}$]$ nebular emission lines. It is sometimes characterized as the ``PAH" band of {\it WISE}, because the center of the band (11.6\micron) is close to the 11.3\micron\ emission feature, but it is clearly far more complex. PAH fractions are high in regions of active star formation and their abundance suggests that they are produced in molecular clouds \citep{Sand10}, likely growing on dust grains, but are destroyed by the relatively hard interstellar radiation field of the diffuse interstellar medium (ISM) and that produced by AGN \citep{Smith07}. 

In this paper we build on the work of \citet{Clu14} exploring the use of the \wise 12\micron\ and 23\micron\ bands as SFR indicators. In order to be independent of uncertainties associated with dust extinction corrections for calibrators in the UV and optical \citep[for example, see comparison of 12 SFR metrics in][]{Dav16}, we exploit the {\it Spitzer} and {\it Herschel} photometry of the combined SINGS \citep[{\it Spitzer} Infrared Nearby Galaxy Survey;][]{sing} and KINGFISH \citep[Key Insights on Nearby Galaxies: a Far-Infrared Survey with {\it Herschel};][]{king} samples to calibrate star formation determined from $L_{\rm TIR}$.  In addition to this sample of ``typical" galaxies from which we derive new SFR relations, we also compare with \wise measurements of the most luminous infrared galaxies.

In this work, all monochromatic luminosities are defined as $\nu L_{\nu}$, for example, $L_{12\micron} \equiv \nu L_{\nu} (12\micron)$. We adopt a \citet{Kroup02} initial mass function (IMF) throughout this work, where the IMF has the slope $\alpha = 2.3$ for stellar masses 0.5$-$100 M$_\odot$ and a shallower slope $\alpha = 1.3$ for the mass range 0.1$-$0.5 M$_\odot$. The cosmology adopted throughout this paper is $H_0 =70$ km s$^{-1}$ Mpc$^{-1}$, $h = H_0/100$, $\Omega_M = 0.3$ and $\Omega_\Lambda = 0.7$. All magnitudes are in the Vega system accordant to the calibration adopted by the \wise survey \citep[as described in][]{Jar11}.  All linear fits are performed using the Hyper-Fit package \citep{hyper} which incorporates heteroscedastic errors and outputs a measure of intrinsic scatter, as well as a parameter covariance matrix.

\section{Data}

\subsection{Sample}

The SINGS survey \citep{sing} was carried out as a Legacy Project on the {\it Spitzer Space Telescope} in order to study the physics of the star-forming ISM. A sample of 75 galaxies was drawn from the nearby universe (d$< 30$ Mpc) to provide spatially-resolved {\it Spitzer} images and spectra of a diverse set of typical local galaxies. Combined with comprehensive ancillary multi-wavelength data, a key science goal of the project was to develop improved diagnostics of SFRs in galaxies.

Building on the successes of the SINGS study, the KINGFISH project on the {\it Herschel Space Observatory} \citep{king} obtained far-infrared imaging and spectral line maps to enable the detailed characterisation of the ISM, as well as the heating and cooling of gaseous and dust components. The KINGFISH sample consists of a subset (57 galaxies) of the SINGS sample, with the addition of 4 galaxies: IC 0342, NGC 2146, NGC 3077 and M101 (NGC 5457). The combined SINGS/KINGFISH sample therefore consists of 79 galaxies.

\begin{figure*}[!p]
\plotone{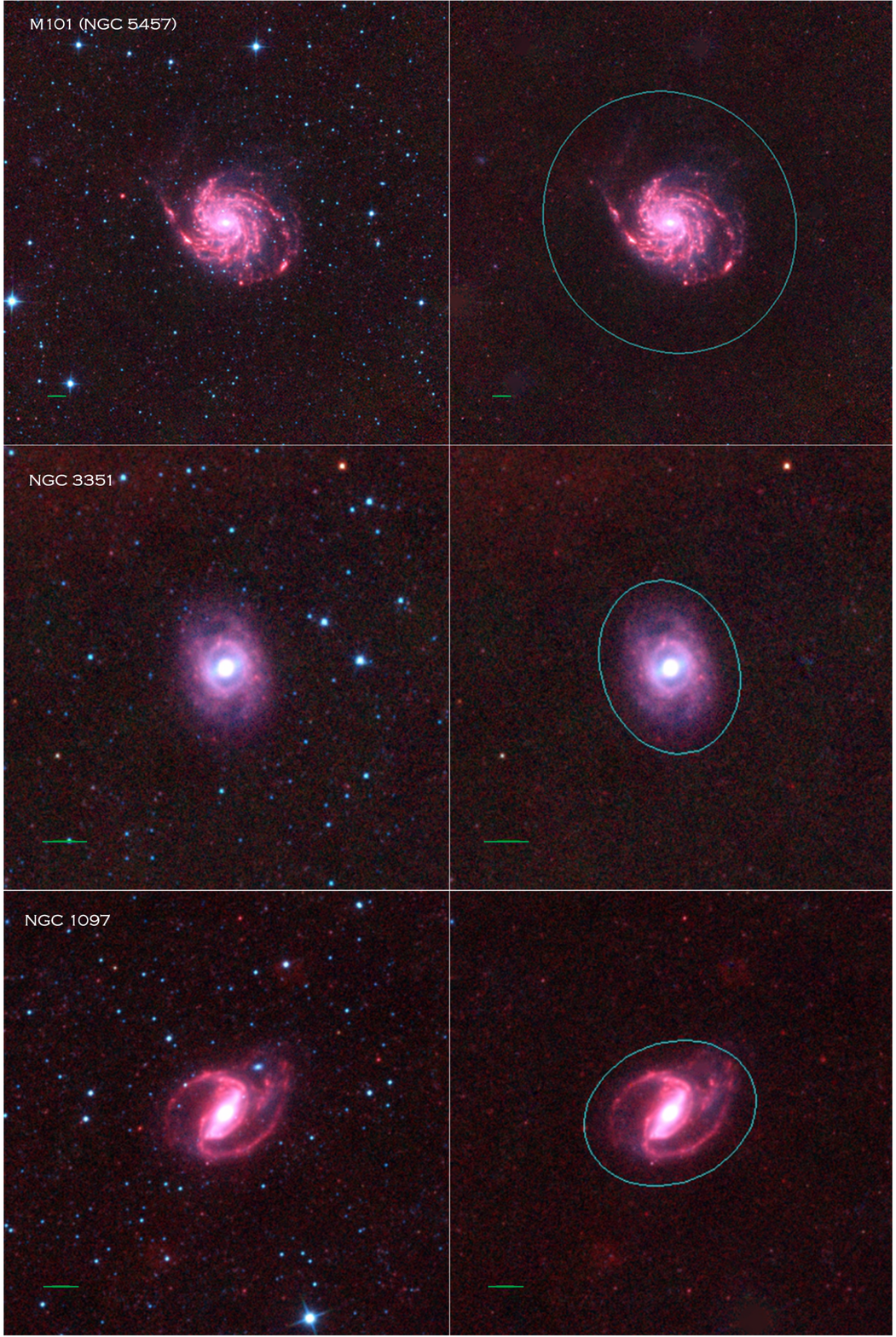}       
\caption{\wise three-color (combined 3.4, 4.6, and 12 \micron) images of three galaxies in the SINGS/KINGFISH sample showing the galaxies before (left) and after (right) star and background source removal. The cyan ellipses in the right panels indicate the 1-$\sigma$ isophotal apertures. In this representation, light from evolved stars is blue, while active star formation has a red color.  A scale of 2$\arcmin$ is indicated with the green horizontal line, bottom left, while North is upwards and East is to the left. \label{fig:f0}}               
\end{figure*}

\subsection{\wise Photometry}

The \wise telescope was launched in December 2009 and completed its nominal mission of an all-sky survey at 3.4, 4.6, 12 and 23\micron\footnote{We adopt the W4 calibration from \citet{Br14b}, in which the W4 response is redder, with a 
central wavelength is 22.8$\, \mu$m and the magnitude-to-flux conversion factor is 7.871 Jy.} before depleting its cryogen in October 2010. \wise was placed in hibernation in February 2011, but was successfully reactivated in September 2013 and continues to observe at wavelengths of 3.4 and 4.6\micron\ as part of the NEOWISE program \citep{amy14}.

With a 40cm diameter primary mirror, the native image resolution of single exposure \wise frames is $\approx$ 6\arcsec\ in W1 (3.4\micron) and W2 (4.6\micron).
To preserve the native resolution and improve sensitivity by adding the latest NEOWISE imaging, we 
have constructed new mosaics employing the ICORE software package \citep{Mas13}; resulting mosaics achieve a spatial resolution of 5.9\arcsec (W1), 6.5\arcsec (W2), 7.0\arcsec (W3), and 12.4\arcsec (W4) \citep{Jar12}.  Consideration of the mosaic size was also of primary importance because of the angular extent of the SINGS galaxies, capturing both the galaxy, as well as the local environment to properly assess the background levels.

Target galaxies are measured using custom software optimised for performing aperture photometry on resolved galaxies \citep{Jar13, Clu14, Jar17}.  Standard \wise photometry consists of integrated fluxes down to a 1$\sigma$ of the sky surface brightness (typically $\sim$23 mag arcsec$^{-2}$) for all four bands, which is referred to as the isophotal photometry (see Figure \ref{fig:f0}).   Colors are determined using matched apertures -- where apertures are all matched to the W1 isophote, except in cases where the emission is considerably less in the other bands; e.g., stellar-dominated cases, such as elliptical galaxies, with the Rayleigh-Jeans emission in W1, W2 and very little emission in W3 and W4.  In those cases, we match to the smaller aperture (usually W2), and similarly for the W2$-$W3 color.  However, for the majority of the cases, because the SINGS sample is so bright in the mid-infrared, all four bands have matched apertures.  For a few rare cases (notably the dwarf galaxies), the source is not resolved in the W3 or W4 apertures, in which case we use the standard point source photometry from the ALLWISE catalogue\footnote{http://irsa.ipac.caltech.edu/cgi-bin/Gator/nph-dd} \citep{Cut13}.

The photometry pipeline also attempts to measure the total flux per band.  This is carried out by modeling the axi-symmetric radial surface brightness using a S\'{e}rsic Function that fits the inner bulge light and the outer disk light.   In this way, the model is integrated beyond the 1$\sigma$ isophote, extending to three disk scale lengths.  The resulting total flux is typically within 5\% of the isophotal flux for the W1 and W2 bands, but may be 10 to 25\% larger for the W3 and W4 bands because of their relative insensitivity \citep{Jar17}.  Hence, when the galaxy surface brightness is well modeled,  we use the total fluxes to represent the W3 and W4 integrated fluxes used for the star formation relations.

Both the W3 and W4 bands are excellent tracers of ISM emission, but they also have contributions from the evolved stellar populations.  To estimate and remove this contribution, we use the method of \citet{Hel04} in which the near-infrared band is used as a proxy for the stellar emission. Based on the SED template for a 13 Gyr galaxy \citep{Sil98} and the \wise filter transmission \citep{Jar11}, we determine that 15.8\% of the W1 light is contained in the W3 band, and 5.9\% of the W1 light is in the W4 band.   So, after scaling the W1 integrated flux densities by these factors and subtracting from the W3 and W4 total fluxes, respectively, we arrive at the ISM emission, accordingly:  W3$_{\rm PAH}$ and W4$_{\rm dust}$.   Converting from flux density to luminosity then follows from 4$\pi$d$^2$ and scaling by the band center frequency, the so-called 
spectral luminosity,  $L_{12\, \mu \rm m}$ and $L_{23\, \mu \rm m}$, normalized by the solar luminosity ($L_\odot$).

For the aggregate stellar mass, we use the GAMA-derived $log_{10}\, M_{\star}/L_{\rm W1}$  relations from \citep{Clu14}, which include a W1$-$W2 color dependence.   Here $L_{\rm W1_{\rm Sun}}$ represents the luminosity relative to the Sun, in W1, also referred to as the ``in-band" luminosity \citep{Jar13}.  It is important not to confuse the spectral luminosity, $\nu L_{\nu}$, used in the SFR relations, with the in-band luminosity which is the convention for stellar mass relations.

In Table 1, we present the \wise measurements of 76 galaxies in the combined SINGS/KINGFISH sample, i.e. all except HoI, HoII and HoIX. These dwarf, irregular galaxies are too faint and diffuse to measure reliably with {\it WISE}.  The table features the W1 and W2 1$\sigma$ isophotal mags (indicated by a flag value $=$ 0) for most galaxies indicating that they are resolved, or in some isolated cases, the point-source mags (flag value $=$ 1);   and for the W3 and W4 fluxes, the ``total" mags  (flag value $=$ 10) are shown when radial surface brightness solutions were possible.  The brightest galaxy in the mid-infrared bands is the starburst galaxy, M82 (NGC3034), and the faintest detected is the dwarf, M81DwB.

\begin{figure*}[!t]
\plotone{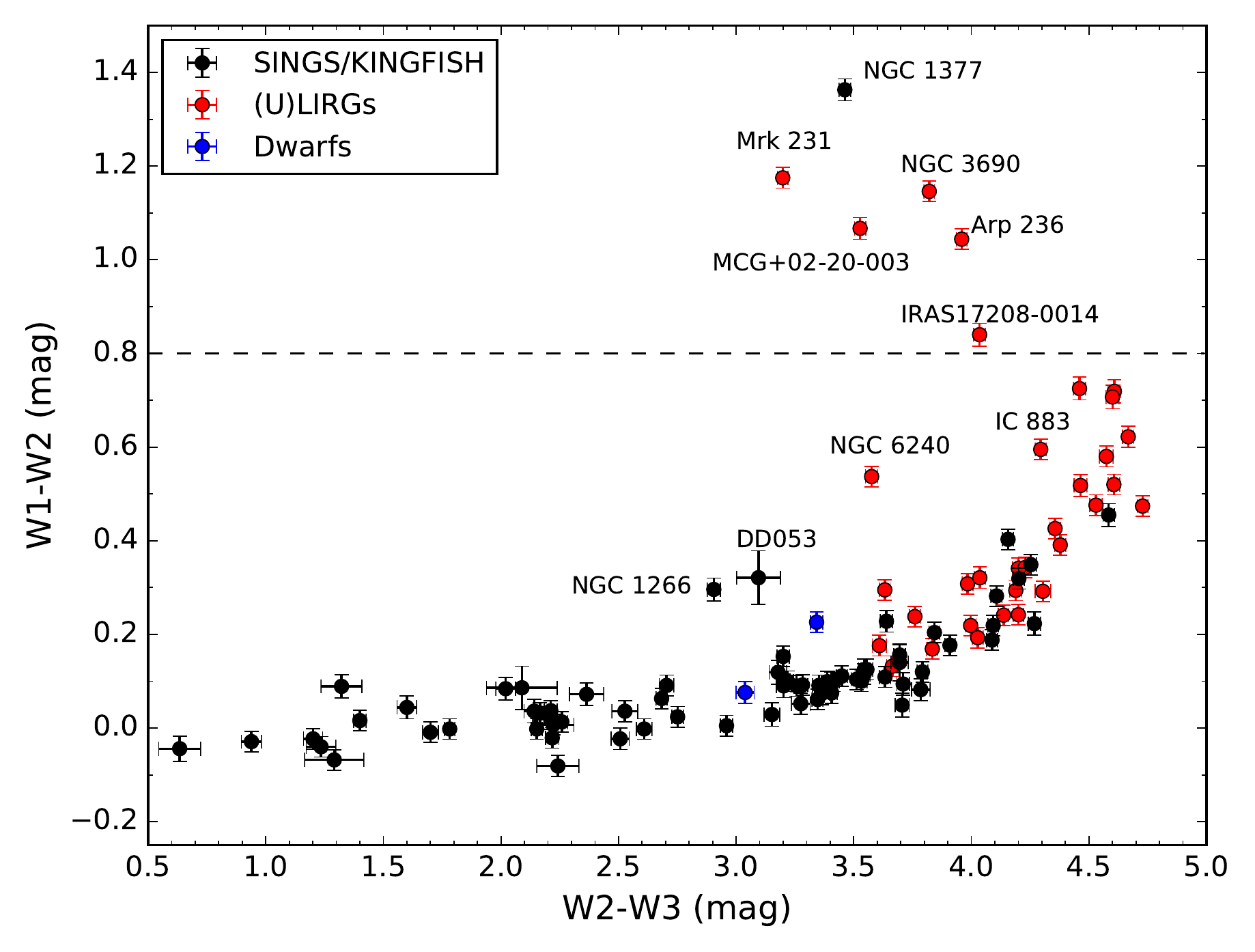}       
\caption{\wise color-color diagram of the SINGS/KINGFISH sample, as well as additional dwarfs and (U)LIRGs from the literature. Early-type galaxies with little star formation are expected to lie in the bottom left of this diagram (W2$-$W3 $<2$), while star-forming disk galaxies occupy the right-hand side of the diagram (W2$-$W3 $>3.5$). The dashed line indicates the threshold of the \citet{Stern12} AGN color space; above this line, the mid-infrared emission is dominated by the heating from dusty AGN. Star-forming galaxies with little hot dust (as traced by W1$-$W2) form a tight sequence in \wise color-color space. Galaxies lying above this trend, as well as those within the AGN color space, are labeled. \label{fig:f1}}               
\end{figure*}

Table 2 contains the rest-frame derived properties of the SINGS/KINGFISH sample and includes \Ltir determined using Equation 5 of \citet{Dale14} i.e. combining 8\micron, 24\micron, 70\micron, and 160\micron, and the most recent SINGS/KINGFISH photometry \citep{Dale17}.  Here the rest frame fluxes are computed using a set of galaxy templates \citep{Br14a} and SED fitting; see \citet{Jar17} for more details.  W3$_{\rm PAH}$ and W4$_{\rm dust}$ represent the ISM emission in the W3 and W4 bands (with the stellar emission subtracted).  Spectral luminosities, used for SFRs,  are indicated by $\nu L_{band}$;  and the W1 in-band luminosity, $L_{\rm W1_{\rm Sun}}$, is used for the stellar mass estimation.  The most luminous galaxy in our SINGS/KINGFISH sample is the peculiar NGC 2146, which is indeed classified as a luminous infrared galaxy (LIRG).

\subsection{Additional Galaxies}

We supplement our sample with several dwarf galaxies and (Ultra) Luminous Infrared Galaxies, (U)LIRGs, drawn from {\it IRAS} studies (see Table 4 references), to better explore the behavior at the extremes of star-forming systems. Table 3 lists thirty-two (U)LIRGS and their \wise measurements. The \Ltir values from the literature, restframe and derived properties are given in Table 4.  A mix of star-forming, AGN, and hybrids thereof comprise this sample of (U)LIRGS. The requirement of dwarf galaxies to be well-measured in \wise, and have reliable \Ltir from the literature, limits the sample to the two listed in Table 5, with their derived quantities given in Table 6. 

\section{Results}

\subsection{\wise colors}

The \wise color-color diagram, shown in Figure \ref{fig:f1}, is a useful diagnostic tool for determining the underlying characteristics of a given sample (Wright et al. 2010; Jarrett et al. 2011).  As in Figure 11 of \citet{Jar17}, the W2$-$W3 color can be used as a broad proxy for morphology: galaxies with a color $<2$ are typically spheroids (with little star formation), while star-forming disks usually have a color $>3.5$. On the other hand, the W1$-$W2 color acts as a proxy for AGN activity; elevated W1$-$W2 colors compared to the intrinsically tight trend typically indicate the presence of an AGN causing an excess of hot dust emission \citep{Stern12}. In the case of our sample, the colors clearly define a star-forming sequence from dust-free (early types) to active star-formers (late-types).  

Deviations arise from extreme activity, either star formation or AGN, or a combination of both (ULIRGs stand out in this way).  For example, NGC 1266 is known to harbor an AGN \citep{Al11} and this is likely why it sits slightly above the general trend in color space. Pointing to a separate issue, the elevated color of the dwarf galaxy DD053 is likely the result of contamination from a background source, located at 08h43m07.1s +66\arcdeg 10\arcmin 57\arcsec, which has the characteristic color of an AGN. The source of its W1$-$W2 color is therefore ambiguous. 

\begin{figure*}[!t]
\gridline{\leftfig{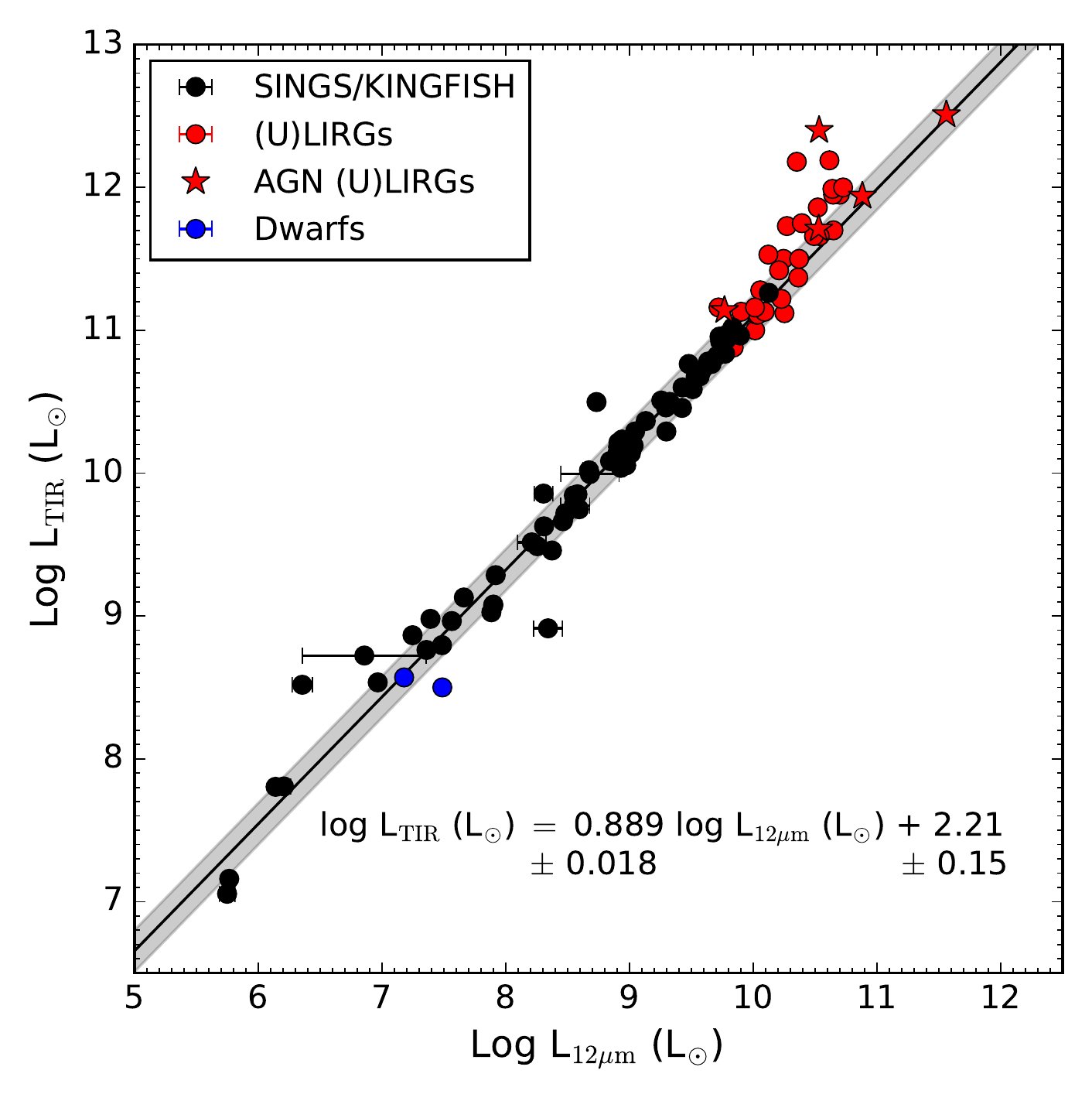}{0.5\textwidth}{(a) W3} 
              \rightfig{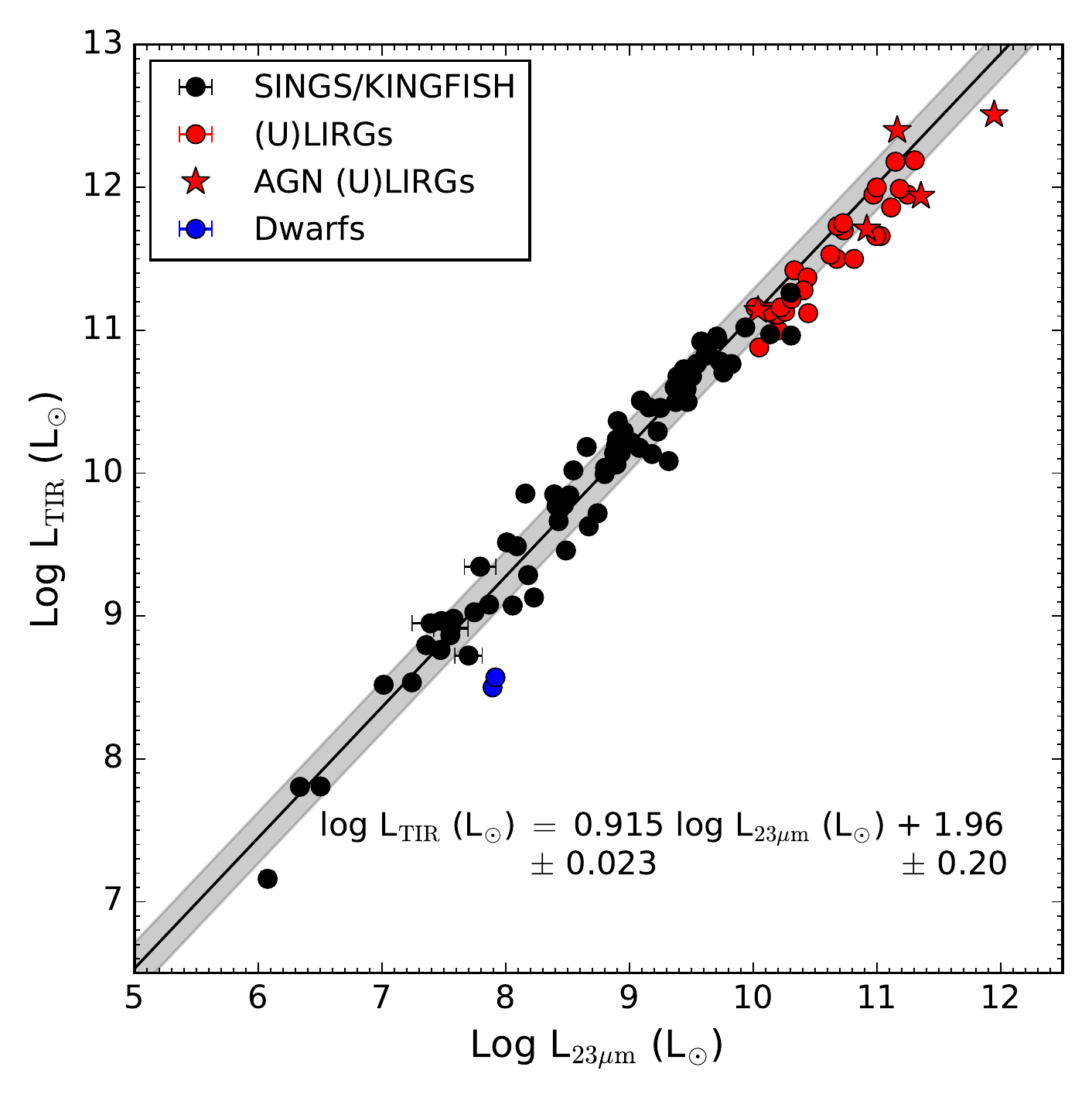}{0.5\textwidth}{(b) W4}}        
\caption{\Ltir as a function of a) $L_{12\micron}$ and b) $L_{23\micron}$ for the SINGS/KINGFISH sample, as well as the added (U)LIRG and dwarf sample (see text). For most points, the error on $L_{12\micron}$ is smaller than the data point. The fit to the SINGS/KINGFISH sample is shown as the solid black line with the 1-$\sigma$ scatter indicated by the shaded region (0.15 for $L_{12\micron}$ and 0.18 for $L_{23\micron}$). \label{fig:f2}}               
\end{figure*}

\begin{figure*}[!t]
\gridline{\leftfig{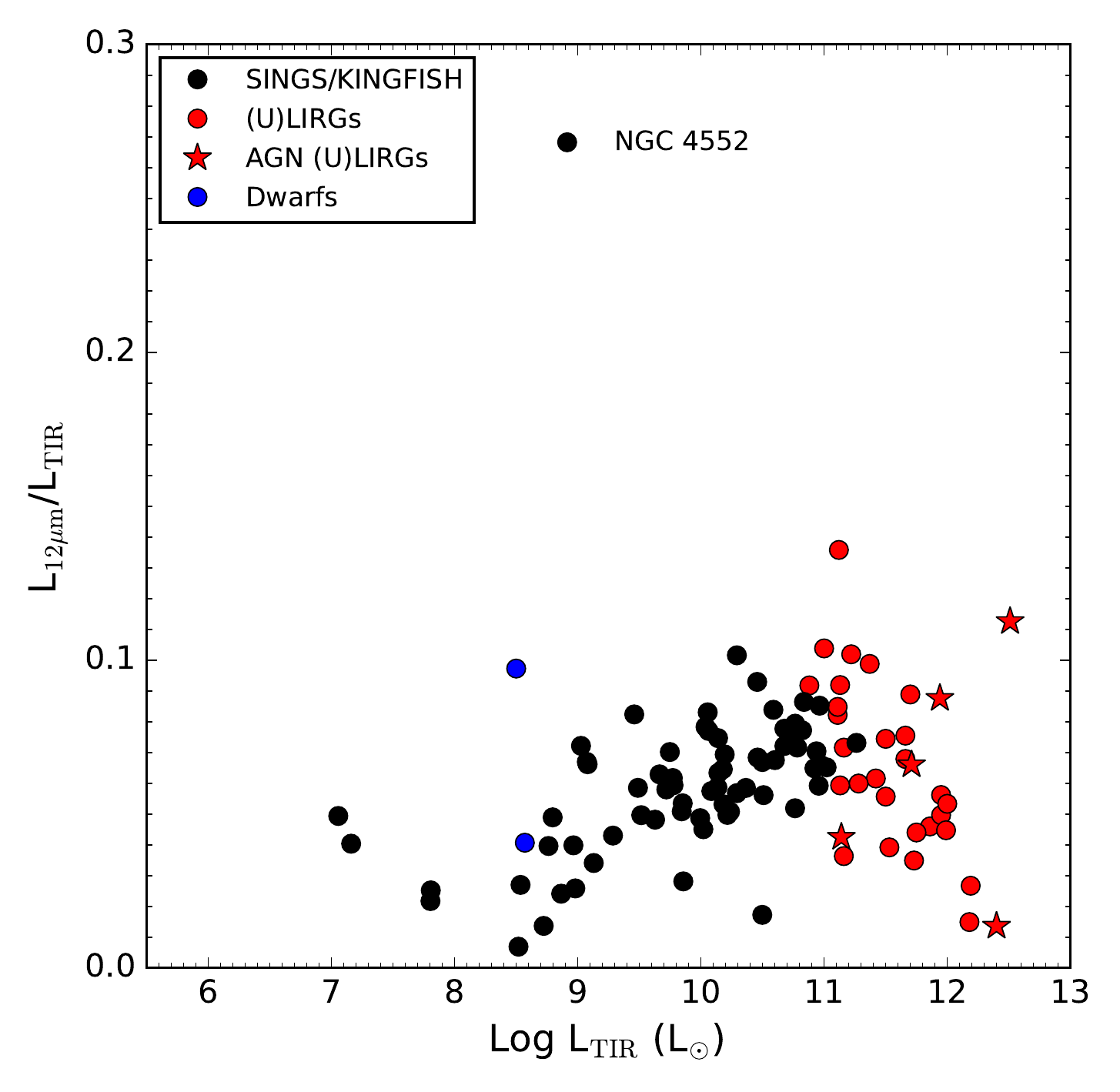}{0.5\textwidth}{(a) W3} 
              \rightfig{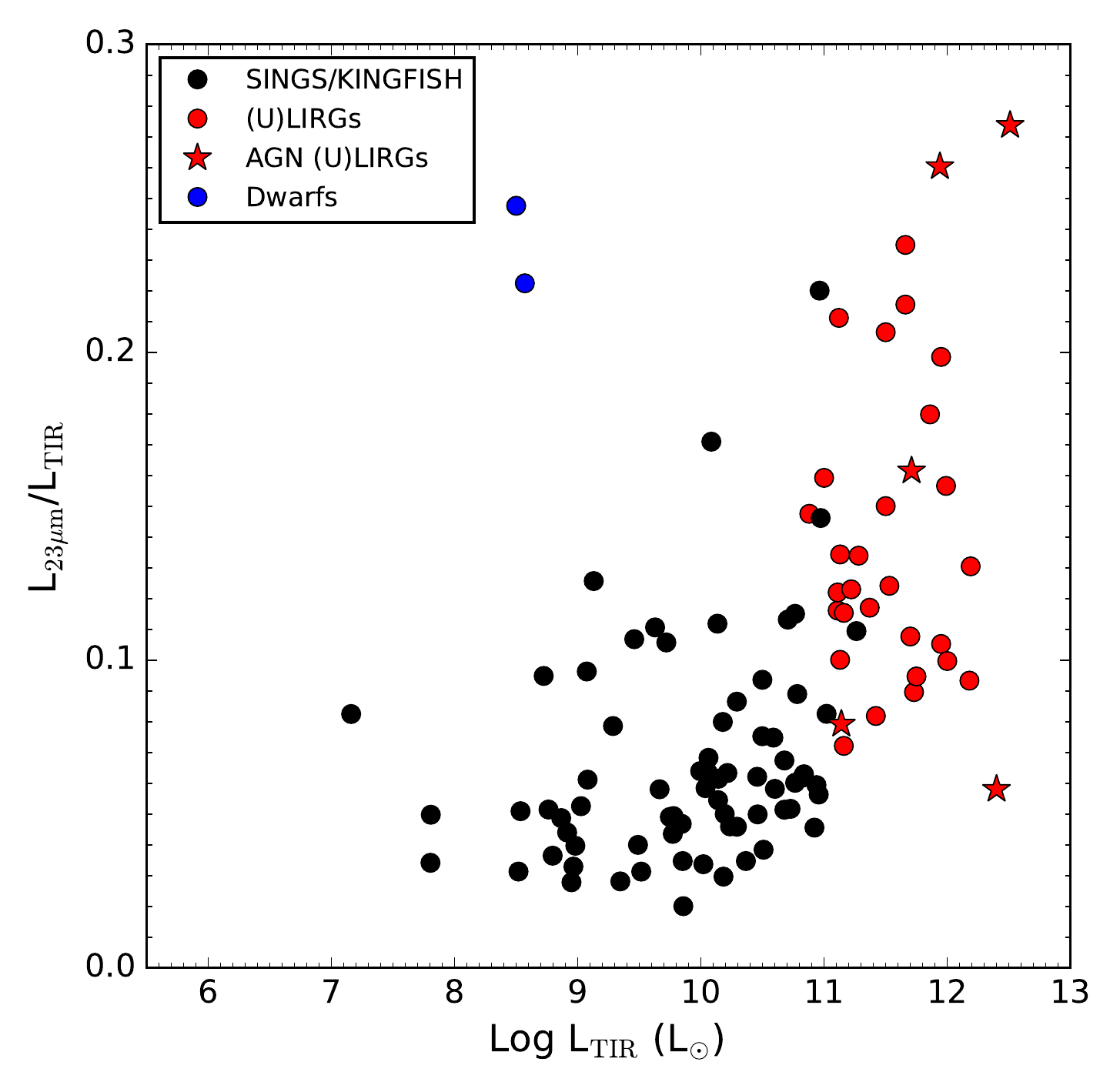}{0.5\textwidth}{(b) W4}}        
\caption{The ratio of a) $L_{12\micron}/L_{\rm TIR}$ and b) $L_{23\micron}/L_{\rm TIR}$ as a function of $L_{\rm TIR}$ shows the variation of the contribution from a) W3, and b) W4 as the total infrared luminosity increases. The W3 luminosity shows a more constant response compared to the W4 luminosity\label{fig:f2b}}               
\end{figure*}

An interesting case, the massive lenticular NGC 1377 is the only galaxy in the SINGS/KINGFISH sample with W1$-$W2 $>0.8$ indicating a global color dominated by emission from the dusty torus of an AGN.  
The W1$-$W2 color is itself not a definitive discriminator of the presence or absence of an AGN \citep{Gur14}, but elevated W1$-$W2 colors indicate excess hot dust emission from an AGN with W1$-$W2 $>0.8$ implying global galaxy colors dominated by this emission. 
We would therefore expect the W3 and W4 luminosity, sensitive to star formation, to be contaminated by AGN emission. 

NGC 1377 is an extreme far-infrared excess galaxy \citep{Rous03} and has to date been considered to be a young, dusty starburst \citep{Rous06} due to being an extreme (low) outlier in the radio-infrared correlation. However, recent ALMA, JVLA and {\it Chandra} observations suggest it is harbouring an extremely obscured AGN with a molecular jet \citep{Aal16,Cos16}.  Moreover, {\it Spitzer} IRS spectroscopy shows a steep, strong continuum with the largest silicate opacity of any SINGS source \citep{Smith07}. 
In this instance, the \wise colors of the integrated emission supports the scenario of an obscured AGN. 

The underlying heating source in cases such as these, where mid-infrared diagnostic emission lines are swamped by the continuum, will be ambiguous in the absence of other diagnostics -- in this case \wise colors.  For this reason, NGC 1377 is excluded from the star formation calibration determination in the next Section.

Unsurprisingly, several local (U)LIRGs show evidence of  AGN-dominated (e.g. Mrk 231) or AGN-contaminated (e.g. NGC 6240) colors. For the remainder of this study we refer to the AGN-dominated \wise color (U)LIRG systems, i.e. with W1$-$W2 colors above the \citet{Stern12} line, as ``AGN (U)LIRGs", with the caveat that we cannot determine if extreme star-formation or AGN-heating is the dominant emission mechanism for these sources.

\begin{figure*}[!t]
\plotone{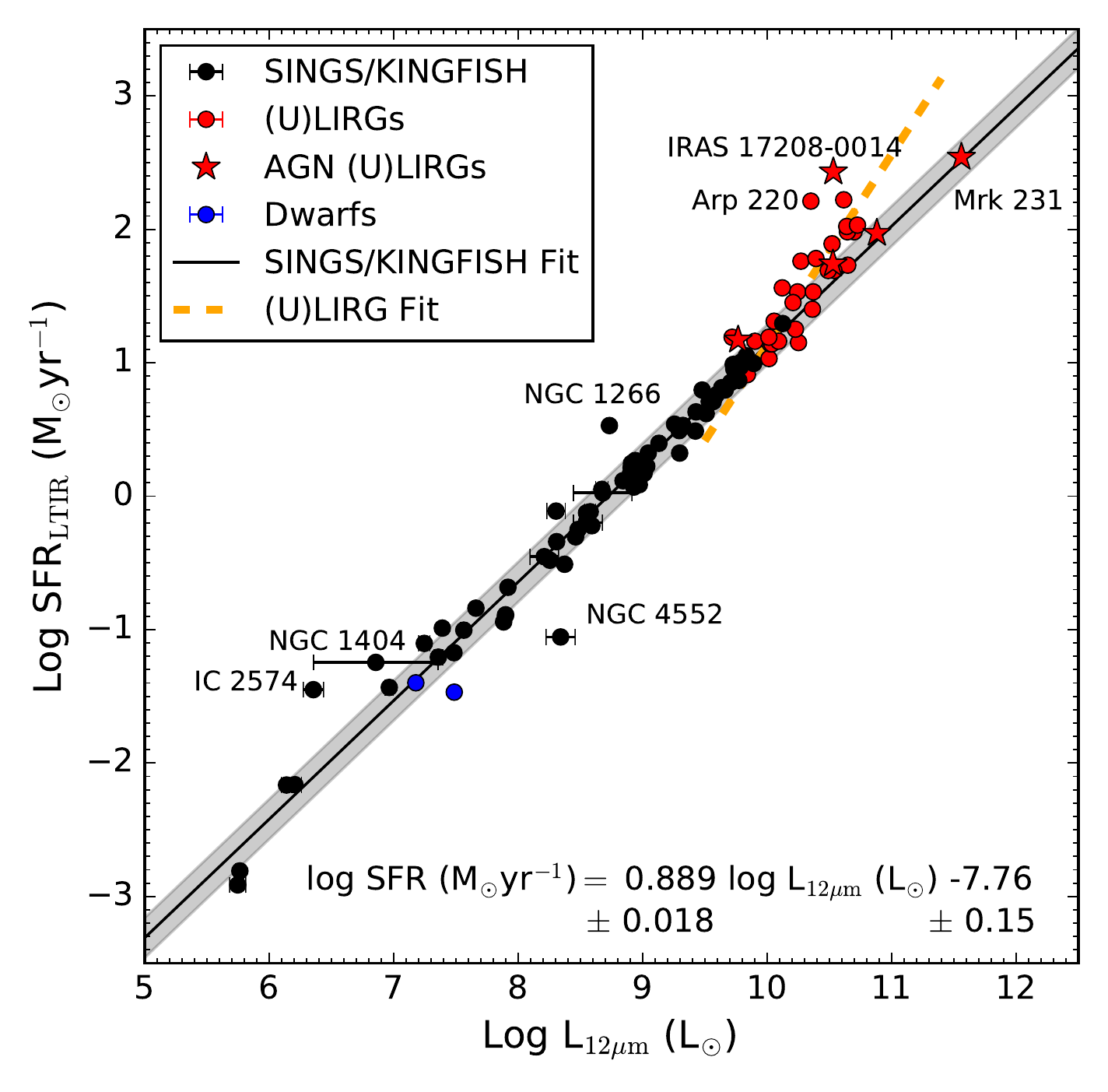}       
\caption{\Ltir-derived star formation rates are plotted against the \wise W3 12\micron\ luminosity and the best fit line to the SINGS/KINGFISH sample is shown (solid line); the 1-$\sigma$ scatter is indicated by the shaded region (0.15 dex). For comparison, a sample of (U)LIRGs and dwarf galaxies are also shown. A fit to the (U)LIRG sample is shown by the dashed line and is given by: $ {\rm log\, SFR}\ (\rm M_{\odot}\, {\rm yr}^{-1}) =  (1.430\pm 0.161)\, {\rm log}\, L_{12\micron} (L_\odot) - (13.17\pm1.66)$ with a 1-$\sigma$ scatter of 0.22 dex. The trend suggests that \Ltir gives a higher SFR compared to the W3-derived value for high luminosity sources. \label{fig:f3}}               
\end{figure*}


\subsection{\wise Star Formation Rates: Relations and Performance}\label{SFR}

In this Section we explore the behavior of the \wise W3 (12\micron) and W4 (23\micron) luminosities as a measure of star formation for the SINGS/KINGFISH sample. Figure \ref{fig:f2} shows the \Ltir values plotted as a function of the $L_{12\micron}$ and $L_{23\micron}$ luminosities, respectively (here and throughout, $L_{12\micron}$ and $L_{23\micron}$ refer to $\nu L_{\nu}(12\, \mu \rm m)$ and $\nu L_{\nu}(23\, \mu \rm m)$, respectively). The location of the (U)LIRG sample and additionally measured dwarf galaxies is also shown. For the  \Ltir vs $L_{12\micron}$ diagram, a tight linear trend is observed, even down to low luminosities ($\sim 10^{6} L_\odot$). A best fit is given by:

\begin{multline}
{\rm log}\, L_{\rm TIR} (L_\odot)  (\rm M_{\odot}\, {\rm yr}^{-1}) = \\ (0.889\pm0.018)\, {\rm log}\, L_{12\micron} (L_\odot) + (2.21\pm 0.15), \\
\end{multline}
with a 1-$\sigma$ scatter of 0.15 dex.

At the high luminosity ($> 10^{10.5}  L_\odot$) end, however, the (U)LIRG sample tends towards higher values of $L_{\rm TIR}$, i.e. the W3 band is under-luminous. This suggests additional heating not being traced by the W3 band, or alternatively, the effect of strong silicate absorption in the W3 band.  Curiously, the most luminous source in the entire ensemble, Mrk 231, a dusty and gas-rich broad-line QSO, falls exactly on the relation derived from star-forming galaxies several orders of magnitude less luminous.   

In contrast to the W3 band, comparison of the $L_{23\micron}$ luminosities to \Ltir shows increased scatter, and the opposite behavior with extreme galaxies, i.e. the $L_{23\micron}$ luminosity is over-luminous compared to the \Ltir values for (U)LIRGs. This suggests that the W4 band has enhanced continuum emission relative to the far-infrared, likely boosted by dust-obscured AGN. The best fit relation for the SINGS/KINGFISH sample is: 

\begin{multline}
{\rm log}\, L_{\rm TIR} (L_\odot)  (\rm M_{\odot}\, {\rm yr}^{-1}) = \\ (0.915\pm0.023)\, {\rm log}\, L_{23\micron} (L_\odot) + (1.96\pm 0.20), \\
\end{multline}
with a 1-$\sigma$ scatter of 0.18 dex.

Figure \ref{fig:f2b} examines the ratio of the \wise bands to \Ltir as a function of increasing $L_{\rm TIR}$. In Figure \ref{fig:f2b}a, NGC 4552 (Messier 89) is a clear outlier; it is an elliptical galaxy in the Virgo Cluster with little or no star formation \citep[e.g.][]{Shap10}, but with an infrared excess at 12\micron\ relative to a dust-free elliptical \citep{Br14a}, detected by {\it Herschel}  \citep{Ser13}, and containing a ``mini-AGN" \citep{Capp99}. More distant galaxies of this type would not have a W3 detection. 

Comparing Figure \ref{fig:f2b}a and b, the W3 band shows a more constant response compared to W4, where higher \Ltir produces correspondingly more 23\micron\ emission. For W3, we see the (U)LIRG population turning over, unlike what is observed for W4 where the $L_{23\micron}/L_{\rm TIR}$ ratio appears to {\it increase}. This could be the result of silicate absorption in W3 preventing the continuum at 12\micron\ from gaining power relative to $L_{\rm TIR}$.

Figure \ref{fig:f2b}b suggests that for \Ltir $>10^{11} L_{\odot}$, the fractional power traced by W4 (the heating of small dust grains) compared to \Ltir increases rapidly.

\begin{figure*}[!t]
\plotone{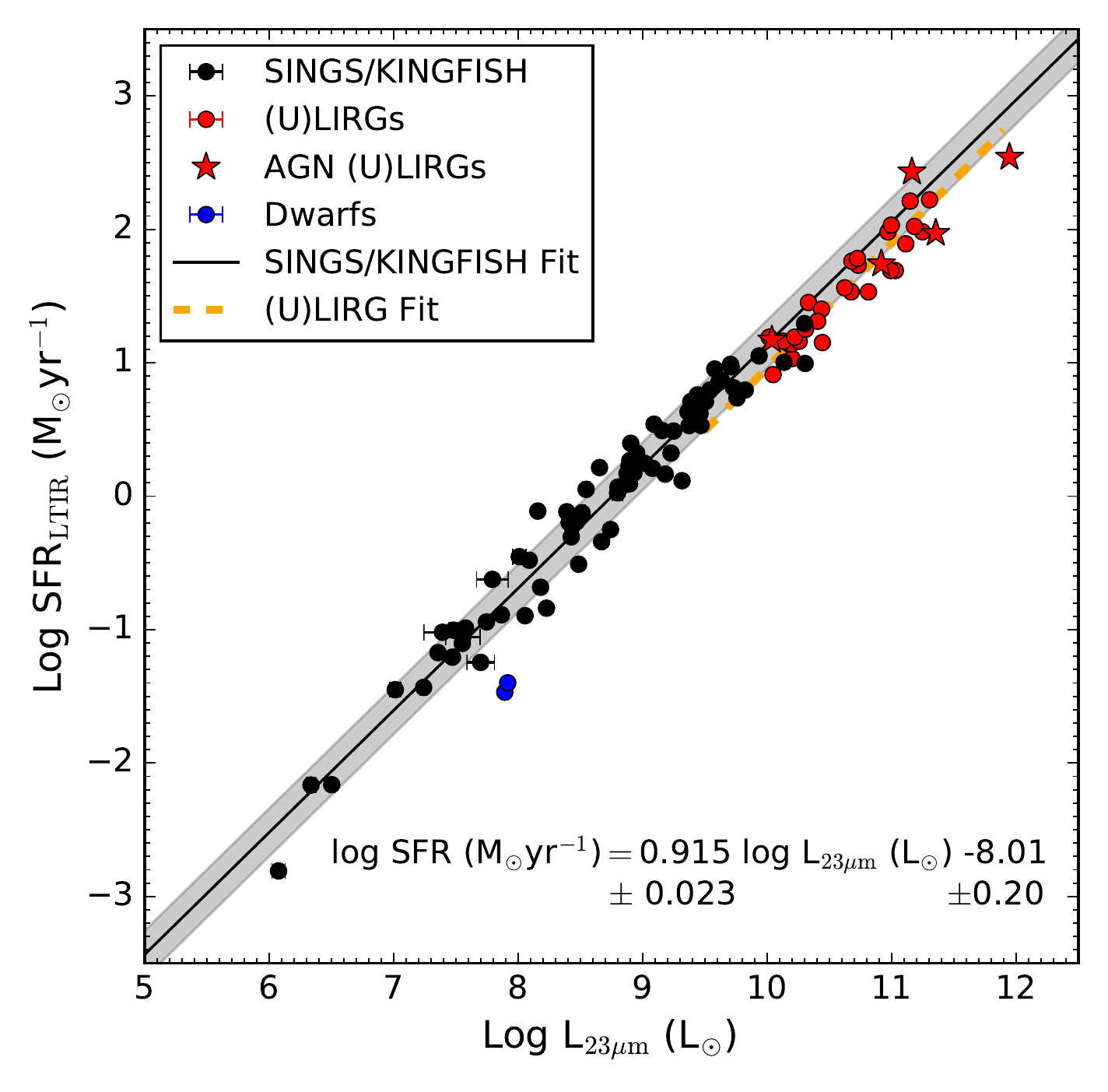}       
\caption{\Ltir-derived star formation rates  are plotted as a function of \wise W4 23\micron\ luminosity. The best fit line to the SINGS/KINGFISH sample is shown as a solid line and has a 1-$\sigma$ scatter of 0.18 dex (indicated by the shaded region). A  fit to the (U)LIRG sample (dashed line) is given by:
$ {\rm log\, SFR}\ (\rm M_{\odot}\, {\rm yr}^{-1}) =  (0.938\pm 0.062)\, {\rm log}\, L_{23\micron} (L_\odot) - (8.42\pm0.66)$ with a 1$\sigma$ scatter of 0.13 dex. This suggest that W4 is marginally overrestimating the SFR compared to \Ltir at high luminosity. \label{fig:f4}}               
\end{figure*}

In the next step, we calibrate the W3 and W4 spectral luminosities of the SINGS/KINGFISH sample to star formation rates (SFRs) determined from the total infrared luminosity, $L_{\rm TIR}$. The SFR is determined using Equation 1.3 in \citet{Cal13}, which uses the Starburst99 \citep{Star99} models assuming solar metallicity, constant star formation over $\tau= 100$Myr, and a Kroupa IMF, as follows: 

\begin{equation}
{\rm SFR} (M_{\odot}\, {\rm yr}^{-1}) = 2.8 \times 10^{-44}\, L_{\rm TIR} ({\rm erg.s}^{-1}) \\
\end{equation}

We note that this calibration assumes that the stellar emission, most notably the UV component, is entirely absorbed and re-radiated at infrared wavelengths, which constitutes an upper limit to what would occur in a real galaxy. Comparison of this relation to that of \citet{Murph11}, who also use the Starburst99 models, yields a negligible difference. 

In Figure \ref{fig:f3} we relate the SFR derived from \Ltir (using Equation 1) to the $L_{12\micron}$ luminosity and show the best fit to the SINGS/KINGFISH sample. The resulting fit is given by:

\begin{multline}
 {\rm log\, SFR}\ (\rm M_{\odot}\, {\rm yr}^{-1}) = \\ (0.889\pm 0.018)\, {\rm log}\, L_{12\micron} (L_\odot) - (7.76\pm 0.15), \\
\end{multline}
with a 1-$\sigma$ scatter of 0.15 dex.

The overall trend is tight, with the exception of the outliers: NGC 1266, NGC 4552, and IC 2574. In the case of NGC 1266, the AGN activity is likely producing excess dust heating compared to what is traced in W3. The offset location of the dwarf irregular, IC 2574, could be due to the strong variations in dust temperature and characteristics found across the galaxy due to triggered star formation \citep{Can05}. 

The (U)LIRGs overplotted on Figure \ref{fig:f3} show a trend to higher SFRs than what is seen for the fit to the SINGS/KINGFISH sample. A best fit to only the (U)LIRG sample is given by:

\begin{multline}
 {\rm log\, SFR}\ (\rm M_{\odot}\, {\rm yr}^{-1}) = \\ (1.430\pm 0.161)\, {\rm log}\, L_{12\micron} (L_\odot) - (13.17\pm 1.66), \\
\end{multline}
with a 1-$\sigma$ scatter of 0.22 dex. 

This suggests that the 12\micron\ luminosity is underestimated compared to \Ltir (see also Figure \ref{fig:f2}a) for the (U)LIRGs. The W3 band is sensitive to the silicate absorption feature expected to be common in the dusty, embedded starbursts powering the infrared emission of the (U)LIRGs and this likely diminishes the $L_{12\micron}$ compared to $L_{\rm TIR}$. 

Although the center of the W3 band is close to the 11.3\micron\ PAH feature, the breadth of the band samples several features, as well as the continuum from warm, large grains and stochastically heated grains. On average, the total PAH emission in the band only accounts for $\sim 34\%$ of the 12\micron\ flux (see Appendix A).  Although associated with star formation, the strong radiation fields associated with nuclear starbursts and/or AGN would likely suppress PAH emission close in, where molecules are less shielded than they would be in a star-forming disk, however the hot dust would boost the continuum being traced by W3. Therefore the advantage of the 12\micron\ {\it WISE} band as a SFR indicator appears to be its breadth thus sampling a mix of PAHs, nebular lines and continuum.

In Figure \ref{fig:f4}, a fit is derived between the \Ltir-derived SFR and the $L_{23\micron}$ luminosities of the SINGS/KINGFISH sample. The resulting fit is given by:

\begin{multline}
{\rm log\, SFR}\ (\rm M_{\odot}\, {\rm yr}^{-1}) = \\ (0.915\pm0.023)\, {\rm log}\, L_{23\micron} (L_\odot) - (8.01\pm 0.20), \\
\end{multline}
with a 1-$\sigma$ scatter of 0.18 dex.

In the case of 23\micron\ luminosity, a fit to the (U)LIRGs is given by:
\begin{multline}
 {\rm log\, SFR}\ (\rm M_{\odot}\, {\rm yr}^{-1}) = \\ (0.938\pm 0.062)\, {\rm log}\, L_{23\micron} (L_\odot) - (8.42\pm0.66),\\
\end{multline}
with a 1-$\sigma$ scatter of 0.13 dex. This indicates a marginal trend to lower SFRs than what is seen for the fit to the SINGS/KINGFISH sample, possibly due to increased 23\micron\ emission due to obscured AGN activity or embedded star formation. 

Comparing the fits of Figure \ref{fig:f3} and \ref{fig:f4}, the 12\micron\ relation appears to have less scatter and hold over 5 orders of magnitude of 12\micron\ luminosity. 
The W4 band, which is dominated by the warm dust continuum, shows increased scatter and appears more curved (or at least, a strong break) at the extreme (LIRG) end, which agrees with Figure \ref{fig:f2b}b.

\begin{figure}[!t]
\plotone{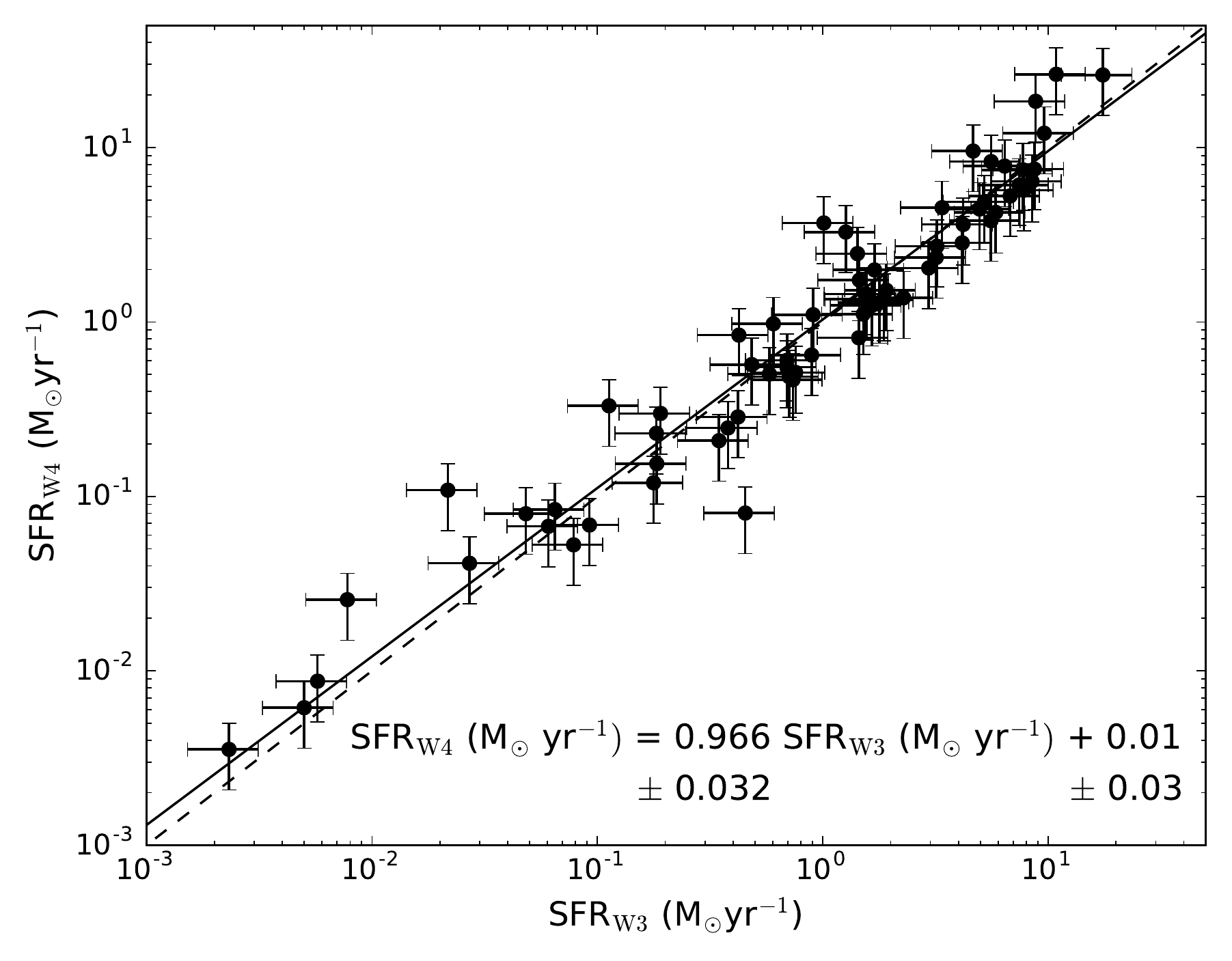}       
\caption{W3-derived SFR compared to the W4-derived SFR for the SINGS/KINGFISH sample indicates that overall the mid-infrared SFRs are consistent. A fit to the points is given by the solid line (with 1-$\sigma$ scatter of 0.0001 dex) and a one-to-one (dashed) shown for comparison. \label{fig:f5}}               
\end{figure}

In Figure \ref{fig:f5} we compare the SFRs for the SINGS/KINGFISH sample determined using Equations 2 and 3, respectively. Although the points show some scatter, overall they are consistent with respect to the one-to-one relation over several orders of magnitude of SFR. A fit to the points is given by:

\begin{multline}
 {\rm SFR}_{\rm W4}\ (\rm M_{\odot}\, {\rm yr}^{-1}) = \\ (0.966\pm 0.032)\,  {\rm SFR}_{\rm W3}\ (\rm M_{\odot}\, {\rm yr}^{-1}) + (0.01\pm0.03),\\
\end{multline}
with a 1-$\sigma$ scatter of 0.0001 dex.

\begin{figure*}
\gridline{\leftfig{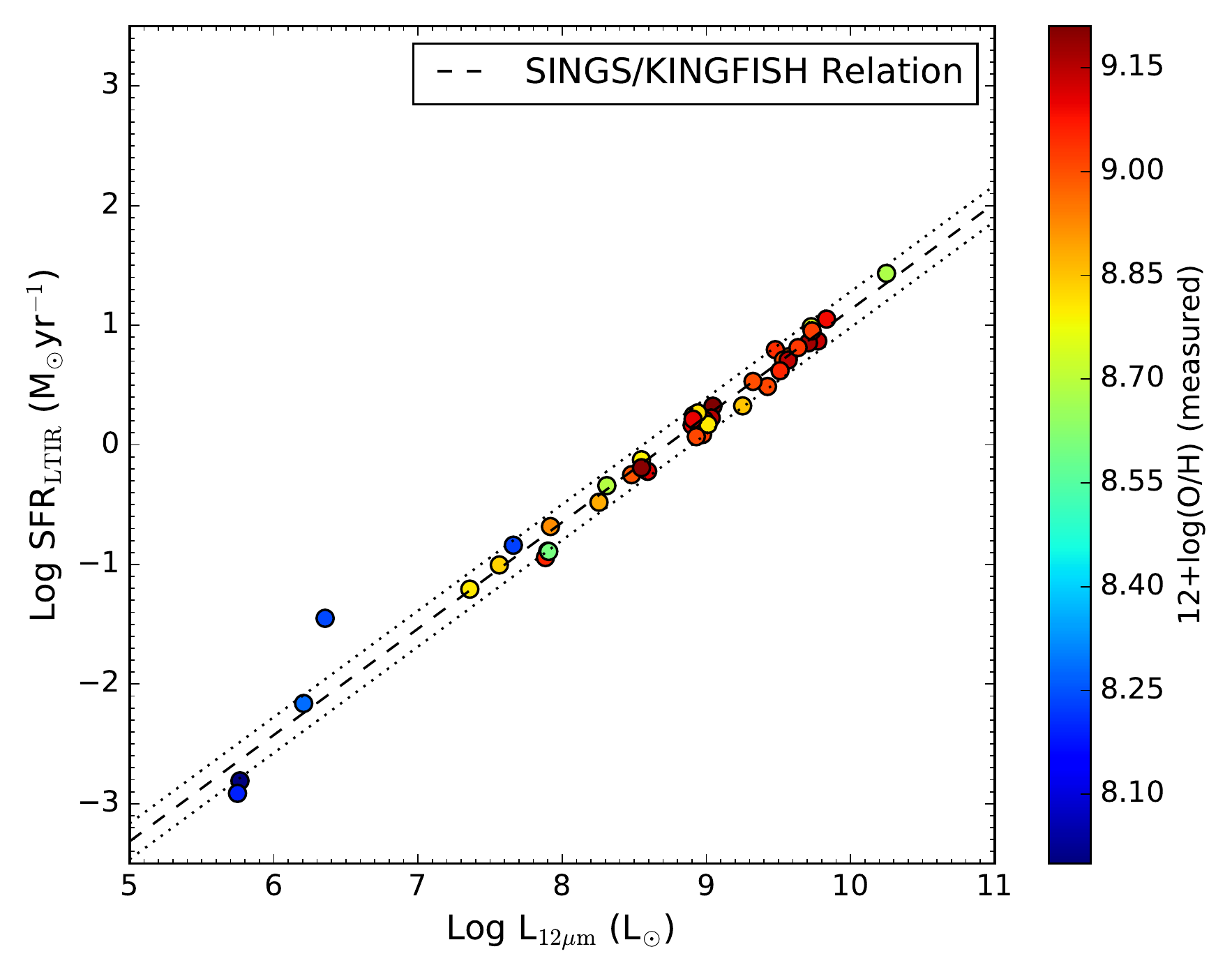}{0.5\textwidth}{(a)} 
              \rightfig{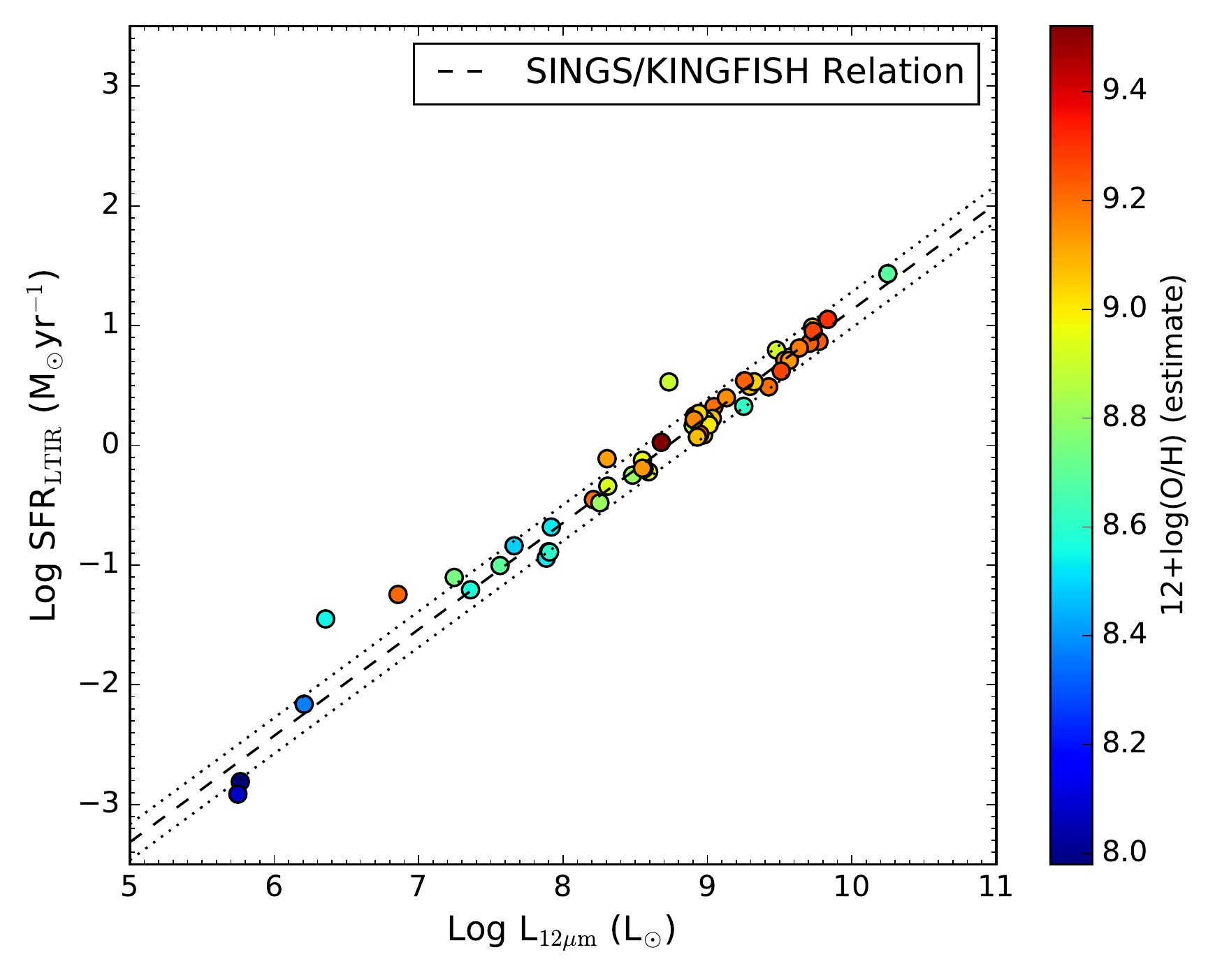}{0.5\textwidth}{(b)}}        
\caption{SINGS/KINGFISH galaxies color-coded by metallicity: in a) showing abundances determined using characteristic (global) abundances from \citet{Mous10} using the \citet{KK04} calibration, and b) metallicity estimate from \citet{Mous10} derived from the B-band luminosity-metallicity relation, using the \citet{KK04} calibration. The dotted lines show the 1-$\sigma$ scatter around the SFR relation (dashed line); the W3 band ($L_{12\micron}$) does not appear sensitive to the metallicity of the galaxy.  \label{fig:f6}}               
\end{figure*}

Finally, we explore the effect of metallicity on, in particular, the $L_{12\micron}$ luminosity relation. Since PAH features can be affected by the strong radiation fields associated with low metallicity systems \citep{Smith07} and given the contribution of PAHs to the W3 band, we color code the galaxies in the SINGS/KINGFISH sample according to their oxygen [O/H] metallicities, as given by \citet{Mous10}. In Figure \ref{fig:f6}a the abundances are calculated (where possible) using the \citet{KK04} calibration, whereas in \ref{fig:f7}b, the metallicity is estimated from the B-band luminosity (and is therefore available for all galaxies in the SINGS/KINGFISH sample). From Figure \ref{fig:f6} it is evident that for the metallicities probed in this sample, there does not appear to be a systematic effect on the SINGS/KINGFISH $L_{12\micron}$ SFR relation. However, we do not expect this SFR calibration to hold for lower metallicity environments.

\subsection{Comparison to other \wise SFR relations}

\begin{figure*}
\gridline{\leftfig{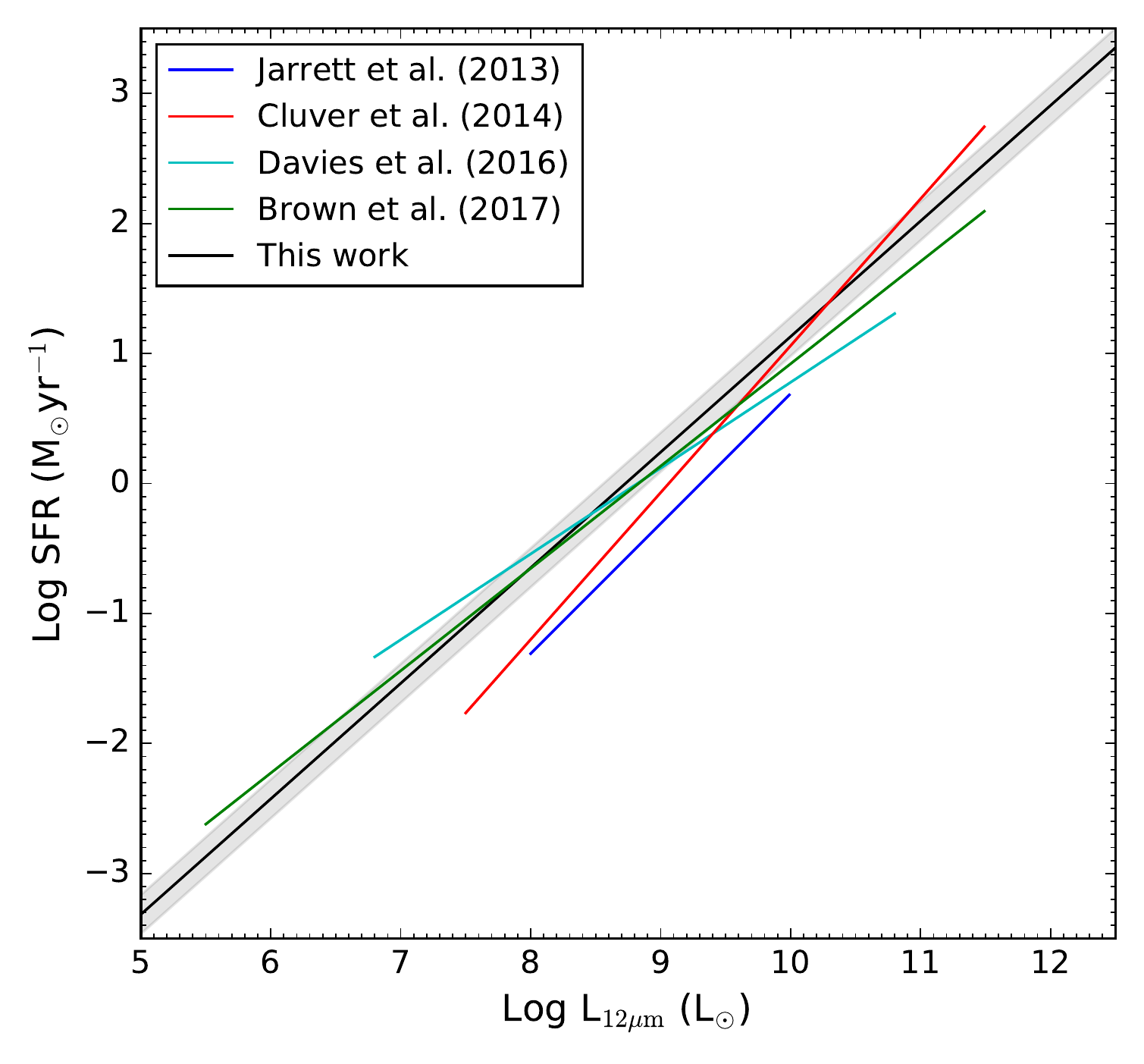}{0.5\textwidth}{(a)} 
              \rightfig{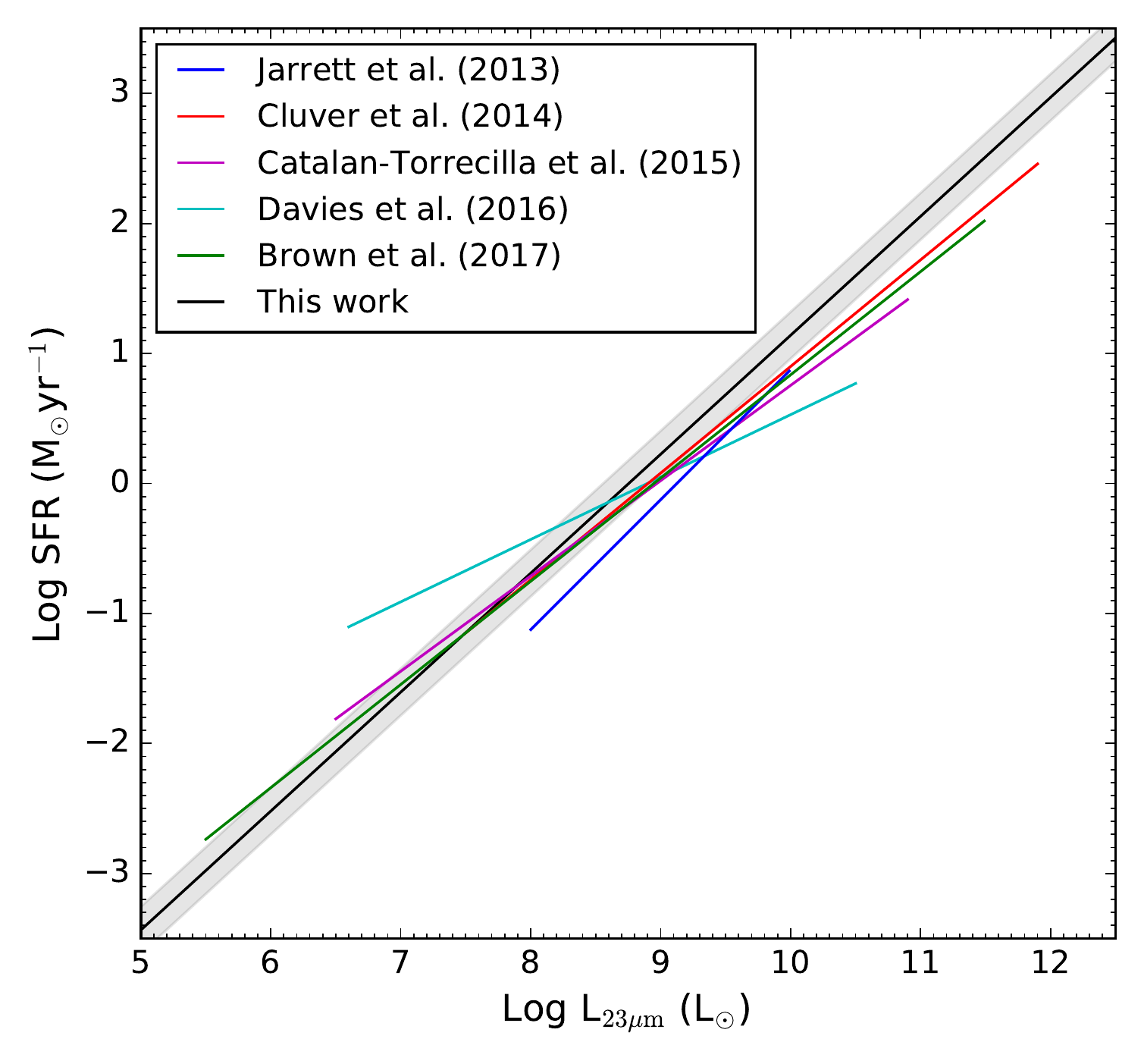}{0.5\textwidth}{(b)}}        
\caption{The a) $L_{12\micron}$ (W3) and b) $L_{23\micron}$ (W4) SFR relation from this work compared to existing relations from the literature using different calibrators and IMFs (see Table \ref{tab:fitscomp1} and Table \ref{tab:fitscomp2}). This plot illustrates the (often hidden) complexities in applying SFR relations that differ in methodology and sample selection. \label{fig:f12}}               
\end{figure*}

In Figure \ref{fig:f12} we compare the SFR relations derived in Section \ref{SFR} to those in the literature that similarly make use of \wise resolved source photometry. We list in Table \ref{tab:fitscomp1} and Table \ref{tab:fitscomp2}, the source of these relations for W3 and W4, respectively, the calibrators used and the adopted IMF. Apart from the \citet{Br17} relation where a conversion to a \citet{Kroup02} IMF has been applied, we have made no attempt to take into account differences in IMF and calibrator in order to illustrate the breadth of uncertainty in any given SFR, depending on the chosen relation. See \citet{Br17} for a broader listing of SFR relations from the literature.

The differences due to using either the W3 or W4 band are also apparent. For W4, relations calibrated against H$_\alpha$-derived SFRs \citep{Clu14, Cat15, Br17} appear to agree within the expected scatter. However, the W3 relations show little consistency, suggesting that the choice of sample and calibrator may cause significant variation.

\begin{deluxetable*}{lcll}
\tablenum{7}
\tablecaption{Comparison of W3 SFR Relations\label{tab:fitscomp1}}
\tablewidth{0pt}
\tablehead{
\colhead{Reference} & \colhead{Calibrator} & \colhead{SFR Conversion}  & \colhead{Adopted IMF}\\
}

\startdata
\citet{Jar13} & 24\micron\ & \citet{Riek09} & \citet{Riek93} \\
\citet{Clu14} & H$_\alpha$ & \citet{Wij11}  &  \citet{Bal03})\\
\citet{Dav16} & Radiative Transfer & \citet{Groot17}  &  \citet{Ch03}  \\
\citet{Br17} & H$_\alpha$ & \citet{Kenn09}  & \citet{Ch03}  \\
\enddata
\end{deluxetable*}

\begin{deluxetable*}{lcll}
\tablenum{8}
\tablecaption{Comparison of W4 SFR Relations\label{tab:fitscomp2}}
\tablewidth{0pt}
\tablehead{
\colhead{Reference} & \colhead{Calibrator} & \colhead{SFR Conversion} & \colhead{Adopted IMF}\\
}

\startdata
\citet{Jar13} & 24\micron\ & \citet{Riek09} & \citet{Riek93} \\
\citet{Clu14} & H$_\alpha$ & \citet{Wij11}  & \citet{Bal03}\\
\citet{Cat15}$^a$  & H$_\alpha$ & \citet{Kenn09} & \citet{Kroup02} \\
\citet{Dav16} & Radiative Transfer & \citet{Groot17} &  \citet{Ch03}  \\
\citet{Br17} & H$_\alpha$ & \citet{Kenn09}  & \citet{Ch03}  \\
\enddata
\tablecomments{$^a$ The log-scale fit is used here for comparison.}
\end{deluxetable*}

\subsection{Comparison to other wavelengths}

In this Section we compare the 12\micron- and 23\micron-derived SFRs, using Equations 2 and 3, respectively, to SFRs derived using H$\alpha+$24\micron, radio continuum and UV$+$IR measurements.

In \citet{Kenn09} the authors provide a hybrid calibration that combines observed H$\alpha$ and 24\micron\ luminosities as a proxy for dust attenuation-corrected H$\alpha$. We use the H$\alpha$ fluxes given in \citet{Kenn09} for the SINGS galaxies and combine them with the {\it Spitzer} MIPS 24\micron\ fluxes given by \citet{Dale17}. In addition, we use the H$\alpha$ spectro-photometric fluxes for 15 SINGS galaxies included in \citet{MK06}.  Employing the mid-infrared \citet{Kenn09} coefficient that corrects H$\alpha$ for attenuation, 
in Figure \ref{fig:fnew2}a we compare the combined H$\alpha$ and 24\micron\ luminosities to the 12\micron\ luminosities and find a linear fit for the H$\alpha$ sample given by:

\begin{multline}
{\rm log}\, (L_{\rm H\alpha}+ 0.02*L_{24 \mu m})\  {\rm erg}.{\rm s}^{-1}  = \\ (0.686\pm0.018)\, {\rm log}\, L_{12\mu m} ({\rm erg}.{\rm s}^{-1}) + (11.92\pm 1.52), \\
\end{multline}
with a 1-$\sigma$ scatter of 0.26 dex.
We note that although there is considerable scatter, the $WISE$ W3 luminosity is proportional to the H$\alpha$ luminosity over nearly 3 orders of magnitude. Comparison of this relation to that of \citet{Br17}, which compares $L_{12\micron}$ to that of Balmer Decrement extinction-corrected H$\alpha$, we find that our relation is marginally flatter, but broadly consistent.  

\begin{figure*}
\gridline{\leftfig{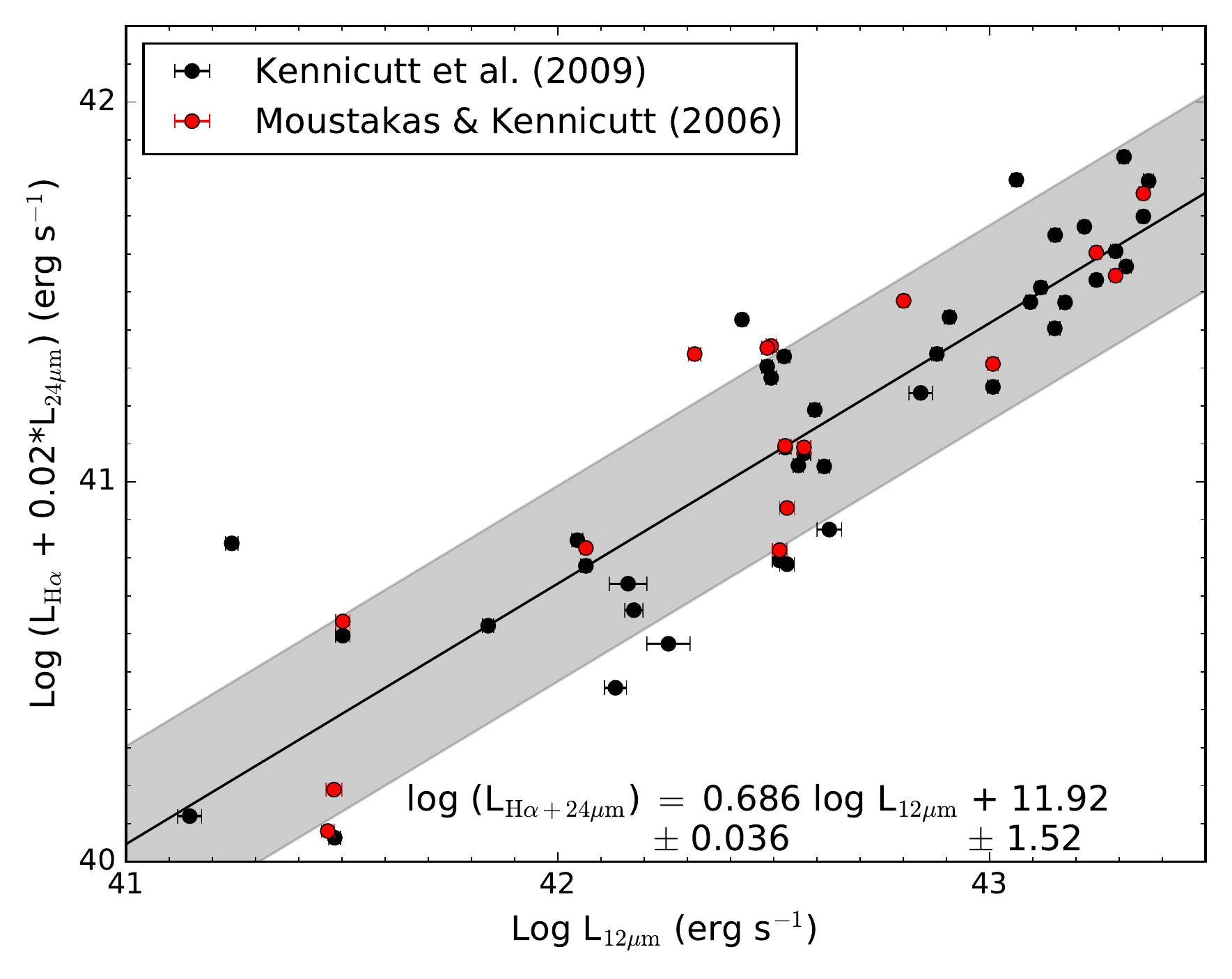}{0.5\textwidth}{(a) } 
              \rightfig{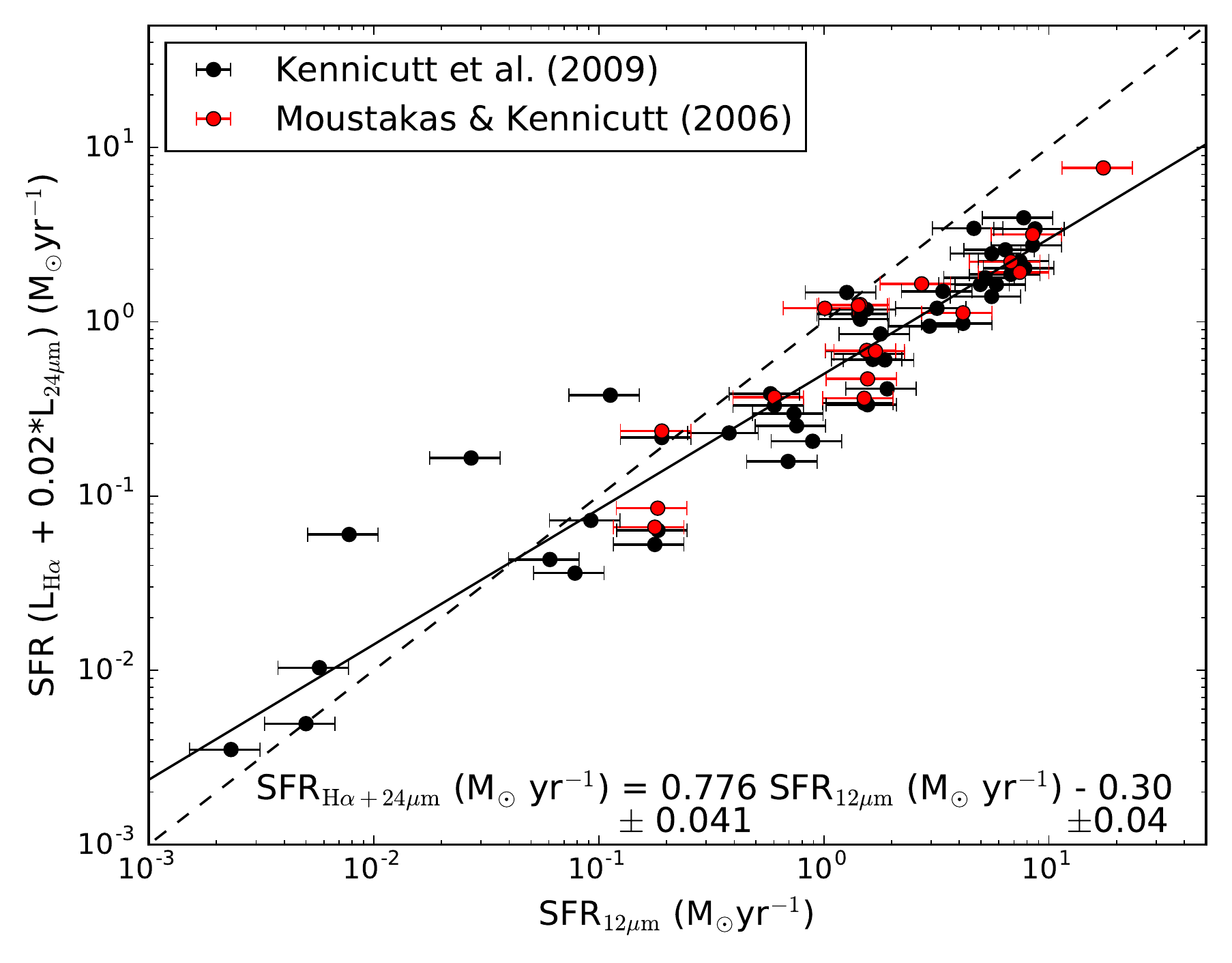}{0.5\textwidth}{(b) }}        
\caption{Comparing H$\alpha + 24 \mu$m luminosities to $L_{12\micron}$ in (a) shows a linear trend for galaxies using H$\alpha$ measurements from \citet{Kenn09} (black points) with a 1-$\sigma$ scatter of 0.26 dex. Using galaxies where the H$\alpha$ measurements are drawn from \citet{MK06} (red points) follows a similar distribution. In (b) we compare the SFRs derived using H$\alpha + 24 \mu$m to those derived using the 12\micron\ relation given by Equation (3). A one-to-one relation (dashed line) indicates the 12\micron-derived SFRs are systematically higher than those from H$\alpha + 24 \mu$m for SFR $>0.1$ M$_{\odot}$yr$^{-1}$. The 1-$\sigma$ scatter of the best-fit relation (solid line) is 0.23 dex. \label{fig:fnew}}               
\end{figure*}

\begin{figure*}
\gridline{\leftfig{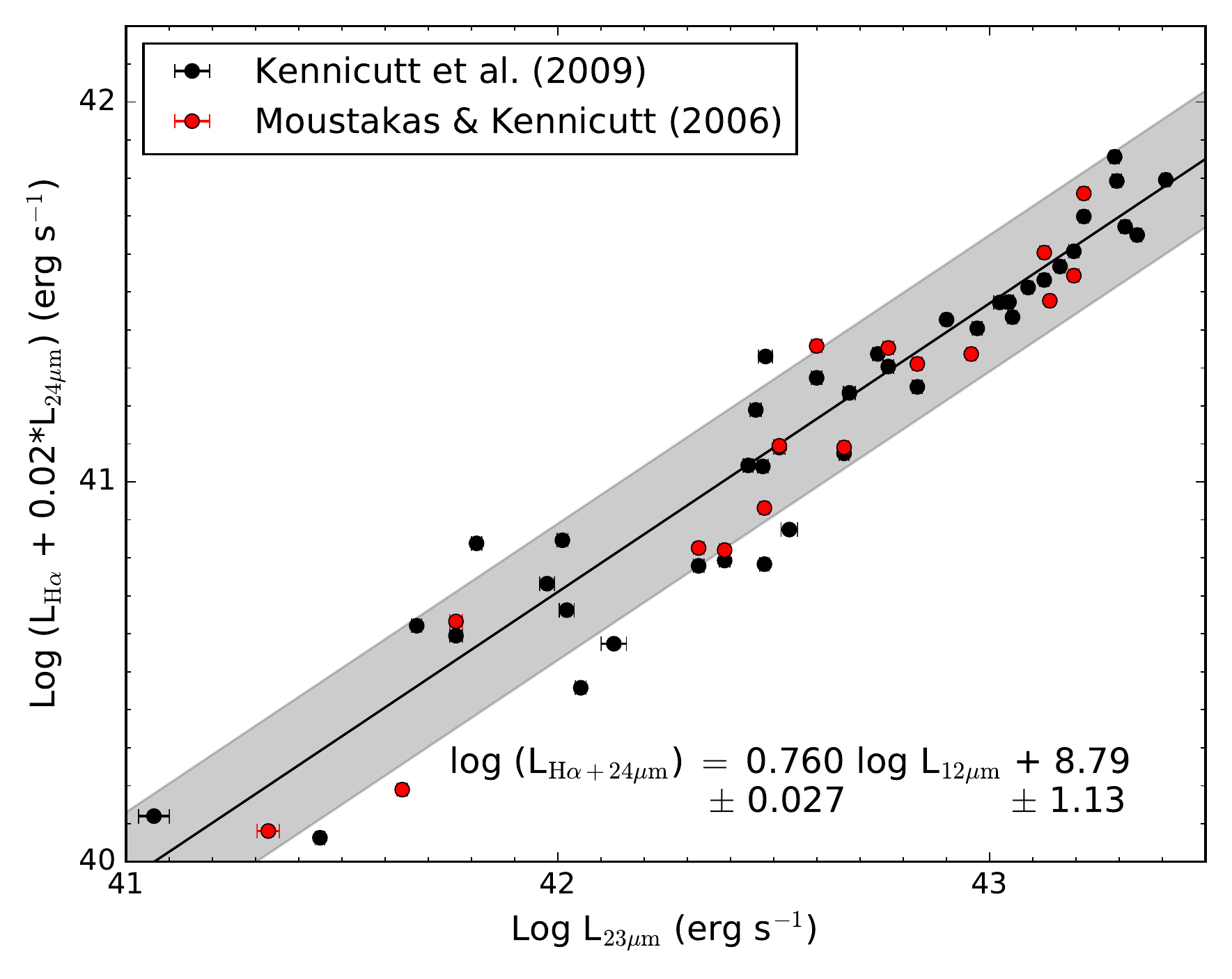}{0.5\textwidth}{(a) } 
              \rightfig{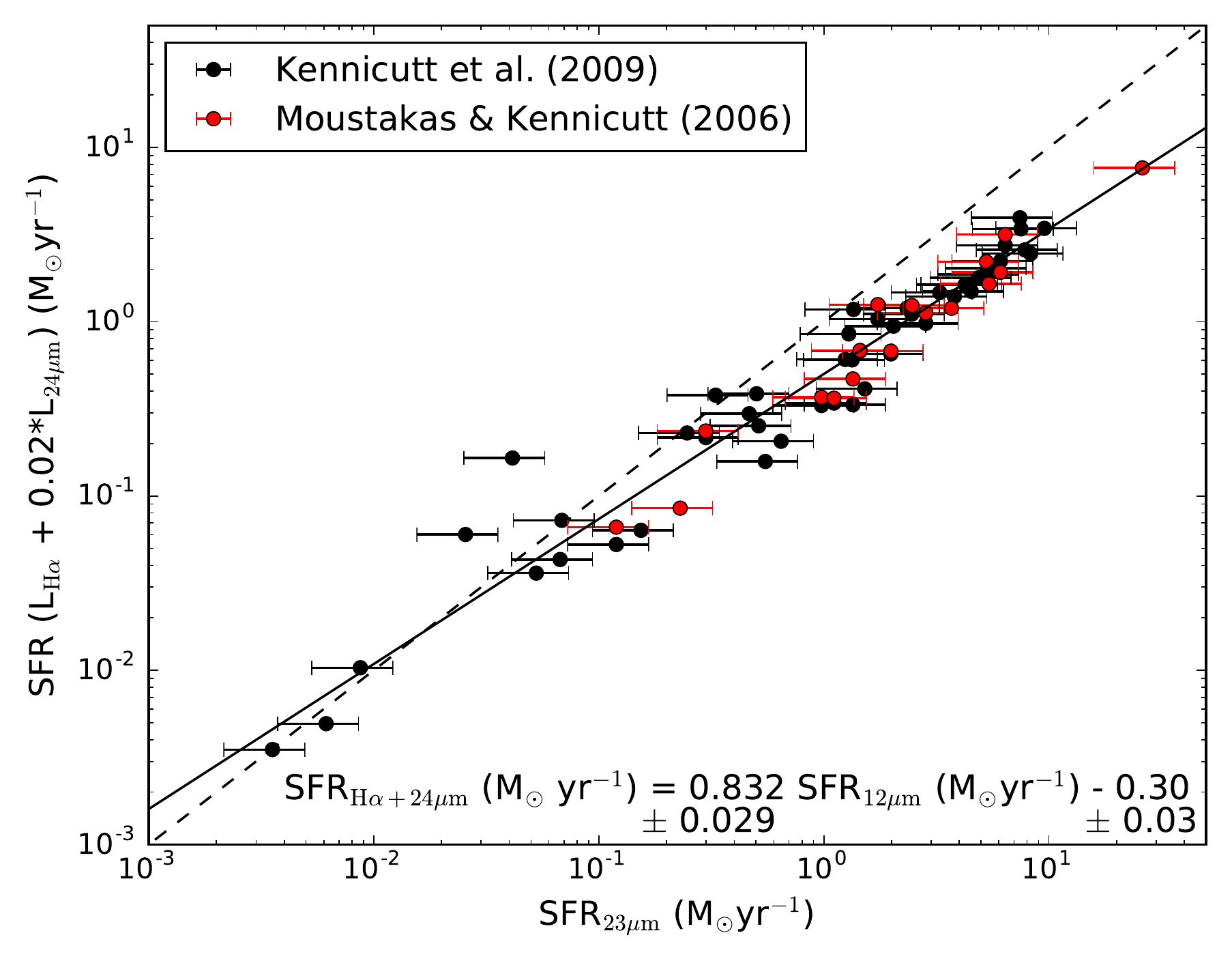}{0.5\textwidth}{(b) }}        
\caption{Same as Figure \ref{fig:fnew}, but now comparing to $L_{23\micron}$ in (a) and 23\micron-derived SFRs (Equation 4) in (b). The linear trend in (a) has a 1-$\sigma$ scatter of 0.18 and indicates a tighter correlation than what was found for $L_{12\micron}$. Similarly, the SFR relation in (b) has a 1-$\sigma$ scatter of only 0.10 dex. Given that the \wise 23\micron\ band is so similar to the MIPS 24\micron\ band, it is probably not surprising and some of the tightness in the relation arises from the x and y axes not being independent. Compared to the one-to-one relation (dashed line) we observe the same trend as in Figure \ref{fig:fnew}b where the $WISE$-derived SFR is systematically high for SFR $>0.1$ M$_{\odot}$yr$^{-1}$ \label{fig:fnew2}}               
\end{figure*}

Using the conversion to SFR, assuming a Kroupa IMF, given by \citet{Kenn09}, i.e. 

\begin{equation}
{\rm SFR} (M_{\odot}\, {\rm yr}^{-1}) = 5.5 \times 10^{-42}\, [L({\rm H\alpha}_{\rm obs}) + 0.02* L_{24 \mu m}] ({\rm erg.s}^{-1}) \\
\end{equation}

allows us to compare SFR relations (Figure \ref{fig:fnew}). The linear best-fit is given by:

\begin{multline}
{\rm SFR}(L({\rm H\alpha}_{\rm obs}) + 0.02* L_{24 \mu m}) (\rm M_{\odot}\, {\rm yr}^{-1}) = \\ (0.776\pm 0.041)\,  {\rm SFR}_{12\micron}\  (\rm M_{\odot}\, {\rm yr}^{-1}) - (0.30\pm0.04),\\
\end{multline}
with a 1-$\sigma$ scatter of 0.23 dex. 

The distribution of points and best-fit relation shows a systematic effect where the 12\micron-derived SFRs (using Equation 2) are higher than those given by H$\alpha + 24 \mu$m for SFRs $>0.1$ M$_{\odot}$yr$^{-1}$).

In Figure \ref{fig:fnew2} we show the comparison with 23\micron\ instead of 12\micron. The H$\alpha + 24 \mu$m luminosities as a function of $L_{23\micron}$ (Figure \ref{fig:fnew2}a) show a clear linear trend given by:

\begin{multline}
{\rm log}\, (L_{\rm H\alpha}+ 0.02*L_{24 \mu m})\  {\rm erg}.{\rm s}^{-1}  = \\ (0.760\pm0.027)\, {\rm log}\, L_{12\mu m} ({\rm erg}.{\rm s}^{-1}) + (8.79\pm 1.13), \\
\end{multline}
with a 1-$\sigma$ scatter of 0.18 dex.  Note the much tighter relation of W4 
with H$\alpha + 24 \mu$m 
as compared to that with W3.

Using the same SFR conversion for $L_{\rm H\alpha}+ 0.02*L_{24 \mu m}$ as above, and the \wise 23\micron\ relation given by Equation (4), we find that the SFRs are related by the equation:

\begin{multline}
{\rm SFR}(L({\rm H\alpha}_{\rm obs}) + 0.02* L_{24 \mu m}) (\rm M_{\odot}\, {\rm yr}^{-1}) = \\ (0.832\pm 0.029)\,  {\rm SFR}_{12\micron}\  (\rm M_{\odot}\, {\rm yr}^{-1}) - (0.30\pm0.03),\\
\end{multline}
with a 1-$\sigma$ scatter of 0.10 dex. 

The tightness in the relation is likely at least partially due to the similarities between the \wise 23\micron\ and the MIPS 24\micron\ band such that the x and y axes are not fully independent of each other. As in Figure \ref{fig:fnew}b, Figure \ref{fig:fnew2}b shows that the $WISE$-derived SFR is systematically higher than what the H$\alpha + 24 \mu$m predicts (for SFR $>0.1$ M$_{\odot}$yr$^{-1}$). Since the luminosities agree so well, the SFR differences are therefore due to the scaling between luminosity (e.g., $L_{\rm TIR}$) and SFR.

\begin{figure*}
\gridline{\leftfig{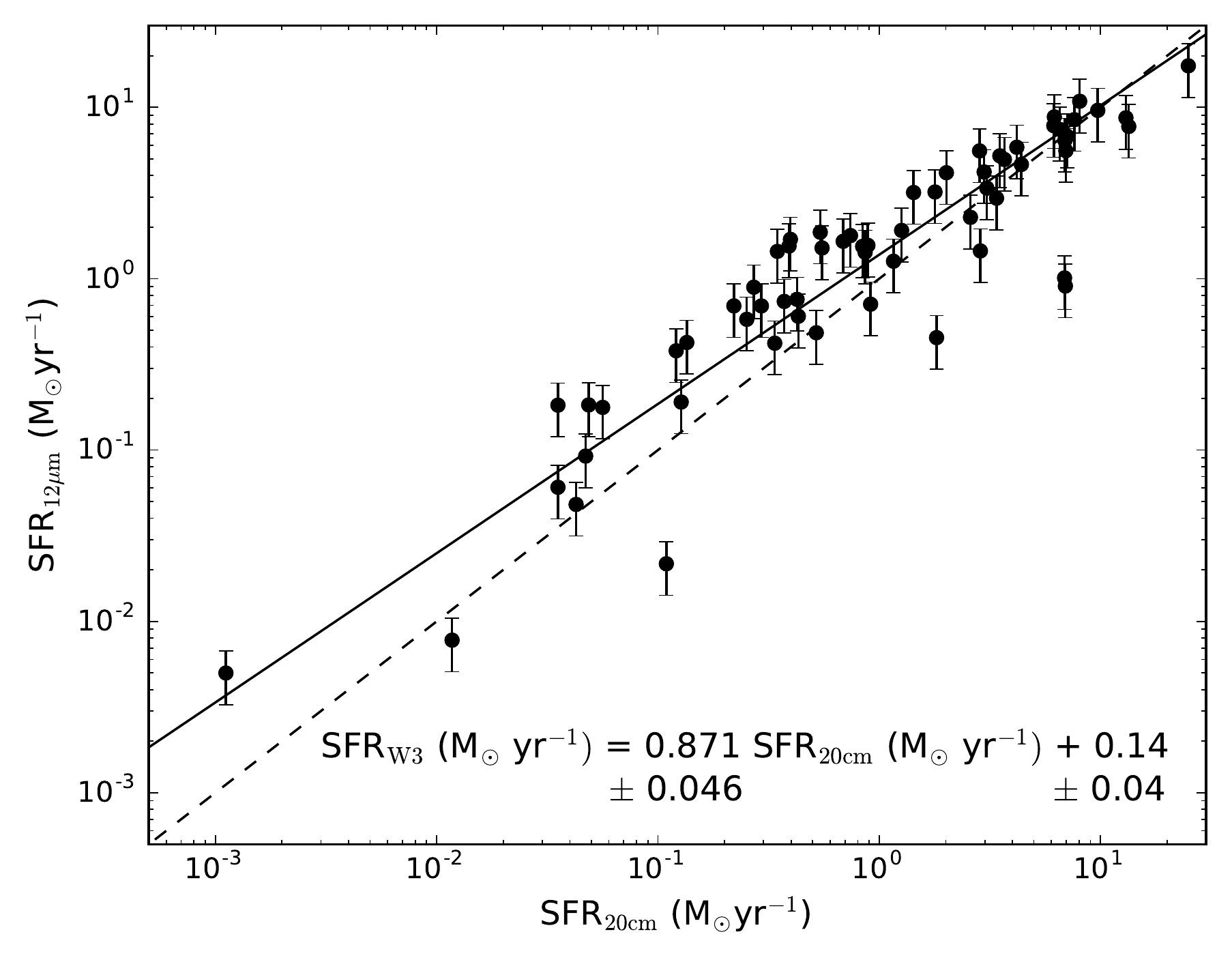}{0.5\textwidth}{(a) 12\micron } 
              \rightfig{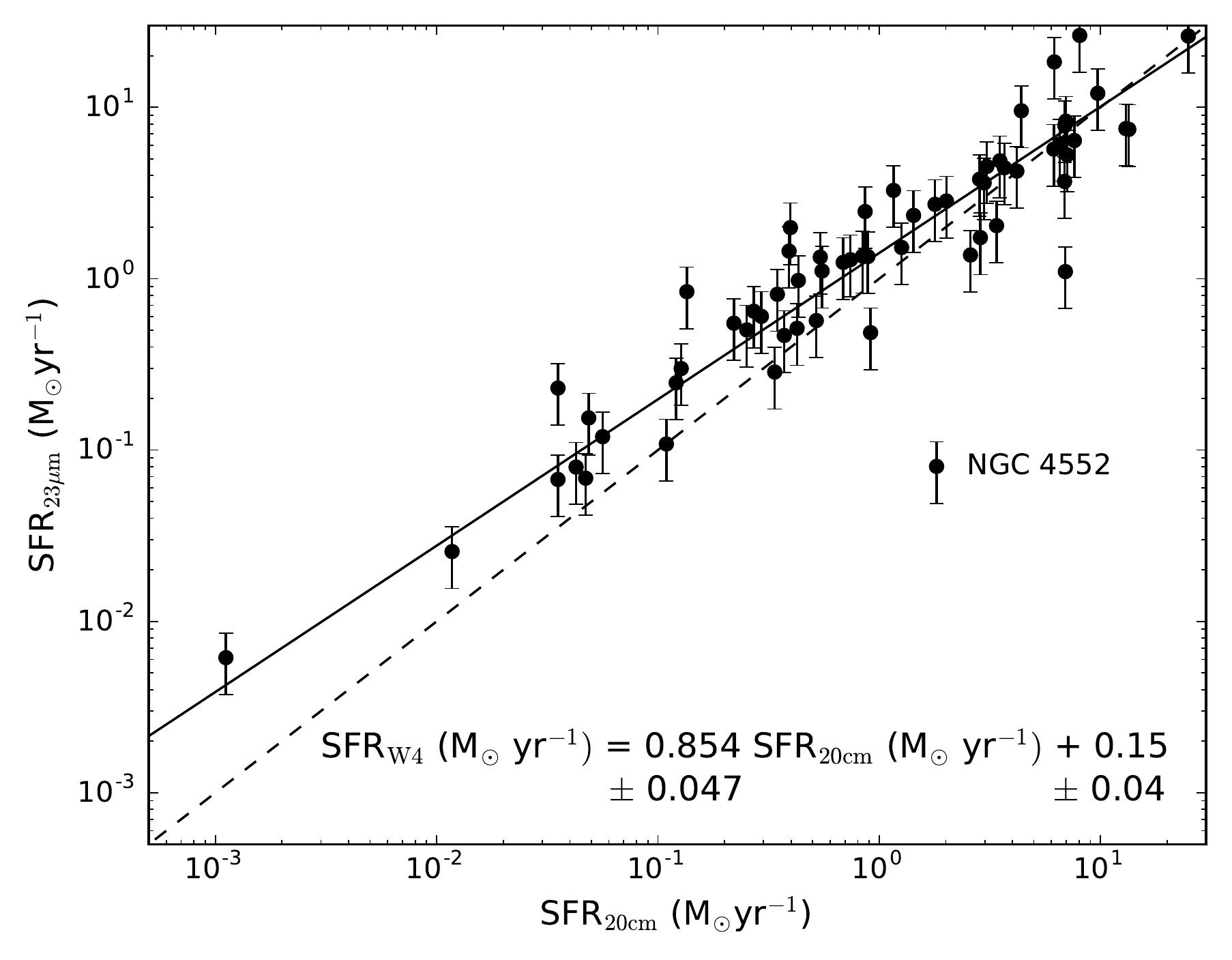}{0.5\textwidth}{(b) 23\micron}}        
\caption{Comparison of the \wise a) 12\micron-derived SFRs and b) 23\micron-derived SFRs compared to the radio continuum (20cm) SFRs. The dashed line shows a one-to-one relation, while the solid line is a fit to the points, clearly showing a trend for both the W3 and W4 SFRs to be higher than the 20 cm-derived values. Although the 23\micron\ comparison appears to have fewer outliers compared to the 12\micron-derived, the fits are very similar with the 12\micron\ and 23\micron\ relations both having a 1-$\sigma$ scatter of 0.26 dex. \label{fig:f7}}               
\end{figure*}

\begin{figure*}
\gridline{\leftfig{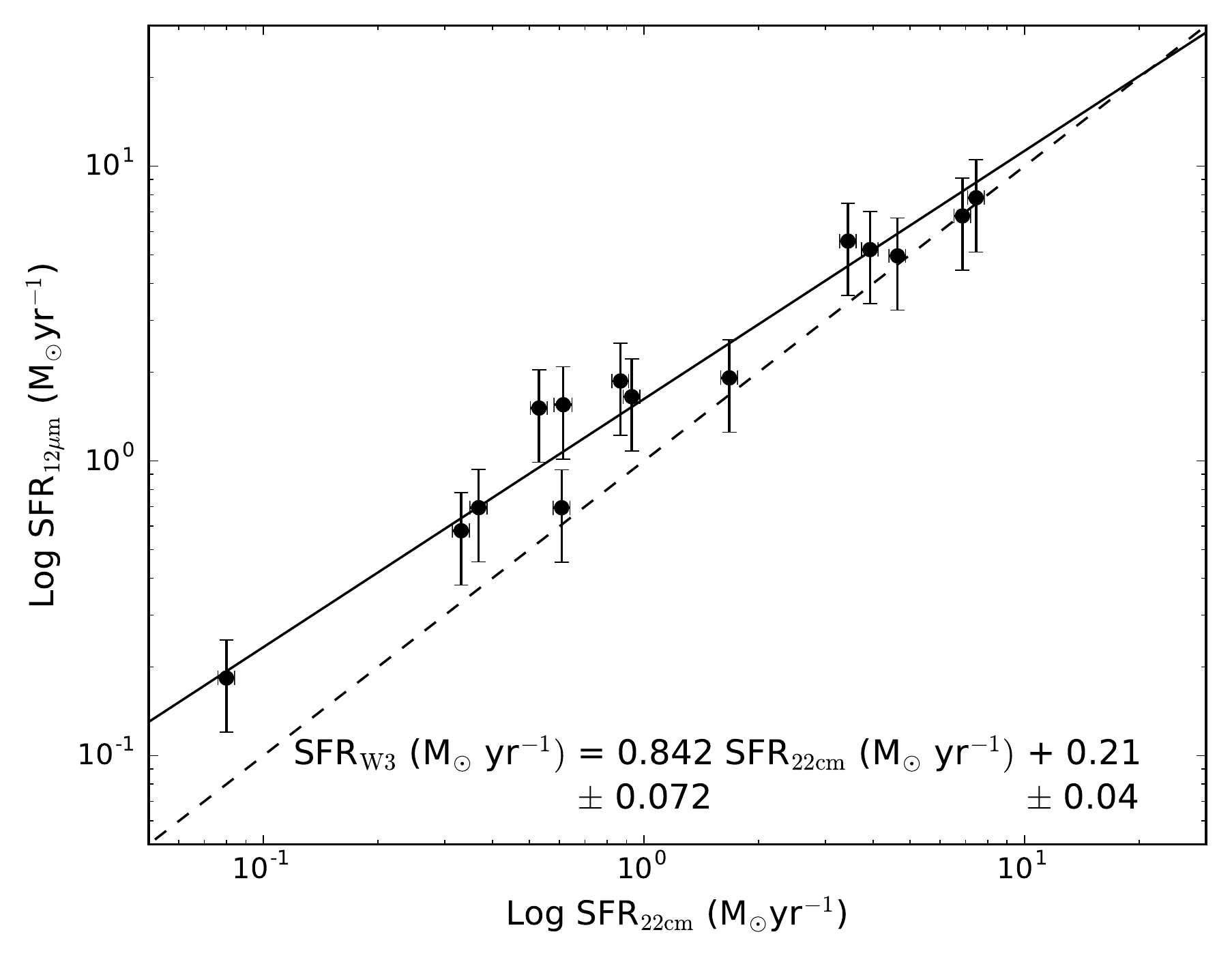}{0.5\textwidth}{(a) 12\micron} 
              \rightfig{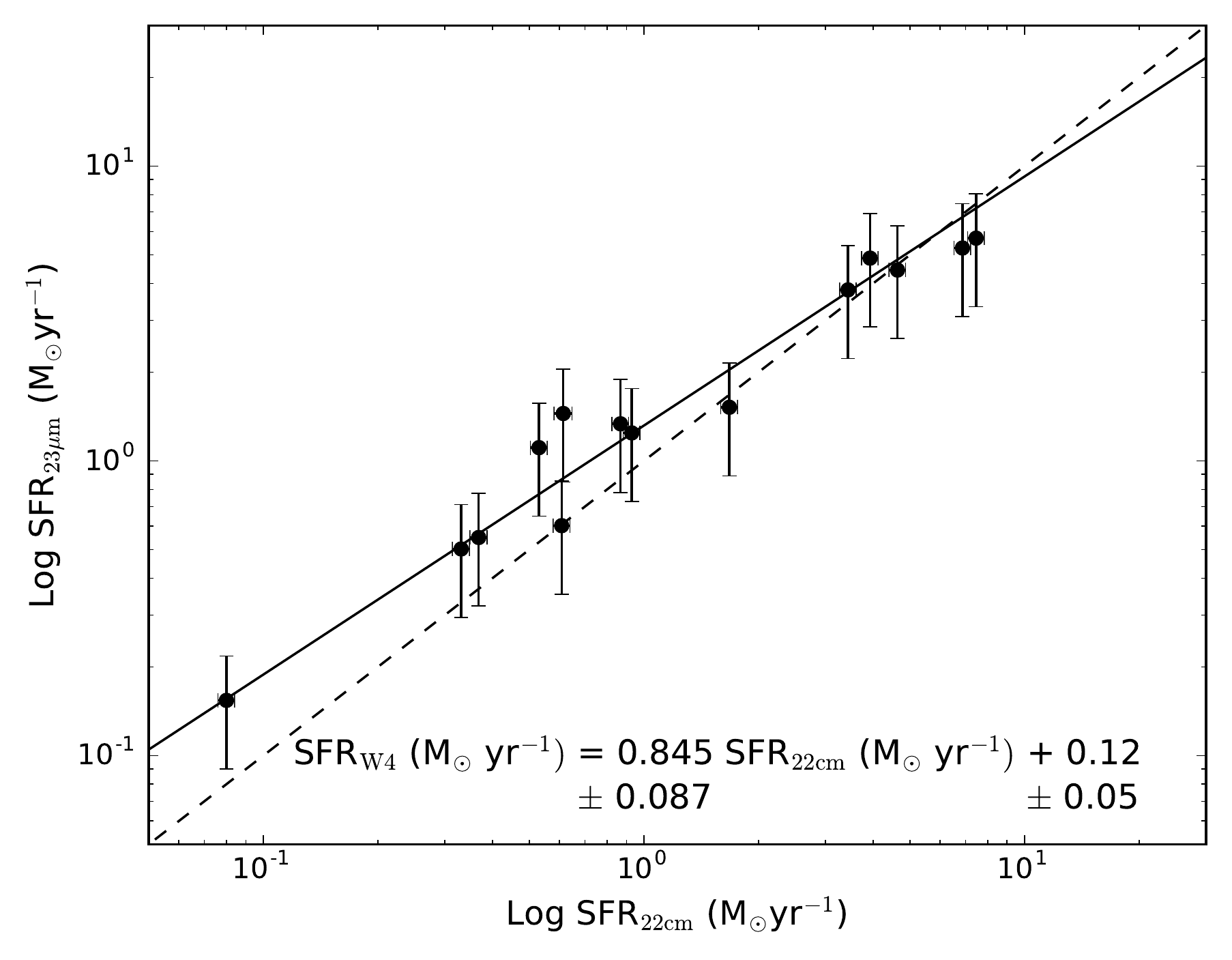}{0.5\textwidth}{(b) 23\micron}}        
\caption{Comparison of the \wise a) 12\micron-derived SFRs and b) 23\micron-derived SFRs compared to the radio continuum (22\,cm) SFRs from \citet{Hees14} with a one-to-one relation given by the dashed line. The best linear fit to the data points has a 1-$\sigma$ scatter of $<0.001$ for both plots. A similar trend is seen as in Figure \ref{fig:f7} where the mid-infrared SFRs tend to be higher than the radio continuum-derived SFRs, particularly at low SFR. \label{fig:f8}}               
\end{figure*}

\begin{figure*}
\gridline{\leftfig{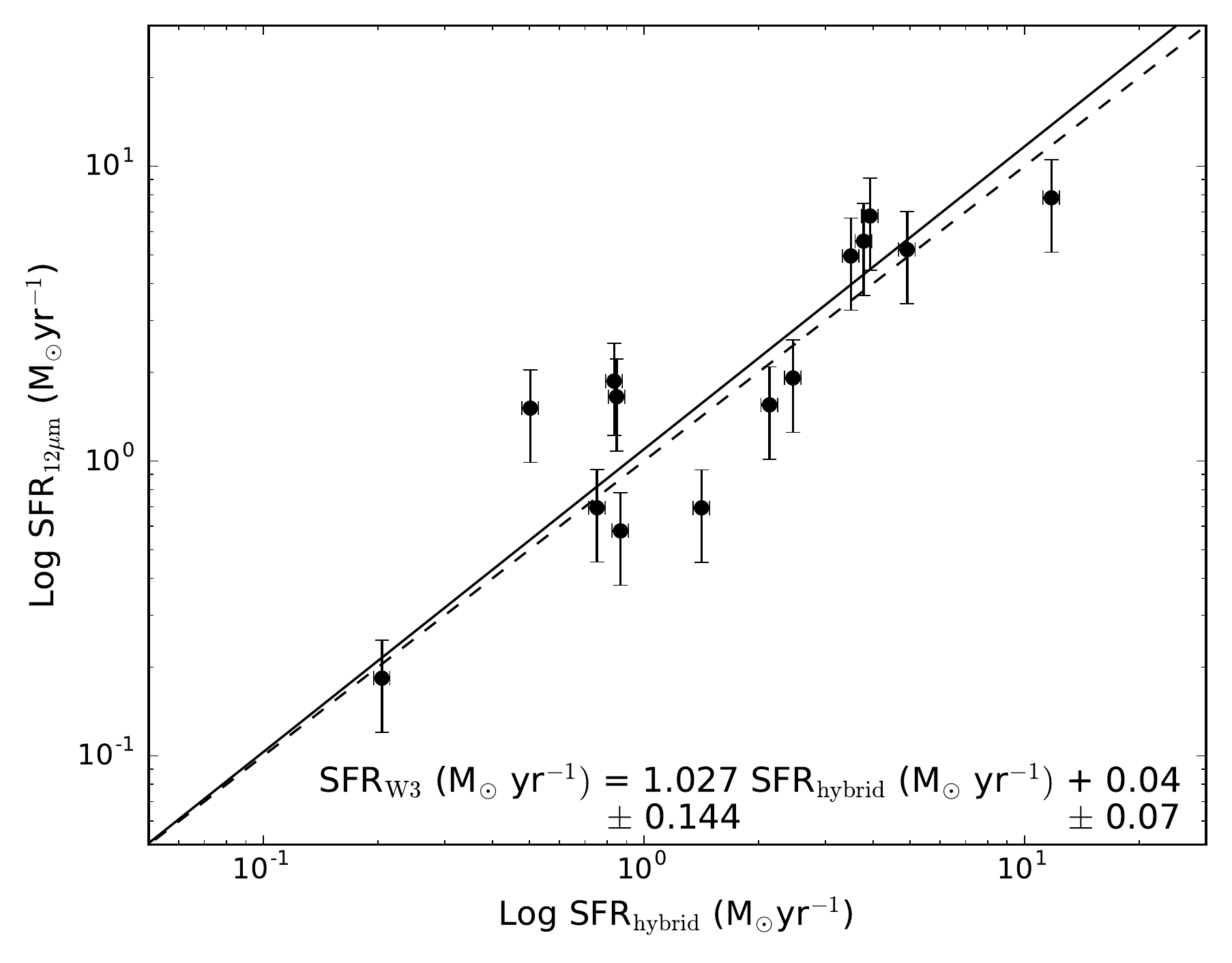}{0.5\textwidth}{(a) 12\micron} 
              \rightfig{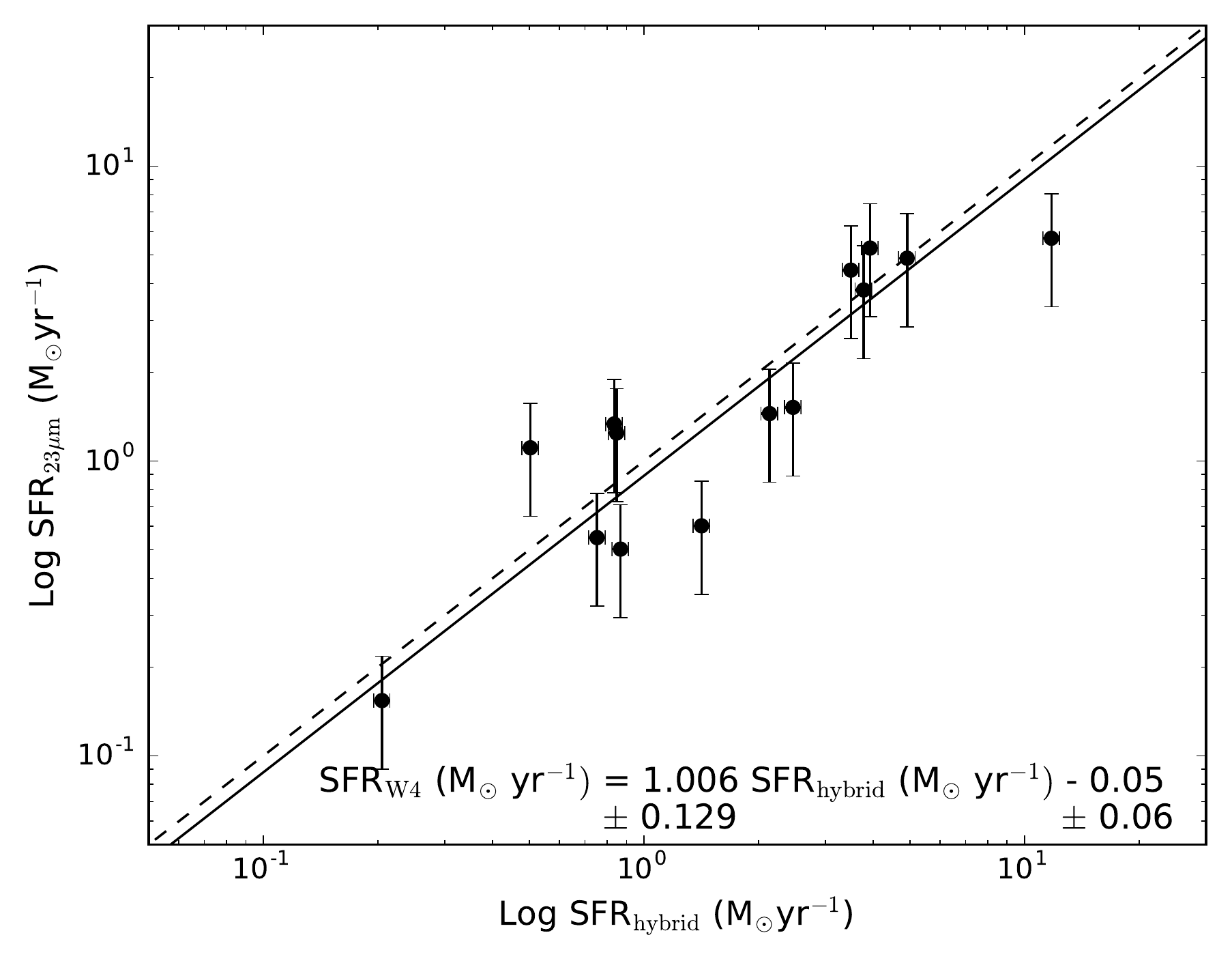}{0.5\textwidth}{(b) 23\micron}}        
\caption{Comparison of the \wise a) 12\micron-derived SFRs and b) 23\micron-derived SFRs compared to the ``hybrid" (FUV$+$24\micron) SFRs from \citet{Hees14}. Both W3- and W4-derived SFRs appear to behave consistently compared to the ``hybrid" SFR, appearing to scatter around a one-to-one relation (dashed line). The best fit line in a) has a scatter of 0.19 dex, while b) has a 1-$\sigma$ scatter of 0.11 dex.\label{fig:f9}}               
\end{figure*}

Next we explore how the $WISE$-derived SFRs compare to those using 20cm radio continuum observations. For the SINGS/KINGFISH sample we use the 20cm fluxes from \citet{Dale17}, where available, except for NGC 584 whose flux is taken from \citet{Br11}, NGC 1512 from \citet{Kor09}, NGC 3077 from \citet{Con98}, Mrk 33 from \citet{Brav04}, and NGC 5195 from \citet{Con02}. To convert to a radio continuum SFR, we employ the relation (Equation 17) of \citet{Murph11}; this also assumes a Kroupa IMF. The comparison is shown in Figure \ref{fig:f7}, with the 12\micron\ SFR comparison in panel (a) and the 23\micron\ comparison in (b). The fit for the 12\micron\ comparison is given by:

\begin{multline}
 {\rm SFR}_{12\micron}\ (\rm M_{\odot}\, {\rm yr}^{-1}) = \\ (0.871\pm 0.046)\,  {\rm SFR}_{\rm 20cm}\ (\rm M_{\odot}\, {\rm yr}^{-1}) + (0.14\pm0.04),\\
\end{multline}
with a 1-$\sigma$ scatter of 0.26 dex. 

Although the distribution of points in Figure \ref{fig:f7}a and b appear different (with a clear outlier of NGC 4552 in Figure \ref{fig:f7}b), we obtain a very similar fit for the 23\micron\ relation, given by:

\begin{multline}
 {\rm SFR}_{23\micron}\ (\rm M_{\odot}\, {\rm yr}^{-1}) = \\ (0.854\pm 0.047)\,  {\rm SFR}_{\rm 20cm}\ (\rm M_{\odot}\, {\rm yr}^{-1}) + (0.15\pm0.04),\\
\end{multline}
with a 1-$\sigma$ scatter of 0.26 dex.

In addition, both Figure \ref{fig:f7}a and b indicate a similar behavior compared to the one-to-one relation (dashed lines) where concordance appears closest for SFRs $>1$ M$_{\odot}$\,yr$^{-1}$ with increasing scatter towards lower SFR. However, with so few data points at low SFRs, comparisons in this regime are unadvisable.  

\citet{Hees14} used 17 THINGS \citep[The \HI\ Nearby Galaxy Survey;][]{Wal08} galaxies, observed as part of the Westerbork Synthesis Radio Telescope (WRST) SINGS sample, \citep{Braun07} to investigate the spatially resolved radio continuum (22\,cm) SFR compared to a ``hybrid" combination of {\it GALEX} FUV and {\it Spitzer} 24\micron\ maps, tracing unobscured and obscured star formation, respectively. We use their integrated star formation rates for galaxies in common to our SINGS/KINGFISH sample (14 galaxies) for comparison and plot the $WISE$-derived SFRs versus radio continuum SFRs in Figure \ref{fig:f8}, and the comparison to the ``hybrid" SFR in Figure \ref{fig:f9}. 

The best fit SFR relation in Figure \ref{fig:f8}a is given by:

\begin{multline}
 {\rm SFR}_{12\micron}\ (\rm M_{\odot}\, {\rm yr}^{-1}) = \\ (0.842\pm 0.072)\,  {\rm SFR}_{\rm 22cm}\ (\rm M_{\odot}\, {\rm yr}^{-1}) + (0.21\pm0.04),\\
\end{multline}
with a 1-$\sigma$ scatter of 3.9e$^{-5}$ dex,

and for Figure \ref{fig:f8}b:

\begin{multline}
 {\rm SFR}_{23\micron}\ (\rm M_{\odot}\, {\rm yr}^{-1}) = \\ (0.845\pm 0.087)\,  {\rm SFR}_{\rm 22cm}\ (\rm M_{\odot}\, {\rm yr}^{-1}) + (0.12\pm0.05),\\
\end{multline}
with a 1-$\sigma$ scatter of 1.5e$^{-5}$ dex.

Comparing these to Equations (14) and (15), respectively, we note the similarity of the $WISE$-radio continuum SFR relations, the only marked difference being the very small scatter reflected by Figure \ref{fig:f8} a and b. In addition, we see a remarkably similar behavior when considering the one-to-one relation, with the 12\micron- and 23\micron-derived SFRs systematically tending towards higher values of SFR as the SFR decreases. 

In Figure \ref{fig:f9} we consider the ``hybrid" SFR indicator of FUV$+$24\micron, which shows remarkably close correspondence to a one-to-one relation. 

The best fit SFR relation in Figure \ref{fig:f9}a is given by:

\begin{multline}
 {\rm SFR}_{12\micron}\ (\rm M_{\odot}\, {\rm yr}^{-1}) = \\ (1.027\pm 0.144)\,  {\rm SFR}_{\rm FUV+24\mu m}\ (\rm M_{\odot}\, {\rm yr}^{-1}) + (0.04\pm0.07),\\
\end{multline}
with a 1-$\sigma$ scatter of 0.19 dex,

and for Figure \ref{fig:f9}b:

\begin{multline}
 {\rm SFR}_{23\micron}\ (\rm M_{\odot}\, {\rm yr}^{-1}) = \\ (1.006\pm 0.129)\,  {\rm SFR}_{\rm FUV+ 24\mu m}\ (\rm M_{\odot}\, {\rm yr}^{-1}) - (0.05\pm0.06),\\
\end{multline}
with a 1-$\sigma$ scatter of 0.11 dex.

Considering that in Figure \ref{fig:f9}a we are using entirely independent methods and tracers to compare SFRs, this result illustrates the utility of W3-derived SFRs, although additional data points would allow for a more substantive comparison.

\subsection{Specific Star Formation in SINGS/KINGSFISH}

\begin{figure*}
\gridline{\leftfig{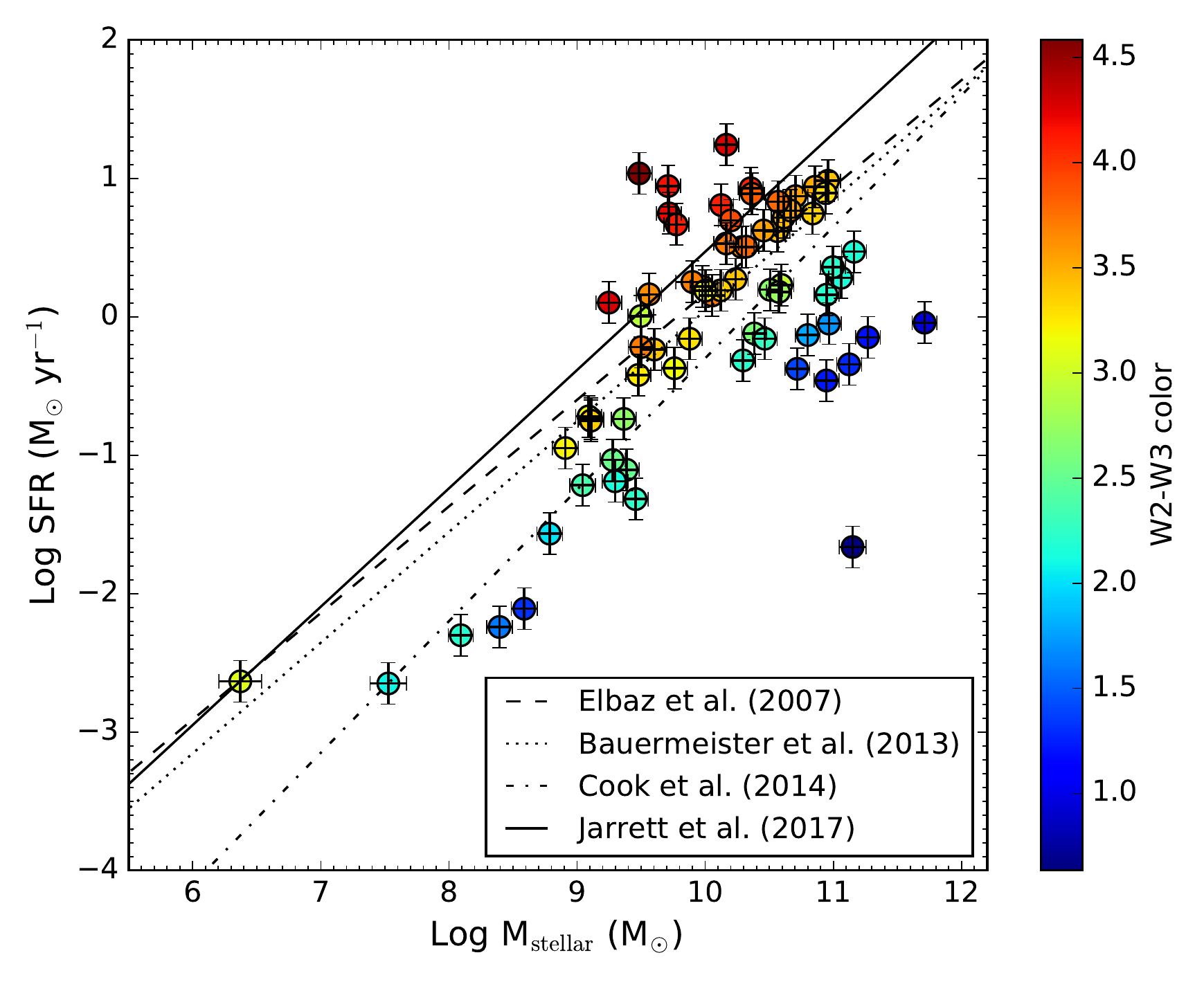}{0.5\textwidth}{(a) SFR vs stellar mass } 
              \rightfig{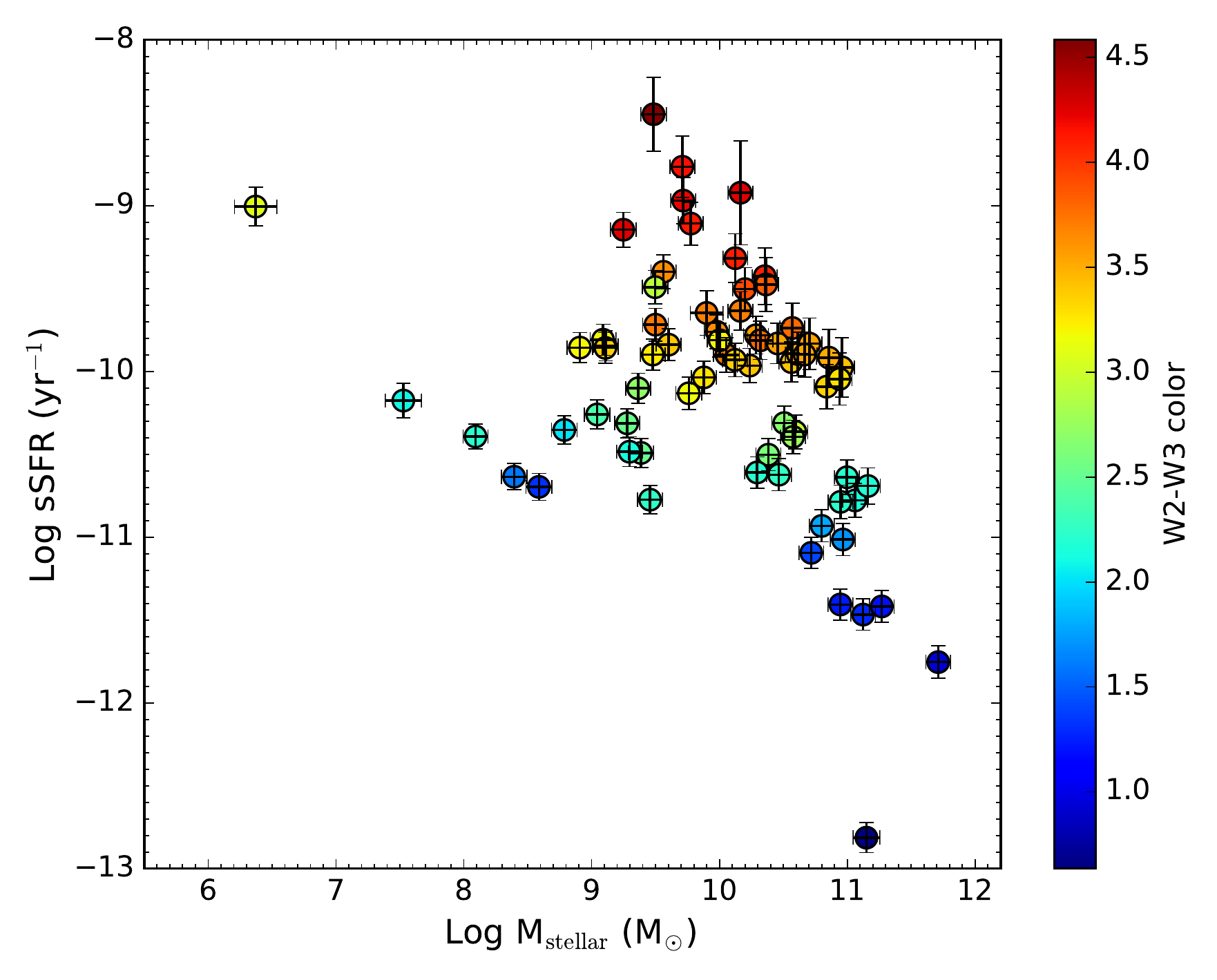}{0.5\textwidth}{(b) Specific Star Formation vs stellar mass }}        
\caption{Plot of a) log SFR and b) log sSFR vs log stellar mass, using the 12\micron\ SFR calibration given by Equation (2), for the SINGS/KINGFISH sample, color-coded by W2$-$W3 color. The ``Main Sequence" relations from \citet{El07}, \citet{Bauer13}, \citet{Coo14}, and \citet{Jar17}, is shown as dashed, dotted, dot-dash, and solid lines, respectively in (a). For $1.5<$ W2$-$W3 $<3$ the LVL relation of \citet{Coo14} is most closely matched, while the other relations do better for $3<$ W2$-$W3 $<4$. \label{fig:f10}}               
\end{figure*}

Here we use the SFR, derived using the W3 luminosity and Equation (2), in combination with the stellar mass relation of \citet{Clu14} to explore star formation in relation to stellar mass in the SINGS/KINGFISH sample. 

In Figure \ref{fig:f10}a we plot SFR as a function of stellar mass, color coding galaxies by their W2$-$W3 color, and include the ``Main Sequence" relations for local galaxies ($z=0$) from \citet{El07} and \citet{Bauer13}, who choose their relation to be consistent with that of \citet{Bou10}, \citet{Kar11} and \citet{El11}.  We also include the fit from the {\it Spitzer} Local Volume Legacy (LVL) study \citep{Coo14} and the GAMA-G12 study of \citet{Jar17}.  As we have color-coded by W2$-$W3 colour, the bluer colors represent low star-forming, spheroidal systems. As indicated by this figure, these systems occupy both the low mass and high mass regions of the diagram: dwarf spheroids lie at the low mass end experiencing relatively low star formation, while elliptical galaxies with high mass and low (quenched) to non-existent star formation, are at the other end. For intermediate W2$-$W3 color, the LVL relation of \citet{Coo14} matches the SINGS/KINGFISH most closely, while the other relations do better for the higher star-forming systems, where $3<$ W2$-$W3 $<4$. 

An alternative view is presented in Figure \ref{fig:f10}b where we show specific SFR (sSFR; SFR divided by stellar mass) versus stellar mass, once again color-coded by W2$-$W3 color. From this diagram we see a clear separation as a function of star formation, where systems with high specific star formation have W2$-$W3 colors $>3$ and stellar masses between 10$^9$ and 10$^{11}$ M$_\odot$. Systems with low specific star formation have W2$-$W3 colors $<1.5$ and typically have masses $>$ 10$^{11}$  M$_\odot$. An intermediate band appears between log sSFR of -10 and -11, exhibiting a broad range of stellar mass and W2$-$W3 colors between 1.5 and 3.

\section{Discussion}

The {\it Spitzer} Space Telescope enabled detailed study of SFR indicators in the mid-infrared and extensively investigated the use of monochromatic tracers of star formation. The MIPS 24\micron\ band measures the warm dust continuum and is relatively uncontaminated by emission line and aromatic features, similar to the \wise 23\micron\ band, and can be related to $L_{\rm TIR}$, and therefore SFR as shown in \citet{Ruj13}. However, this tracer is sensitive to the presence of an AGN, which produces an excess of continuum emission, as well as the dust geometry in relation to the heating source \citep{Far08}.

As shown in Section 3.2, the \wise W3 band shows a tighter correlation with \Ltir SFR in the absence of strong silicate absorption usually associated with embedded starbursts, i.e. coupled to a significant dust column \citep[e.g. local ULIRGs;][]{Des07}. Although the center of this band is close to the 11.3\micron\ PAH feature, as discussed in Appendix A, the fractional contribution of the 11.3\micron\ PAH feature is relatively low -- the largest contribution being 11.6\% for NGC 4559. In terms of total PAH contribution to W3, this varies from 2.9\% (NGC 584) to 52.6\% (NGC 925).

Although PAH features do trace star formation \citep{For04, Jim07, Dia12, Al14, Ship16}, hard radiation fields destroy PAHs, and as a result AGN and low metallicity environments significantly suppress PAH emission \citep{Smith07}. PAH emission varies depending on the physical conditions of a given star-forming region, and is therefore not constant across a galaxy \citep{Smith07, DL07}. 

Figure \ref{fig:f2b}a shows the steady response of the W3 band as \Ltir increases, yielding a tight correlation (Figure \ref{fig:f3}) extending to even relatively extreme star-forming systems. Although we do not probe very low metallicity environments in our sample, the lack of influence of metallicity on the W3 SFR relation is consistent with the fact that the 11.3\micron\ PAH feature and combined PAH contribution to the band is ameliorated by the  
more substantial contributions from the continuum arising from warm, large grains (Appendix A).

The relation between W3 and \Ltir suggests that W3 must be sampling a range of excitation sources to closely mimic the behaviour of the total infrared emission of the galaxy. The warm, large grains and stochastically heated grains, as well as the PAH features \citep{Li02}, will be powered by contributions from both hot and cool stars, similar to $L_{\rm TIR}$. However, we would not expect this relation to hold for sub-regions of a galaxy where variations in dust composition, temperature and PAH characteristics would likely produce large fluctuations.

\section{Conclusions}

In this paper we have derived star formation relations that rely exclusively on the 12\micron\ and 23\micron\ bands of $WISE$, calibrated to $L_{\rm TIR}$. The W3 relation in particular shows a tight correlation, that suggests it could be as reliable as \Ltir as a SFR indicator, over nearly 5 orders of magnitude in star formation and 12\micron\ luminosity, and similarly, stellar mass range, 10$^7$ M$_\odot$ to 10$^{11.5}$ M$_\odot$.
For the most extreme infrared-luminous galaxies, the \wise W3 and W4 relations may be, respectively, under-estimating and over-estimating the star formation activity.

The W3 band appears to be a reliable SFR measure in the absence of strong silicate absorption common in compact, dust-embedded starbursts, and powerful AGN. Taking these caveats into consideration, the benefit of using a SFR tracer unaffected by uncertainties due to dust extinction corrections may well be a key consideration for large area surveys in the future.

The tight relationship we have found between $L_{12\micron}$ and \Ltir, and the known correlation that exists between the mid-infrared 24\micron\ and radio continuum \citep[e.g.][]{app04}, suggests that the W3 band can be used in combination with upcoming sensitive, large-area radio surveys to disentangle star formation and AGN heating. 


\appendix

\section{\wise Spectral Bands}

\begin{figure*}[!t]
\plotone{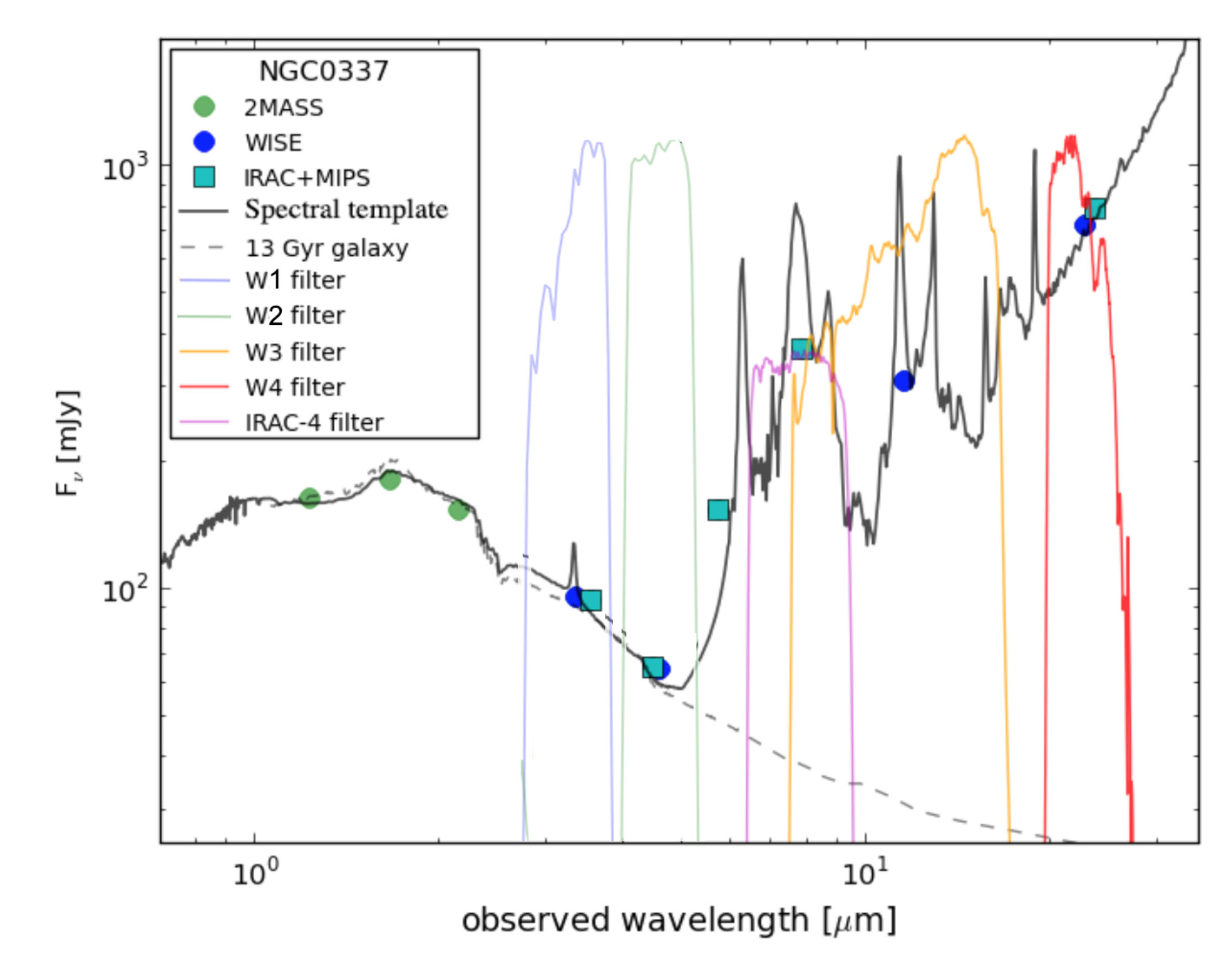}       
\caption{Infrared spectral energy distribution for the galaxy NGC 337.
The global fluxes come from 2MASS (green), WISE (blue) and Spitzer (cyan).  For comparison, two composite templates (Brown et al. 2014a) are shown: late-type that highlights the ISM emission components, 
and an early-type that isolates the stellar (evolved) component.  Note the broad spectral coverage of the \wise W3 (12\micron) band, which overlaps with IRAC-4 (8\micron) band and extends to nearly 16\micron. 
 \label{fig:A1}}               
\end{figure*}

The \wise relative system response curves are given in \citet{Jar11}, comprising four bands centered on 3.4, 4.6, 12 and 23\micron\ in the infrared window.  The first two are essentially the near-infrared, sensitive to the evolved populations in galaxies, while the second two are mid-infrared ISM bands.  W3 is notably broad in spectral coverage, as demonstrated in Fig.\ref{fig:A1}, showing the spectral energy distribution (SED) of the SINGS galaxy, NGC 337.  As a late-type barred spiral, NGC 337 has active star formation, exhibiting strong molecular (PAHs) and thermal (dust) continuum emission.  The W3 band encompasses the 7.7\micron\ and 8.5\micron\ PAH features (typically traced by the IRAC-4 band), the 10\micron\ silicate absorption, 11.3\micron\ PAH band, 12.8$\, \mu$m $[$Ne {\sc ii}$]$ and 15.7\micron\ $[$Ne {\sc iii}$]$ nebular emission. 

Convolving SINGS global spectra with the W3 response function and measuring the fractional contribution of the 11.3\micron\ PAH indicates that it only contributes (on average) 7.5\% to the W3 flux  (J.D.T Smith, private communication). Taking into account all PAH features, this increases to 34\%. The contribution from emission lines is 3.5\% on average. The W3 band is therefore dominated by non-PAH continuum, coming from warm, large grains and stochastically heated grains. 

For example, NGC 337 (Figure \ref{fig:A1}), has a fractional contribution of 45.3\% from all PAH features, 10.4\% from the 11.3\micron\ PAH and 5.0\% from the nebular emission lines. The most IR-luminous galaxy in the sample, NGC 7331, has a fractional contribution of 37.4\% from all PAH features, 8.5\% from the 11.3\micron\ PAH and 2.4\% from the nebular emission lines.

The dwarf galaxy NGC 1705 has, in comparison, a fractional contribution of 21.5\% from all PAH features, 7.6\% from the 11.3\micron\ PAH and 13.3\% from the nebular emission lines

\acknowledgements

We thank the anonymous referee for recommendations that improved the content of this paper. MEC and THJ acknowledge funding from the National Research Foundation under the Research Career Advancement and South African Research Chair Initiative programs, respectively.
This publication makes use of data products from the Wide-field Infrared Survey Explorer, which is a joint project of the University of California, Los Angeles, and the Jet Propulsion Laboratory/California Institute of Technology, funded by the National Aeronautics and Space Administration.

This research has made use of the NASA/IPAC Extragalactic Database (NED) which is operated by the Jet Propulsion Laboratory, California Institute of Technology, under contract with the National Aeronautics and Space Administration.

\newpage
\clearpage

\begin{longrotatetable}
\begin{deluxetable}{lccccccccccc}

\tabletypesize{\scriptsize}


\tablecaption{Measured \wise properties of the SINGS/KINGFISH Sample}

\tablenum{1}


\tablehead{\colhead{Galaxy} & \colhead{W1} & \colhead{W1f} & \colhead{W2} & \colhead{W2f}  & \colhead{W3}  & \colhead{W3f}  & \colhead{W4}  & \colhead{W4f}  & \colhead{Radius}  & \colhead{b/a} & \colhead{P.A.}    \\ 
\colhead{} & \colhead{(mag)} & \colhead{} & \colhead{(mag)} & \colhead{} & \colhead{(mag)} & \colhead{} & \colhead{(mag)} & \colhead{} & \colhead{(arcsec)} & \colhead{} & \colhead{(deg.)}}

\startdata
NGC0024 & 8.66 $\pm$ 0.011 & 0 & 8.682 $\pm$ 0.02  &  0 & 6.176 $\pm$ 0.033 & 10 & 4.49 $\pm$ 0.06 & 10 & 187.97 & 0.296 & 45.1 \\
NGC0337 & 8.777 $\pm$ 0.011 & 0 & 8.552 $\pm$ 0.02 & 0 & 4.912 $\pm$ 0.017 & 10 & 2.549 $\pm$ 0.02 & 10 & 101.98 & 0.622 & 131.8 \\
NGC0584 & 7.19 $\pm$ 0.011 & 0 & 7.247 $\pm$ 0.019 & 0 & 6.781 $\pm$ 0.05 & 10 & 5.757 $\pm$ 0.107 & 10 & 190.32 & 0.784 & 70.6 \\
NGC0628 & 6.442 $\pm$ 0.011 & 0 & 6.332 $\pm$ 0.019 & 0 & 2.701 $\pm$ 0.015 & 10 & 0.882 $\pm$ 0.018 & 10 & 324.47 & 0.819 & 86.4 \\
NGC0855 & 9.665 $\pm$ 0.012 & 0 & 9.592 $\pm$ 0.021 & 0 & 7.229 $\pm$ 0.07 & 10 & 4.976 $\pm$ 0.048 & 10 & 106.47 & 0.446 & 65.0 \\
NGC0925 & 7.57 $\pm$ 0.011 & 0 & 7.517 $\pm$ 0.02 & 0 & 4.243 $\pm$ 0.034 & 10 & 2.252 $\pm$ 0.058 & 10 & 311.71 & 0.501 & 110.1 \\
NGC1097 & 5.983 $\pm$ 0.011 & 0 & 5.882 $\pm$ 0.019 & 0 & 2.469 $\pm$ 0.013 & 10 & 0.131 $\pm$ 0.014 & 10 & 322.65 & 0.792 & 112.9 \\
NGC1266 & 9.385 $\pm$ 0.012 & 0 & 9.091 $\pm$ 0.021 & 0 & 6.199 $\pm$ 0.018 & 10 & 2.557 $\pm$ 0.016 & 10 & 65.89 & 0.733 & 115.4 \\
NGC1291 & 5.35 $\pm$ 0.011 & 0 & 5.391 $\pm$ 0.019 & 0 & 4.156 $\pm$ 0.06 & 10 & 3.043 $\pm$ 0.102 & 10 & 431.86 & 0.987 & 55.1 \\
NGC1316 & 5.079 $\pm$ 0.011 & 0 & 5.102 $\pm$ 0.019 & 0 & 4.17 $\pm$ 0.038 & 10 & 2.783 $\pm$ 0.047 & 10 & 768.93 & 0.645 & 34.7 \\
NGC1377 & 9.507 $\pm$ 0.012 & 0 & 8.153 $\pm$ 0.02 & 0 & 4.682 $\pm$ 0.014 & 10 & 1.712 $\pm$ 0.013 & 10 & 72.06 & 0.636 & 92.0 \\
NGC1404 & 6.611 $\pm$ 0.014 & 0 & 6.651 $\pm$ 0.023 & 0 & 6.021 $\pm$ 0.086 & 10 & 5.049 $\pm$ 0.067 & 10 & 216.32 & 0.88 & 160.0 \\
NGC1482 & 7.938 $\pm$ 0.011 & 0 & 7.629 $\pm$ 0.019 & 0 & 3.418 $\pm$ 0.014 & 10 & 0.825 $\pm$ 0.014 & 10 & 135.91 & 0.799 & 112.7 \\
NGC1512 & 7.245 $\pm$ 0.011 & 0 & 7.246 $\pm$ 0.019 & 0 & 4.639 $\pm$ 0.027 & 10 & 3.005 $\pm$ 0.036 & 10 & 291.06 & 0.644 & 45.3 \\
NGC1566 & 6.568 $\pm$ 0.011 & 0 & 6.465 $\pm$ 0.019 & 0 & 2.956 $\pm$ 0.013 & 10 & 1.048 $\pm$ 0.015 & 10 & 333.13 & 0.735 & 27.1 \\
NGC1705 & 10.085 $\pm$ 0.012 & 0 & 10.001 $\pm$ 0.021 & 0 & 7.983 $\pm$ 0.036 & 10 & 5.407 $\pm$ 0.064 & 10 & 94.69 & 0.595 & 44.2 \\
M81DwA & 16.989 $\pm$ 0.138 & 1 & 16.543 $\pm$ 0.329 & 1 & -- $\pm$ -- & 0 & 11.362 $\pm$ 2.735 & 0 & 6.01 & 1.0 & 98.8 \\
DDO053 & 12.598 $\pm$ 0.022 & 0 & 12.277 $\pm$ 0.053 & 0 & 9.182 $\pm$ 0.077 & 0 & 6.324 $\pm$ 0.13 & 10 & 49.3 & 0.789 & 138.5 \\
NGC2403 & 5.689 $\pm$ 0.011 & 0 & 5.59 $\pm$ 0.019 & 0 & 2.203 $\pm$ 0.016 & 10 & 0.195 $\pm$ 0.023 & 10 & 656.59 & 0.528 & 124.1 \\
NGC2798 & 8.658 $\pm$ 0.011 & 0 & 8.384 $\pm$ 0.019 & 0 & 4.278 $\pm$ 0.014 & 10 & 1.302 $\pm$ 0.013 & 10 & 94.13 & 0.522 & 158.4 \\
NGC2841 & 5.89 $\pm$ 0.011 & 0 & 5.892 $\pm$ 0.019 & 0 & 3.742 $\pm$ 0.014 & 10 & 2.038 $\pm$ 0.04 & 10 & 384.35 & 0.47 & 151.4 \\
NGC2915 & 9.427 $\pm$ 0.012 & 0 & 9.383 $\pm$ 0.021 & 0 & 7.783 $\pm$ 0.036 & 10 & 5.331 $\pm$ 0.052 & 10 & 123.78 & 0.493 & 129.7 \\
NGC2976 & 7.155 $\pm$ 0.011 & 0 & 7.055 $\pm$ 0.019 & 0 & 3.837 $\pm$ 0.017 & 10 & 1.838 $\pm$ 0.017 & 10 & 237.09 & 0.589 & 144.3 \\
NGC3031 & 3.606 $\pm$ 0.011 & 0 & 3.608 $\pm$ 0.019 & 0 & 1.826 $\pm$ 0.012 & 10 & 0.426 $\pm$ 0.021 & 10 & 868.41 & 0.557 & 155.9 \\
NGC3049 & 9.774 $\pm$ 0.012 & 0 & 9.727 $\pm$ 0.022 & 0 & 6.027 $\pm$ 0.02 & 10 & 3.237 $\pm$ 0.021 & 10 & 78.19 & 0.624 & 26.0 \\
M81DwB & 12.046 $\pm$ 0.017 & 0 & 11.96 $\pm$ 0.043 & 0 & 9.873 $\pm$ 0.144 & 0 & 7.615 $\pm$ 0.405 & 0 & 66.28 & 0.501 & 139.9 \\
NGC3190 & 7.298 $\pm$ 0.011 & 0 & 7.258 $\pm$ 0.019 & 0 & 5.054 $\pm$ 0.019 & 10 & 3.64 $\pm$ 0.029 & 10 & 178.48 & 0.45 & 121.1 \\
NGC3184 & 6.987 $\pm$ 0.011 & 0 & 6.911 $\pm$ 0.019 & 0 & 3.508 $\pm$ 0.015 & 10 & 1.769 $\pm$ 0.021 & 10 & 227.06 & 0.934 & 106.7 \\
NGC3198 & 7.609 $\pm$ 0.011 & 0 & 7.52 $\pm$ 0.019 & 0 & 4.171 $\pm$ 0.02 & 10 & 2.116 $\pm$ 0.025 & 10 & 279.97 & 0.349 & 35.3 \\
IC2574 & 8.681 $\pm$ 0.012 & 0 & 8.592 $\pm$ 0.022 & 0 & 7.27 $\pm$ 0.084 & 10 & 4.1 $\pm$ 0.108 & 0 & 326.09 & 0.49 & 39.6 \\
NGC3265 & 10.216 $\pm$ 0.012 & 0 & 10.125 $\pm$ 0.021 & 0 & 6.415 $\pm$ 0.016 & 10 & 3.67 $\pm$ 0.022 & 10 & 36.53 & 0.898 & 68.9 \\
Mrk33 & 10.23 $\pm$ 0.012 & 0 & 10.006 $\pm$ 0.022 & 0 & 5.74 $\pm$ 0.013 & 10 & 2.471 $\pm$ 0.014 & 10 & 56.68 & 0.784 & 125.4 \\
NGC3351 & 6.464 $\pm$ 0.011 & 0 & 6.459 $\pm$ 0.019 & 0 & 3.502 $\pm$ 0.018 & 10 & 1.226 $\pm$ 0.016 & 10 & 245.14 & 0.78 & 17.2 \\
NGC3521 & 5.439 $\pm$ 0.011 & 0 & 5.342 $\pm$ 0.019 & 0 & 1.938 $\pm$ 0.014 & 10 & 0.276 $\pm$ 0.018 & 10 & 427.58 & 0.621 & 165.4 \\
NGC3621 & 6.38 $\pm$ 0.02 & 0 & 6.241 $\pm$ 0.034 & 0 & 2.546 $\pm$ 0.014 & 10 & 0.775 $\pm$ 0.023 & 10 & 349.92 & 0.478 & 162.4 \\
NGC3627 & 5.556 $\pm$ 0.011 & 0 & 5.446 $\pm$ 0.019 & 0 & 1.997 $\pm$ 0.013 & 10 & -0.025 $\pm$ 0.015 & 10 & 395.56 & 0.532 & 177.3 \\
NGC3773 & 10.464 $\pm$ 0.012 & 0 & 10.346 $\pm$ 0.022 & 0 & 7.171 $\pm$ 0.03 & 10 & 4.45 $\pm$ 0.029 & 10 & 55.96 & 0.834 & 173.7 \\
NGC3938 & 7.496 $\pm$ 0.011 & 0 & 7.371 $\pm$ 0.019 & 0 & 3.816 $\pm$ 0.016 & 10 & 2.057 $\pm$ 0.019 & 10 & 182.74 & 0.823 & 49.5 \\
NGC4125 & 6.501 $\pm$ 0.011 & 0 & 6.535 $\pm$ 0.019 & 0 & 6.089 $\pm$ 0.052 & 10 & 5.014 $\pm$ 0.067 & 10 & 336.66 & 0.71 & 83.9 \\
NGC4236 & 7.91 $\pm$ 0.011 & 0 & 7.99 $\pm$ 0.02 & 0 & 5.749 $\pm$ 0.087 & 10 & 3.091 $\pm$ 0.047 & 0 & 519.32 & 0.299 & 158.0 \\
NGC4254 & 6.65 $\pm$ 0.011 & 0 & 6.464 $\pm$ 0.019 & 0 & 2.375 $\pm$ 0.013 & 10 & 0.6 $\pm$ 0.014 & 10 & 174.03 & 0.926 & 58.3 \\
NGC4321 & 6.247 $\pm$ 0.011 & 0 & 6.122 $\pm$ 0.019 & 0 & 2.578 $\pm$ 0.015 & 10 & 0.722 $\pm$ 0.02 & 10 & 322.35 & 0.889 & 55.2 \\
NGC4450 & 6.883 $\pm$ 0.011 & 0 & 6.892 $\pm$ 0.019 & 0 & 5.198 $\pm$ 0.028 & 10 & 3.693 $\pm$ 0.063 & 10 & 216.86 & 0.674 & 172.3 \\
NGC4536 & 7.183 $\pm$ 0.011 & 0 & 7.027 $\pm$ 0.019 & 0 & 3.332 $\pm$ 0.016 & 10 & 0.866 $\pm$ 0.016 & 10 & 262.54 & 0.388 & 122.9 \\
NGC4552 & 6.359 $\pm$ 0.011 & 0 & 6.425 $\pm$ 0.019 & 0 & 5.134 $\pm$ 0.124 & 10 & 4.856 $\pm$ 0.166 & 10 & 321.51 & 0.882 & 134.4 \\
NGC4559 & 7.389 $\pm$ 0.011 & 0 & 7.235 $\pm$ 0.019 & 0 & 4.037 $\pm$ 0.018 & 10 & 2.063 $\pm$ 0.034 & 10 & 297.59 & 0.379 & 149.5 \\
NGC4569 & 6.49 $\pm$ 0.011 & 0 & 6.427 $\pm$ 0.019 & 0 & 3.746 $\pm$ 0.014 & 10 & 1.845 $\pm$ 0.015 & 10 & 323.95 & 0.404 & 25.3 \\
NGC4579 & 6.325 $\pm$ 0.011 & 0 & 6.29 $\pm$ 0.019 & 0 & 4.133 $\pm$ 0.014 & 10 & 2.584 $\pm$ 0.021 & 10 & 242.89 & 0.794 & 97.4 \\
NGC4594 & 4.617 $\pm$ 0.011 & 0 & 4.639 $\pm$ 0.019 & 0 & 3.439 $\pm$ 0.036 & 10 & 2.248 $\pm$ 0.04 & 10 & 660.83 & 0.455 & 90.7 \\
NGC4625 & 9.592 $\pm$ 0.012 & 0 & 9.517 $\pm$ 0.021 & 0 & 6.152 $\pm$ 0.024 & 10 & 4.404 $\pm$ 0.061 & 10 & 64.37 & 0.9 & 177.5 \\
NGC4631 & 6.005 $\pm$ 0.011 & 0 & 5.802 $\pm$ 0.019 & 0 & 1.959 $\pm$ 0.013 & 10 & -0.1 $\pm$ 0.014 & 10 & 496.31 & 0.272 & 83.2 \\
NGC4725 & 6.086 $\pm$ 0.011 & 0 & 6.106 $\pm$ 0.019 & 0 & 3.891 $\pm$ 0.021 & 10 & 2.54 $\pm$ 0.026 & 10 & 386.33 & 0.634 & 40.0 \\
NGC4736 & 4.767 $\pm$ 0.011 & 0 & 4.742 $\pm$ 0.019 & 0 & 1.992 $\pm$ 0.015 & 10 & 0.27 $\pm$ 0.02 & 10 & 579.39 & 0.89 & 105.6 \\
DDO154 & 12.334 $\pm$ 0.022 & 0 & 12.533 $\pm$ 0.077 & 0 & 11.19 $\pm$ 0.591 & 0 & -- $\pm$ -- & 0 & 60.6 & 0.575 & 39.5 \\
NGC4826 & 5.183 $\pm$ 0.011 & 0 & 5.169 $\pm$ 0.019 & 0 & 2.911 $\pm$ 0.016 & 10 & 1.15 $\pm$ 0.018 & 10 & 369.88 & 0.535 & 114.7 \\
DDO165 & 11.125 $\pm$ 0.014 & 0 & 11.143 $\pm$ 0.035 & 0 & --  $\pm$ -- & 0 & 8.729 $\pm$ 1.558 & 0 & 114.65 & 0.521 & 98.2 \\
NGC5033 & 6.719 $\pm$ 0.011 & 0 & 6.618 $\pm$ 0.019 & 0 & 3.086 $\pm$ 0.014 & 10 & 1.365 $\pm$ 0.028 & 10 & 283.42 & 0.574 & 172.5 \\
NGC5055 & 5.257 $\pm$ 0.011 & 0 & 5.195 $\pm$ 0.019 & 0 & 1.85 $\pm$ 0.013 & 10 & 0.204 $\pm$ 0.016 & 10 & 523.49 & 0.601 & 103.4 \\
NGC5194 & 5.087 $\pm$ 0.011 & 0 & 4.968 $\pm$ 0.019 & 0 & 1.176 $\pm$ 0.012 & 10 & -0.638 $\pm$ 0.014 & 10 & 463.09 & 0.666 & 17.6 \\
NGC5195 & 6.451 $\pm$ 0.011 & 0 & 6.432 $\pm$ 0.019 & 0 & 4.223 $\pm$ 0.012 & 10 & 2.009 $\pm$ 0.013 & 10 & 174.12 & 0.776 & 106.9 \\
NGC5408 & 10.006 $\pm$ 0.012 & 0 & 9.915 $\pm$ 0.022 & 0 & 6.714 $\pm$ 0.022 & 0 & 3.233 $\pm$ 0.021 & 10 & 83.72 & 0.601 & 60.5 \\
NGC5474 & 8.646 $\pm$ 0.011 & 0 & 8.609 $\pm$ 0.02 & 0 & 6.083 $\pm$ 0.051 & 10 & 4.28 $\pm$ 0.086 & 10 & 155.1 & 0.85 & 1.1 \\
NGC5713 & 7.966 $\pm$ 0.011 & 0 & 7.75 $\pm$ 0.019 & 0 & 3.656 $\pm$ 0.013 & 10 & 1.31 $\pm$ 0.013 & 10 & 112.14 & 0.883 & 9.2 \\
NGC5866 & 6.62 $\pm$ 0.011 & 0 & 6.603 $\pm$ 0.019 & 0 & 5.205 $\pm$ 0.019 & 10 & 3.864 $\pm$ 0.036 & 10 & 283.19 & 0.555 & 122.3 \\
IC4710 & 9.369 $\pm$ 0.012 & 0 & 9.333 $\pm$ 0.022 & 0 & 7.194 $\pm$ 0.037 & 0 & 4.815 $\pm$ 0.083 & 0 & 130.61 & 0.75 & 115.1 \\
NGC6822 & 5.85 $\pm$ 0.011 & 0 & 5.843 $\pm$ 0.019 & 0 & 3.621 $\pm$ 0.086 & 10 & 1.203 $\pm$ 0.092 & 10 & 545.48 & 0.789 & 136.9 \\
NGC6946 & 5.015 $\pm$ 0.011 & 0 & 4.839 $\pm$ 0.019 & 0 & 0.928 $\pm$ 0.011 & 10 & -1.066 $\pm$ 0.012 & 10 & 446.82 & 0.909 & 51.5 \\
NGC7331 & 5.702 $\pm$ 0.011 & 0 & 5.61 $\pm$ 0.019 & 0 & 2.329 $\pm$ 0.012 & 10 & 0.627 $\pm$ 0.015 & 10 & 373.35 & 0.426 & 172.0 \\
NGC7552 & 7.132 $\pm$ 0.011 & 0 & 6.733 $\pm$ 0.019 & 0 & 2.577 $\pm$ 0.013 & 10 & -0.408 $\pm$ 0.012 & 10 & 152.04 & 0.8 & 104.8 \\
NGC7793 & 6.518 $\pm$ 0.011 & 0 & 6.428 $\pm$ 0.019 & 0 & 3.171 $\pm$ 0.014 & 10 & 1.492 $\pm$ 0.018 & 10 & 336.27 & 0.617 & 97.9 \\
IC0342 & 3.974 $\pm$ 0.012 & 0 & 3.892 $\pm$ 0.02 & 0 & 0.107 $\pm$ 0.034 & 10 & -1.833 $\pm$ 0.012 & 10 & 778.51 & 0.946 & 75.5 \\
M101 & 5.215 $\pm$ 0.011 & 0 & 5.1 $\pm$ 0.019 & 0 & 1.557 $\pm$ 0.017 & 10 & -0.384 $\pm$ 0.024 & 10 & 870.2 & 0.924 & 33.3 \\
NGC5398 & 9.772 $\pm$ 0.012 & 0 & 9.744 $\pm$ 0.022 & 0 & 6.598 $\pm$ 0.025 & 10 & 3.644 $\pm$ 0.031 & 10 & 101.89 & 0.642 & 178.4 \\
NGC3077 & 6.797 $\pm$ 0.011 & 0 & 6.706 $\pm$ 0.019 & 0 & 4.002 $\pm$ 0.024 & 10 & 1.58 $\pm$ 0.02 & 10 & 351.94 & 0.734 & 38.5 \\
NGC2146 & 6.383 $\pm$ 0.011 & 0 & 6.041 $\pm$ 0.019 & 0 & 1.781 $\pm$ 0.014 & 10 & -0.778 $\pm$ 0.011 & 10 & 250.61 & 0.552 & 126.2 \\
NGC3034 & 4.066 $\pm$ 0.012 & 0 & 3.611 $\pm$ 0.021 & 0 & -0.973 $\pm$ 0.011 & 10 & -4.142 $\pm$ 0.011 & 10 & 363.32 & 0.787 & 54.1 \\
\enddata


\tablecomments{Integrated fluxes. The flag for each band indicates if the measurement was from:  (0) $1-\sigma$ isophotal aperture of semi-major axis "Radius", axis ratio and position angle,  or (10) the total flux combining the isophotal with the radial SB extrapolation, or (1) the point source measurement from the ALLWISE catalogue.
 }


\end{deluxetable}
\end{longrotatetable}

\begin{longrotatetable}
\begin{deluxetable}{lccccccccccc}




\tablecaption{Derived \wise properties of the SINGS/KINGFISH Sample}

\tablenum{2}

\tablehead{\colhead{Galaxy} & \colhead{$L_{\rm TIR}$} & \colhead{D$_{\rm lum}$} & \colhead{W3$_{\rm PAH}$} & \colhead{W4$_{\rm dust}$} & \colhead{W1$-$W2} & \colhead{W2$-$W3} & \colhead{log\,$L_{\rm W1}$}  & \colhead{log\,$L_{\rm W2}$}  & \colhead{log\,$L_{\rm W3}$} &  \colhead{log\,$L_{\rm W4}$}  & \colhead{log\,$L_{\rm W1_{\rm Sun}}$}  \\ 
\colhead{} & \colhead{(Wm$^{-2}$)} & \colhead{(Mpc)} & \colhead{(mJy)} & \colhead{(mJy)} & \colhead{(mag)} & \colhead{(mag)} & \colhead{}  & \colhead{}  & \colhead{} & \colhead{} & \colhead{} } 

\startdata
NGC0024 & -12.5666 & 6.9 & 79.4136 & 116.7656 & -0.023 & 2.507 & 8.1386 $\pm$   0.010 & 7.7329 $\pm$   0.012 & 7.4851 $\pm$   0.024 & 7.3575 $\pm$   0.026 & 9.4979 $\pm$   0.014 \\
NGC0337 & -12.0051 & 18.52 & 293.9644 & 738.5792 & 0.228 & 3.639 & 8.9513 $\pm$   0.010 & 8.6461 $\pm$   0.012 & 8.9112 $\pm$   0.012 & 9.0164 $\pm$   0.012 & 10.3106 $\pm$   0.014 \\
NGC0584 & -13.3539 & 20.37 & -- & 14.5342 & -0.059 & 0.469 & 9.6692 $\pm$   0.010 & 9.2491 $\pm$   0.012 & -- $\pm$ -- & 7.3931 $\pm$   0.149 & 11.0286 $\pm$   0.014 \\
NGC0628 & -11.3481 & 7.23 & 2227.4197 & 3362.2595 & 0.109 & 3.633 & 9.0668 $\pm$   0.010 & 8.7139 $\pm$   0.012 & 8.9741 $\pm$   0.012 & 8.8579 $\pm$   0.011 & 10.4262 $\pm$   0.014 \\
NGC0855 & -12.8997 & 9.74 & 29.8283 & 76.437 & 0.072 & 2.364 & 8.0365 $\pm$   0.010 & 7.6688 $\pm$   0.012 & 7.3598 $\pm$   0.035 & 7.4735 $\pm$   0.021 & 9.3959 $\pm$   0.014 \\
NGC0925 & -11.7646 & 9.16 & 523.5233 & 949.6427 & 0.052 & 3.275 & 8.8208 $\pm$   0.010 & 8.445 $\pm$   0.012 & 8.5503 $\pm$   0.018 & 8.514 $\pm$   0.025 & 10.1801 $\pm$   0.014 \\
NGC1097 & -11.1533 & 17.55 & 2738.437 & 6841.18 & 0.1 & 3.415 & 10.0233 $\pm$   0.010 & 9.6667 $\pm$   0.012 & 9.8339 $\pm$   0.012 & 9.9366 $\pm$   0.010 & 11.3826 $\pm$   0.014 \\
NGC1266 & -12.0784 & 27.91 & 86.0285 & 741.4937 & 0.296 & 2.905 & 9.059 $\pm$   0.010 & 8.781 $\pm$   0.012 & 8.7339 $\pm$   0.014 & 9.3744 $\pm$   0.011 & 10.4183 $\pm$   0.014 \\
NGC1291 & -12.0395 & 8.61 & 270.7171 & 336.6199 & -0.04 & 1.235 & 9.6552 $\pm$   0.010 & 9.2428 $\pm$   0.012 & 8.2108 $\pm$   0.114 & 8.0104 $\pm$   0.053 & 11.0145 $\pm$   0.014 \\
NGC1316 & -12.2435 & 18.89 & 166.0631 & 430.4736 & -0.029 & 0.939 & 10.4484 $\pm$   0.010 & 10.0405 $\pm$   0.012 & 8.6807 $\pm$   0.235 & 8.7994 $\pm$   0.040 & 11.8077 $\pm$   0.014 \\
NGC1377 & -12.1891 & 23.35 & 374.3989 & 1610.5756 & 1.363 & 3.462 & 8.8622 $\pm$   0.010 & 9.0109 $\pm$   0.012 & 9.2179 $\pm$   0.010 & 9.5565 $\pm$   0.010 & 10.2215 $\pm$   0.014 \\
NGC1404 & -13.5278 & 19.19 & 2.4194 & 33.2088 & -0.044 & 0.634 & 9.8486 $\pm$   0.010 & 9.4344 $\pm$   0.013 & 6.8576 $\pm$   0.500 & 7.7001 $\pm$   0.100 & 11.208 $\pm$   0.014 \\
NGC1482 & -11.5645 & 19.61 & 1191.4984 & 3640.1294 & 0.319 & 4.202 & 9.3376 $\pm$   0.010 & 9.0686 $\pm$   0.012 & 9.5688 $\pm$   0.011 & 9.7588 $\pm$   0.010 & 10.6969 $\pm$   0.014 \\
NGC1512 & -12.0965 & 12.01 & 336.0567 & 463.3411 & -0.002 & 2.609 & 9.188 $\pm$   0.010 & 8.7906 $\pm$   0.012 & 8.5932 $\pm$   0.021 & 8.4377 $\pm$   0.017 & 10.5473 $\pm$   0.014 \\
NGC1566 & -11.3787 & 20.7 & 1759.6826 & 2928.889 & 0.104 & 3.512 & 9.9314 $\pm$   0.010 & 9.5766 $\pm$   0.012 & 9.7853 $\pm$   0.012 & 9.7116 $\pm$   0.011 & 11.2907 $\pm$   0.014 \\
NGC1705 & -13.0695 & 9.12 & 13.7508 & 51.2158 & 0.084 & 2.02 & 7.8104 $\pm$   0.010 & 7.4475 $\pm$   0.012 & 6.9658 $\pm$   0.033 & 7.2419 $\pm$   0.027 & 9.1697 $\pm$   0.014 \\
M81DwA & -13.9309 & 3.55 & -- & -- & -- & -- & 4.241 $\pm$   0.056 & 4.0229 $\pm$   0.132 & -- $\pm$ -- & -- $\pm$ -- & 5.6004 $\pm$   0.057 \\
DDO053 & -13.6254 & 3.55 & 5.7238 & 23.0834 & 0.321 & 3.095 & 5.9974 $\pm$   0.012 & 5.7294 $\pm$   0.023 & 5.7655 $\pm$   0.033 & 6.0761 $\pm$   0.053 & 7.3568 $\pm$   0.016 \\
NGC2403 & -11.0229 & 3.17 & 3559.5686 & 6480.167 & 0.099 & 3.387 & 8.6642 $\pm$   0.010 & 8.3073 $\pm$   0.012 & 8.4624 $\pm$   0.013 & 8.4276 $\pm$   0.013 & 10.0236 $\pm$   0.014 \\
NGC2798 & -11.7634 & 26.39 & 535.0079 & 2340.4434 & 0.282 & 4.107 & 9.3021 $\pm$   0.010 & 9.0184 $\pm$   0.012 & 9.479 $\pm$   0.011 & 9.825 $\pm$   0.010 & 10.6615 $\pm$   0.014 \\
NGC2841 & -11.689 & 14.06 & 694.688 & 1106.7286 & -0.002 & 2.152 & 9.8663 $\pm$   0.010 & 9.469 $\pm$   0.012 & 9.0458 $\pm$   0.028 & 8.9531 $\pm$   0.019 & 11.2256 $\pm$   0.014 \\
NGC2915 & -13.0286 & 3.76 & 14.0903 & 54.9298 & 0.044 & 1.6 & 7.3165 $\pm$   0.010 & 6.9376 $\pm$   0.012 & 6.2074 $\pm$   0.054 & 6.5033 $\pm$   0.023 & 8.6758 $\pm$   0.014 \\
NGC2976 & -11.7055 & 3.55 & 780.672 & 1423.0991 & 0.1 & 3.218 & 8.1747 $\pm$   0.010 & 7.8182 $\pm$   0.012 & 7.9003 $\pm$   0.013 & 7.8661 $\pm$   0.011 & 9.534 $\pm$   0.014 \\
NGC3031 & -10.9432 & 3.59 & 3637.6533 & 4656.822 & -0.002 & 1.782 & 9.6053 $\pm$   0.010 & 9.208 $\pm$   0.012 & 8.5797 $\pm$   0.044 & 8.392 $\pm$   0.017 & 10.9646 $\pm$   0.014 \\
NGC3049 & -12.5271 & 30.77 & 104.157 & 391.8682 & 0.049 & 3.706 & 8.9893 $\pm$   0.010 & 8.6125 $\pm$   0.012 & 8.9019 $\pm$   0.013 & 9.1824 $\pm$   0.012 & 10.3487 $\pm$   0.014 \\
M81DwB & -14.0797 & 5.31 & 2.4606 & -99.0 & 0.086 & 2.089 & 6.556 $\pm$   0.011 & 6.1939 $\pm$   0.019 & 5.7485 $\pm$   0.064 & -- $\pm$ -- & 7.9153 $\pm$   0.015 \\
NGC3190 & -12.2179 & 28.12 & 211.8346 & 248.0571 & 0.037 & 2.212 & 9.9017 $\pm$   0.010 & 9.5198 $\pm$   0.012 & 9.1317 $\pm$   0.026 & 8.9053 $\pm$   0.016 & 11.261 $\pm$   0.014 \\
NGC3184 & -11.5971 & 11.29 & 1048.2959 & 1489.8738 & 0.075 & 3.406 & 9.2379 $\pm$   0.010 & 8.8714 $\pm$   0.012 & 9.034 $\pm$   0.012 & 8.8917 $\pm$   0.012 & 10.5972 $\pm$   0.014 \\
NGC3198 & -11.8235 & 13.81 & 568.9117 & 1088.1012 & 0.091 & 3.352 & 9.1635 $\pm$   0.010 & 8.8035 $\pm$   0.012 & 8.9434 $\pm$   0.014 & 8.93 $\pm$   0.013 & 10.5228 $\pm$   0.014 \\
IC2574 & -12.3258 & 3.8 & 19.4162 & 174.1573 & 0.089 & 1.322 & 7.6244 $\pm$   0.010 & 7.2635 $\pm$   0.012 & 6.3561 $\pm$   0.082 & 7.0139 $\pm$   0.044 & 8.9837 $\pm$   0.014 \\
NGC3265 & -12.6741 & 22.6 & 73.4912 & 264.3272 & 0.094 & 3.71 & 8.5486 $\pm$   0.010 & 8.1899 $\pm$   0.012 & 8.4821 $\pm$   0.012 & 8.743 $\pm$   0.012 & 9.908 $\pm$   0.014 \\
Mark33 & -12.3931 & 24.91 & 138.9836 & 815.8638 & 0.223 & 4.268 & 8.6247 $\pm$   0.010 & 8.3174 $\pm$   0.012 & 8.8436 $\pm$   0.010 & 9.3173 $\pm$   0.010 & 9.984 $\pm$   0.014 \\
NGC3351 & -11.5837 & 10.92 & 1007.0319 & 2460.0928 & 0.005 & 2.959 & 9.4176 $\pm$   0.010 & 9.0231 $\pm$   0.012 & 8.9871 $\pm$   0.016 & 9.08 $\pm$   0.011 & 10.777 $\pm$   0.014 \\
NGC3521 & -11.0981 & 8.61 & 4429.864 & 5838.9404 & 0.097 & 3.404 & 9.6201 $\pm$   0.010 & 9.2624 $\pm$   0.012 & 9.4247 $\pm$   0.012 & 9.2496 $\pm$   0.011 & 10.9794 $\pm$   0.014 \\
NGC3621 & -11.2383 & 7.02 & 2578.5344 & 3712.6533 & 0.14 & 3.695 & 9.0657 $\pm$   0.012 & 8.7252 $\pm$   0.016 & 9.0121 $\pm$   0.011 & 8.8755 $\pm$   0.013 & 10.425 $\pm$   0.015 \\
NGC3627 & -11.0106 & 10.04 & 4211.5283 & 7762.558 & 0.111 & 3.449 & 9.7056 $\pm$   0.010 & 9.3536 $\pm$   0.012 & 9.5355 $\pm$   0.012 & 9.5061 $\pm$   0.011 & 11.0649 $\pm$   0.014 \\
NGC3773 & -12.8598 & 17.0 & 35.5029 & 128.0237 & 0.119 & 3.178 & 8.2024 $\pm$   0.010 & 7.8535 $\pm$   0.012 & 7.9191 $\pm$   0.017 & 8.1811 $\pm$   0.014 & 9.5617 $\pm$   0.014 \\
NGC3938 & -11.7086 & 17.46 & 799.0945 & 1150.3269 & 0.125 & 3.556 & 9.4134 $\pm$   0.010 & 9.0669 $\pm$   0.012 & 9.2948 $\pm$   0.012 & 9.158 $\pm$   0.012 & 10.7727 $\pm$   0.014 \\
NGC4125 & -13.0226 & 21.96 & -99.0 & 31.459 & -0.037 & 0.452 & 10.01 $\pm$   0.010 & 9.5986 $\pm$   0.012 & -- $\pm$ -- & 7.794 $\pm$   0.128 & 11.3693 $\pm$   0.014 \\
NGC4236 & -12.1201 & 4.47 & 109.2715 & 435.3659 & -0.081 & 2.242 & 8.062 $\pm$   0.010 & 7.6331 $\pm$   0.012 & 7.2471 $\pm$   0.044 & 7.5525 $\pm$   0.021 & 9.4213 $\pm$   0.014 \\
NGC4254 & -11.2238 & 15.41 & 3089.4766 & 4435.8027 & 0.188 & 4.088 & 9.6428 $\pm$   0.010 & 9.3216 $\pm$   0.012 & 9.7731 $\pm$   0.010 & 9.6352 $\pm$   0.010 & 11.0022 $\pm$   0.014 \\
NGC4321 & -11.2668 & 15.91 & 2497.652 & 3942.605 & 0.125 & 3.545 & 9.8321 $\pm$   0.010 & 9.4856 $\pm$   0.012 & 9.7088 $\pm$   0.012 & 9.6121 $\pm$   0.012 & 11.1914 $\pm$   0.014 \\
NGC4450 & -12.2443 & 19.48 & 153.4864 & 226.3166 & -0.009 & 1.7 & 9.7531 $\pm$   0.010 & 9.353 $\pm$   0.012 & 8.6732 $\pm$   0.050 & 8.5468 $\pm$   0.029 & 11.1124 $\pm$   0.014 \\
NGC4536 & -11.5026 & 14.4 & 1257.544 & 3471.5278 & 0.156 & 3.695 & 9.371 $\pm$   0.010 & 9.0369 $\pm$   0.012 & 9.3241 $\pm$   0.012 & 9.4701 $\pm$   0.011 & 10.7303 $\pm$   0.014 \\
NGC4552 & -13.1491 & 15.45 & 113.8553 & 36.8238 & -0.068 & 1.291 & 9.7616 $\pm$   0.010 & 9.3378 $\pm$   0.012 & 8.342 $\pm$   0.116 & 7.5568 $\pm$   0.139 & 11.1209 $\pm$   0.014 \\
NGC4559 & -11.6776 & 13.01 & 637.4404 & 1138.5181 & 0.153 & 3.2 & 9.2003 $\pm$   0.010 & 8.8649 $\pm$   0.012 & 8.9411 $\pm$   0.013 & 8.8981 $\pm$   0.016 & 10.5596 $\pm$   0.014 \\
NGC4569 & -11.77 & 11.84 & 782.2782 & 1367.6968 & 0.063 & 2.683 & 9.4776 $\pm$   0.010 & 9.1063 $\pm$   0.012 & 8.9478 $\pm$   0.017 & 8.8955 $\pm$   0.011 & 10.837 $\pm$   0.014 \\
NGC4579 & -11.8328 & 21.29 & 493.3753 & 665.5659 & 0.032 & 2.166 & 10.0534 $\pm$   0.010 & 9.6695 $\pm$   0.012 & 9.2574 $\pm$   0.027 & 9.0924 $\pm$   0.014 & 11.4127 $\pm$   0.014 \\
NGC4594 & -11.8567 & 9.37 & 513.3752 & 714.6739 & -0.023 & 1.201 & 10.0217 $\pm$   0.010 & 9.616 $\pm$   0.012 & 8.5616 $\pm$   0.117 & 8.4103 $\pm$   0.036 & 11.381 $\pm$   0.014 \\
NGC4625 & -12.6767 & 10.21 & 90.8875 & 130.5536 & 0.074 & 3.365 & 8.1063 $\pm$   0.010 & 7.7393 $\pm$   0.012 & 7.8842 $\pm$   0.015 & 7.7465 $\pm$   0.026 & 9.4656 $\pm$   0.014 \\
NGC4631 & -10.8964 & 12.13 & 4491.2637 & 8430.112 & 0.204 & 3.843 & 9.693 $\pm$    0.010 & 9.3781 $\pm$   0.012 & 9.7282 $\pm$   0.011 & 9.7067 $\pm$   0.010 & 11.0523 $\pm$   0.014 \\
NGC4725 & -11.7135 & 12.76 & 614.6294 & 676.2189 & -0.021 & 2.218 & 9.7045 $\pm$   0.010 & 9.2995 $\pm$   0.012 & 8.9083 $\pm$   0.028 & 8.6548 $\pm$   0.016 & 11.0638 $\pm$   0.014 \\
NGC4736 & -11.0764 & 5.18 & 3926.9233 & 5776.309 & 0.024 & 2.751 & 9.4472 $\pm$   0.010 & 9.0603 $\pm$   0.012 & 8.9307 $\pm$   0.017 & 8.8033 $\pm$   0.012 & 10.8066 $\pm$   0.014 \\
DDO154 & -14.0824 & 4.09 & -- & -- & -0.199 & -- & 6.2274 $\pm$   0.012 & 5.7513 $\pm$   0.032 & -- $\pm$ -- & -- $\pm$ -- & 7.5868 $\pm$   0.016 \\
NGC4826 & -11.3643 & 5.35 & 1535.1316 & 2516.038 & 0.013 & 2.259 & 9.3085 $\pm$   0.010 & 8.9172 $\pm$   0.012 & 8.5505 $\pm$   0.025 & 8.4701 $\pm$   0.012 & 10.6678 $\pm$   0.014 \\
DDO165 & -13.9786 & 1.55 & -- & -- & -0.018 & -- & 5.865 $\pm$   0.010 & 5.4613 $\pm$   0.016 & -- $\pm$ -- & -- $\pm$ -- & 7.2243 $\pm$   0.014 \\
NGC5033 & -11.448 & 17.59 & 1561.0349 & 2173.0088 & 0.1 & 3.533 & 9.7308 $\pm$   0.010 & 9.3743 $\pm$   0.012 & 9.5918 $\pm$   0.012 & 9.4405 $\pm$   0.014 & 11.0902 $\pm$   0.014 \\
NGC5055 & -10.9888 & 9.79 & 4772.666 & 6225.7417 & 0.061 & 3.346 & 9.8037 $\pm$   0.010 & 9.4315 $\pm$   0.012 & 9.5679 $\pm$   0.012 & 9.3883 $\pm$   0.011 & 11.163 $\pm$   0.014 \\
NGC5194 & -10.7154 & 7.9 & 9139.244 & 13661.376 & 0.12 & 3.792 & 9.6852 $\pm$   0.010 & 9.3368 $\pm$   0.012 & 9.664 $\pm$   0.011 & 9.5436 $\pm$   0.010 & 11.0445 $\pm$   0.014 \\
NGC5195 & -12.0357 & 8.03 & 454.2555 & 1161.9993 & 0.019 & 2.211 & 9.1533 $\pm$   0.010 & 8.7644 $\pm$   0.012 & 8.3741 $\pm$   0.026 & 8.487 $\pm$   0.011 & 10.5126 $\pm$   0.014 \\
NGC5408 & -12.5801 & 10.29 & 53.6192 & 390.1243 & 0.09 & 3.202 & 7.9475 $\pm$   0.010 & 7.5869 $\pm$   0.012 & 7.6618 $\pm$   0.015 & 8.2287 $\pm$   0.012 & 9.3068 $\pm$   0.014 \\
NGC5474 & -12.4337 & 7.19 & 87.7675 & 143.0601 & 0.036 & 2.527 & 8.1801 $\pm$   0.010 & 7.798 $\pm$   0.012 & 7.5645 $\pm$   0.028 & 7.4817 $\pm$   0.036 & 9.5395 $\pm$   0.014 \\
NGC5713 & -11.657 & 23.82 & 943.7425 & 2313.0437 & 0.219 & 4.093 & 9.4912 $\pm$   0.010 & 9.1825 $\pm$   0.012 & 9.6364 $\pm$   0.010 & 9.7308 $\pm$   0.010 & 10.8505 $\pm$   0.014 \\
NGC5866 & -12.1189 & 13.98 & 127.8423 & 179.8714 & 0.016 & 1.4 & 9.57 $\pm$   0.010 & 9.1798 $\pm$   0.012 & 8.3055 $\pm$   0.074 & 8.1588 $\pm$   0.026 & 10.9293 $\pm$   0.014 \\
IC4710 & -12.7262 & 10.25 & 29.0319 & 88.0063 & 0.036 & 2.141 & 8.1982 $\pm$   0.010 & 7.8162 $\pm$   0.012 & 7.3918 $\pm$   0.031 & 7.5784 $\pm$   0.034 & 9.5576 $\pm$   0.014 \\
NGC6822 & -11.2063 & 0.46 & 810.8865 & 2515.6265 & 0.007 & 2.222 & 6.9202 $\pm$   0.010 & 6.5265 $\pm$   0.012 & 6.1404 $\pm$   0.043 & 6.3371 $\pm$   0.038 & 8.2796 $\pm$   0.014 \\
NGC6946 & -10.637 & 5.89 & 11566.118 & 20351.254 & 0.177 & 3.909 & 9.4601 $\pm$   0.010 & 9.1344 $\pm$   0.012 & 9.5114 $\pm$   0.010 & 9.4619 $\pm$   0.010 & 10.8194 $\pm$   0.014 \\
NGC7331 & -11.0996 & 14.73 & 3081.586 & 4271.3247 & 0.092 & 3.282 & 9.9833 $\pm$   0.010 & 9.6236 $\pm$   0.012 & 9.7333 $\pm$   0.012 & 9.5801 $\pm$   0.011 & 11.3426 $\pm$   0.014 \\
NGC7552 & -11.1823 & 17.17 & 2592.8818 & 11332.939 & 0.403 & 4.156 & 9.5443 $\pm$   0.010 & 9.3091 $\pm$   0.012 & 9.7912 $\pm$   0.010 & 10.1368 $\pm$   0.010 & 10.9036 $\pm$   0.014 \\
NGC7793 & -11.3849 & 3.93 & 1444.9513 & 1946.5873 & 0.09 & 3.257 & 8.5175 $\pm$   0.010 & 8.157 $\pm$   0.012 & 8.2557 $\pm$   0.013 & 8.0902 $\pm$   0.012 & 9.8768 $\pm$   0.014 \\
IC0342 & -10.3843 & 3.13 & 25064.812 & 42112.3 & 0.082 & 3.785 & 9.3383 $\pm$   0.010 & 8.9746 $\pm$   0.012 & 9.2982 $\pm$   0.017 & 9.2285 $\pm$   0.010 & 10.6976 $\pm$   0.014 \\
M101 & -10.8026 & 7.23 & 6359.7734 & 10806.792 & 0.115 & 3.544 & 9.5576 $\pm$   0.010 & 9.2071 $\pm$   0.012 & 9.4297 $\pm$   0.012 & 9.365 $\pm$   0.013 & 10.917 $\pm$   0.014 \\
NGC5398 & -12.6836 & 20.57 & 59.6415 & 270.3057 & 0.029 & 3.152 & 8.6444 $\pm$   0.010 & 8.2596 $\pm$   0.012 & 8.3101 $\pm$   0.016 & 8.6714 $\pm$   0.015 & 10.0038 $\pm$   0.014 \\
NGC3077 & -11.8005 & 3.93 & 634.8691 & 1801.7534 & 0.091 & 2.704 & 8.4059 $\pm$   0.010 & 8.0458 $\pm$   0.012 & 7.8986 $\pm$   0.018 & 8.0566 $\pm$   0.012 & 9.7652 $\pm$   0.014 \\
NGC2146 & -10.9088 & 17.51 & 5397.2617 & 15935.869 & 0.349 & 4.253 & 9.862 $\pm$   0.010 & 9.6049 $\pm$   0.012 & 10.1264 $\pm$   0.011 & 10.3016 $\pm$   0.010 & 11.2213 $\pm$   0.014 \\
NGC3034 & -9.8625 & 3.72 & 70021.49 & 356701.1 & 0.455 & 4.584 & 9.4509 $\pm$   0.010 & 9.2364 $\pm$   0.012 & 9.8937 $\pm$   0.010 & 10.3058 $\pm$   0.010 & 10.8102 $\pm$   0.014 \\
\enddata




\end{deluxetable}
\end{longrotatetable}

\begin{longrotatetable}
\begin{deluxetable}{lccccccccccc}

\tablecaption{Measured \wise properties of (U)LIRG Sample}

\tablenum{3}

\tablehead{\colhead{Galaxy} & \colhead{W1}  & \colhead{W1f} & \colhead{W2}  & \colhead{W2f} & \colhead{W3}  & \colhead{W3f} & \colhead{W4}  & \colhead{W4f} & \colhead{Radius} & \colhead{b/a} & \colhead{P.A.} \\ 
\colhead{} & \colhead{(mag)}  & \colhead{(--)} & \colhead{(mag)}  & \colhead{--} & \colhead{(mag)}  & \colhead{--} & \colhead{(mag)} & \colhead{--} & \colhead{} & \colhead{} & \colhead{} } 

\startdata
Mrk231 & 7.498 $\pm$ 0.011 & 0 & 6.341 $\pm$ 0.019 & 0 & 3.131 $\pm$ 0.012 & 10 & 0.04 $\pm$ 0.012 & 10 & 87.57 & 1.0 & 0.6 \\
IRAS17208-0014 & 10.647 $\pm$ 0.012 & 0 & 10.011 $\pm$ 0.021 & 0 & 5.785 $\pm$ 0.014 & 10 & 2.366 $\pm$ 0.027 & 0 & 34.31 & 1.0 & 174.4 \\
Arp220 & 9.39 $\pm$ 0.011 & 0 & 8.941 $\pm$ 0.019 & 0 & 4.411 $\pm$ 0.018 & 10 & 0.367 $\pm$ 0.013 & 10 & 85.48 & 1.0 & 5.4 \\
NGC6240 & 8.721 $\pm$ 0.011 & 0 & 8.231 $\pm$ 0.019 & 0 & 4.652 $\pm$ 0.018 & 10 & 1.125 $\pm$ 0.013 & 10 & 88.24 & 0.678 & 20.9 \\
NGC0695 & 9.669 $\pm$ 0.011 & 0 & 9.305 $\pm$ 0.019 & 0 & 4.899 $\pm$ 0.013 & 10 & 2.664 $\pm$ 0.013 & 10 & 45.05 & 0.861 & 14.2 \\
Mrk331 & 9.503 $\pm$ 0.011 & 0 & 9.103 $\pm$ 0.019 & 0 & 4.74 $\pm$ 0.012 & 10 & 1.595 $\pm$ 0.012 & 10 & 43.66 & 0.775 & 150.2 \\
UGC12815 & 8.004 $\pm$ 0.011 & 0 & 7.828 $\pm$ 0.019 & 0 & 4.233 $\pm$ 0.023 & 10 & 1.856 $\pm$ 0.019 & 10 & 221.62 & 0.35 & 65.7 \\
HIPASSJ1004-06 & 9.017 $\pm$ 0.011 & 0 & 8.74 $\pm$ 0.019 & 0 & 4.435 $\pm$ 0.027 & 10 & 2.175 $\pm$ 0.016 & 10 & 73.77 & 0.77 & 165.5 \\
NGC2388 & 9.211 $\pm$ 0.011 & 0 & 8.893 $\pm$ 0.019 & 0 & 4.68 $\pm$ 0.015 & 10 & 1.719 $\pm$ 0.013 & 10 & 50.18 & 0.871 & 47.7 \\
NGC2146 & 6.383 $\pm$ 0.011 & 0 & 6.041 $\pm$ 0.019 & 0 & 1.814 $\pm$ 0.014 & 10 & -0.779 $\pm$ 0.011 & 10 & 250.61 & 0.552 & 126.2 \\
NGC1365 & 6.034 $\pm$ 0.011 & 0 & 5.74 $\pm$ 0.019 & 0 & 2.112 $\pm$ 0.011 & 10 & -0.457 $\pm$ 0.011 & 10 & 393.3 & 0.602 & 39.4 \\
NGC4039 & 6.915 $\pm$ 0.011 & 0 & 6.749 $\pm$ 0.019 & 0 & 2.921 $\pm$ 0.014 & 10 & 0.299 $\pm$ 0.014 & 10 & 159.24 & 0.899 & 168.3 \\
UGC89 & 8.568 $\pm$ 0.011 & 0 & 8.436 $\pm$ 0.019 & 0 & 4.786 $\pm$ 0.015 & 10 & 2.347 $\pm$ 0.014 & 10 & 100.03 & 0.647 & 169.3 \\
NGC6701 & 8.87 $\pm$ 0.011 & 0 & 8.68 $\pm$ 0.019 & 0 & 4.667 $\pm$ 0.012 & 10 & 2.179 $\pm$ 0.013 & 10 & 70.72 & 0.978 & 118.7 \\
UGC1845 & 9.476 $\pm$ 0.011 & 0 & 9.173 $\pm$ 0.019 & 0 & 5.203 $\pm$ 0.013 & 10 & 2.541 $\pm$ 0.013 & 10 & 53.24 & 0.653 & 129.1 \\
NGC5936 & 9.04 $\pm$ 0.011 & 0 & 8.823 $\pm$ 0.019 & 0 & 4.68 $\pm$ 0.012 & 10 & 2.188 $\pm$ 0.013 & 10 & 71.02 & 0.95 & 76.5 \\
MCG$+$02-20-003 & 10.347 $\pm$ 0.012 & 0 & 9.294 $\pm$ 0.02 & 0 & 5.757 $\pm$ 0.013 & 10 & 2.948 $\pm$ 0.014 & 10 & 40.16 & 0.657 & 152.6 \\
HIPASSJ0716-62 & 8.157 $\pm$ 0.011 & 0 & 7.925 $\pm$ 0.019 & 0 & 4.18 $\pm$ 0.012 & 10 & 1.569 $\pm$ 0.012 & 10 & 130.24 & 0.489 & 177.5 \\
ESO320-G030 & 8.796 $\pm$ 0.011 & 0 & 8.581 $\pm$ 0.019 & 0 & 4.6 $\pm$ 0.013 & 10 & 1.74 $\pm$ 0.012 & 10 & 95.43 & 0.668 & 126.9 \\
IC5179 & 8.193 $\pm$ 0.011 & 0 & 7.948 $\pm$ 0.019 & 0 & 3.75 $\pm$ 0.012 & 10 & 1.477 $\pm$ 0.015 & 10 & 101.56 & 0.605 & 51.9 \\
MCG$+$12-02-001 & 9.336 $\pm$ 0.011 & 0 & 8.84 $\pm$ 0.019 & 0 & 4.24 $\pm$ 0.015 & 10 & 1.028 $\pm$ 0.012 & 10 & 51.24 & 0.89 & 173.9 \\
F03359$+$1523 & 11.827 $\pm$ 0.012 & 0 & 11.144 $\pm$ 0.022 & 0 & 6.563 $\pm$ 0.016 & 10 & 3.356 $\pm$ 0.016 & 10 & 20.61 & 1.0 & 5.9 \\
NGC1614 & 8.847 $\pm$ 0.011 & 0 & 8.378 $\pm$ 0.019 & 0 & 3.681 $\pm$ 0.013 & 10 & 0.317 $\pm$ 0.026 & 0 & 62.29 & 0.822 & 163.9 \\
UGC2369 & 9.762 $\pm$ 0.011 & 0 & 9.457 $\pm$ 0.019 & 0 & 5.288 $\pm$ 0.017 & 10 & 2.163 $\pm$ 0.012 & 10 & 55.05 & 0.89 & 18.5 \\
Arp236 & 9.223 $\pm$ 0.011 & 0 & 8.208 $\pm$ 0.019 & 0 & 4.224 $\pm$ 0.012 & 10 & 1.159 $\pm$ 0.012 & 10 & 65.34 & 0.867 & 6.9 \\
IC883 & 10.207 $\pm$ 0.011 & 0 & 9.639 $\pm$ 0.019 & 0 & 5.324 $\pm$ 0.013 & 10 & 2.285 $\pm$ 0.012 & 10 & 58.29 & 0.855 & 178.4 \\
UGC4881 & 10.379 $\pm$ 0.011 & 0 & 10.051 $\pm$ 0.02 & 0 & 6.043 $\pm$ 0.013 & 10 & 3.255 $\pm$ 0.015 & 10 & 50.12 & 0.888 & 21.9 \\
NGC3690 & 7.198 $\pm$ 0.011 & 0 & 6.05 $\pm$ 0.019 & 0 & 2.227 $\pm$ 0.011 & 10 & -1.064 $\pm$ 0.011 & 10 & 127.09 & 0.85 & 21.1 \\
F17132$+$5313 & 11.007 $\pm$ 0.011 & 0 & 10.464 $\pm$ 0.019 & 0 & 5.886 $\pm$ 0.022 & 10 & 3.202 $\pm$ 0.012 & 10 & 31.03 & 0.916 & 173.8 \\
MK848 & 10.775 $\pm$ 0.011 & 0 & 10.186 $\pm$ 0.02 & 0 & 5.556 $\pm$ 0.013 & 10 & 2.123 $\pm$ 0.012 & 10 & 34.25 & 0.799 & 150.5 \\
IRAS10565$+$2448W & 10.772 $\pm$ 0.012 & 0 & 10.085 $\pm$ 0.021 & 0 & 5.666 $\pm$ 0.014 & 10 & 2.388 $\pm$ 0.017 & 10 & 33.03 & 1.0 & 174.8 \\
VIIZw31 & 10.735 $\pm$ 0.012 & 0 & 10.239 $\pm$ 0.02 & 0 & 5.793 $\pm$ 0.018 & 10 & 3.163 $\pm$ 0.012 & 10 & 28.24 & 1.0 & 12.6 \\
IRAS23365$+$3604 & 11.693 $\pm$ 0.012 & 0 & 11.026 $\pm$ 0.022 & 0 & 6.51 $\pm$ 0.017 & 10 & 2.946 $\pm$ 0.012 & 10 & 30.84 & 0.779 & 20.4 \\
\enddata




\end{deluxetable}
\end{longrotatetable}

\begin{longrotatetable}
\begin{deluxetable}{lccccccccccc}

\tablecaption{Derived \wise properties of (U)LIRG Sample}

\tablenum{4}

\tablehead{\colhead{Galaxy} & \colhead{$L_{\rm TIR}$} & \colhead{D$_{\rm lum}$} & \colhead{W3$_{\rm PAH}$} & \colhead{W4$_{\rm dust}$} & \colhead{W1$-$W2} & \colhead{W2$-$W3} & \colhead{log\,$L_{\rm W1}$}  & \colhead{log\,$L_{\rm W2}$}  & \colhead{log\,$L_{\rm W3}$} &  \colhead{log\,$L_{\rm W4}$}  & \colhead{log\,$L_{\rm W1_{\rm Sun}}$}  \\ 
\colhead{} & \colhead{(Wm$^{-2}$)} & \colhead{(Mpc)} & \colhead{(mJy)} & \colhead{(mJy)} & \colhead{(mag)} & \colhead{(mag)} & \colhead{}  & \colhead{}  & \colhead{} & \colhead{} & \colhead{} } 

\startdata
Mrk231 & 12.51\tablenotemark{a} & 171.98 & 1522.2544 & 7303.187 & 1.175 & 3.198 & 11.3912 $\pm$   0.010 & 11.4646 $\pm$   0.012 & 11.5613 $\pm$   0.010 & 11.9473 $\pm$   0.010 & 12.7505 $\pm$   0.014 \\
IRAS17208-0014 & 12.4\tablenotemark{b} & 183.01 & 125.8645 & 1061.4515 & 0.84 & 4.035 & 10.1566 $\pm$   0.010 & 10.0962 $\pm$   0.012 & 10.5327 $\pm$   0.010 & 11.1637 $\pm$   0.014 & 11.5159 $\pm$   0.014 \\
Arp220 & 12.18\tablenotemark{c} & 75.99 & 482.0348 & 5964.031 & 0.476 & 4.53 & 9.9231 $\pm$   0.010 & 9.717 $\pm$   0.012 & 10.3525 $\pm$   0.011 & 11.1499 $\pm$   0.010 & 11.2824 $\pm$   0.014 \\
NGC6240 & 11.86\tablenotemark{d} & 104.98 & 373.4446 & 2882.1587 & 0.537 & 3.576 & 10.4599 $\pm$   0.010 & 10.2781 $\pm$   0.012 & 10.5223 $\pm$   0.012 & 11.1148 $\pm$   0.010 & 11.8192 $\pm$   0.014 \\
NGC695 & 11.7\tablenotemark{d} & 139.11 & 284.3924 & 679.9062 & 0.391 & 4.378 & 10.3158 $\pm$   0.010 & 10.0757 $\pm$   0.012 & 10.6485 $\pm$   0.010 & 10.732 $\pm$   0.010 & 11.6751 $\pm$   0.014 \\
Mrk331 & 11.5\tablenotemark{e} & 79.3 & 345.9611 & 1839.5452 & 0.426 & 4.356 & 9.9072 $\pm$   0.010 & 9.6812 $\pm$   0.012 & 10.2454 $\pm$   0.010 & 10.6761 $\pm$   0.010 & 11.2665 $\pm$   0.014 \\
UGC12815 & 11.42\tablenotemark{d} & 61.12 & 535.5026 & 1404.9697 & 0.176 & 3.609 & 10.2838 $\pm$   0.010 & 9.9577 $\pm$   0.012 & 10.209 $\pm$   0.014 & 10.3329 $\pm$   0.012 & 11.6431 $\pm$   0.014 \\
HIPASSJ1004-06 & 11.37\tablenotemark{e} & 79.52 & 452.5499 & 1057.5352 & 0.292 & 4.304 & 10.1026 $\pm$   0.010 & 9.8229 $\pm$   0.012 & 10.3645 $\pm$   0.014 & 10.4382 $\pm$   0.011 & 11.4619 $\pm$   0.014 \\
NGC2388 & 11.28\tablenotemark{e} & 62.09 & 365.9762 & 1613.9646 & 0.341 & 4.2 & 9.8183 $\pm$   0.010 & 9.558 $\pm$   0.012 & 10.0573 $\pm$   0.011 & 10.4068 $\pm$   0.010 & 11.1776 $\pm$   0.014 \\
NGC2146 & 11.12\tablenotemark{e} & 20.56 & 5237.3984 & 16063.61 & 0.343 & 4.229 & 10.0006 $\pm$   0.010 & 9.7413 $\pm$   0.012 & 10.2528 $\pm$   0.011 & 10.4446 $\pm$   0.010 & 11.3599 $\pm$   0.014 \\
NGC1365 & 11.0\tablenotemark{e} & 18.14 & 3901.5818 & 11801.377 & 0.295 & 3.632 & 10.0309 $\pm$   0.010 & 9.7525 $\pm$   0.012 & 10.0161 $\pm$   0.011 & 10.2019 $\pm$   0.010 & 11.3902 $\pm$   0.014 \\
NGC4039 & 10.88\tablenotemark{f} & 21.51 & 1859.7667 & 5897.8604 & 0.169 & 3.834 & 9.8258 $\pm$   0.010 & 9.4971 $\pm$   0.012 & 9.8426 $\pm$   0.011 & 10.0488 $\pm$   0.010 & 11.1851 $\pm$   0.014 \\
UGC89 & 11.11\tablenotemark{g} & 63.89 & 320.6163 & 893.426 & 0.132 & 3.664 & 10.0948 $\pm$   0.010 & 9.7511 $\pm$   0.012 & 10.0247 $\pm$   0.012 & 10.1748 $\pm$   0.010 & 11.4541 $\pm$   0.014 \\
NGC6701 & 11.11\tablenotemark{g} & 60.6 & 367.7937 & 1043.429 & 0.193 & 4.027 & 9.9313 $\pm$   0.010 & 9.6122 $\pm$   0.012 & 10.0383 $\pm$   0.010 & 10.1962 $\pm$   0.010 & 11.2906 $\pm$   0.014 \\
UGC1845 & 11.13\tablenotemark{g} & 66.39 & 224.3041 & 746.7936 & 0.308 & 3.984 & 9.7664 $\pm$   0.010 & 9.493 $\pm$   0.012 & 9.9029 $\pm$   0.010 & 10.1303 $\pm$   0.010 & 11.1257 $\pm$   0.014 \\
NGC5936 & 11.13\tablenotemark{g} & 65.12 & 361.3966 & 1041.8303 & 0.241 & 4.137 & 9.9207 $\pm$   0.010 & 9.6207 $\pm$   0.012 & 10.0932 $\pm$   0.010 & 10.2581 $\pm$   0.010 & 11.2801 $\pm$   0.014 \\
MCG$+$02-20-003 & 11.14\tablenotemark{g} & 72.42 & 138.1558 & 508.2348 & 1.067 & 3.527 & 9.5066 $\pm$   0.010 & 9.5371 $\pm$   0.012 & 9.768 $\pm$   0.010 & 10.0387 $\pm$   0.010 & 10.8659 $\pm$   0.014 \\
HIPASSJ0716-62 & 11.16\tablenotemark{g} & 47.11 & 576.0187 & 1830.9971 & 0.238 & 3.76 & 10.0002 $\pm$   0.010 & 9.6987 $\pm$   0.012 & 10.0146 $\pm$   0.011 & 10.2219 $\pm$   0.010 & 11.3595 $\pm$   0.014 \\
ESO320-G030 & 11.16\tablenotemark{g} & 40.4 & 397.64 & 1557.4791 & 0.219 & 3.997 & 9.6146 $\pm$   0.010 & 9.3059 $\pm$   0.012 & 9.72 $\pm$   0.011 & 10.018 $\pm$   0.010 & 10.9739 $\pm$   0.014 \\
IC5179 & 11.22\tablenotemark{g} & 49.38 & 857.0593 & 2040.4521 & 0.242 & 4.2 & 10.0294 $\pm$   0.010 & 9.7295 $\pm$   0.012 & 10.228 $\pm$   0.010 & 10.3098 $\pm$   0.011 & 11.3887 $\pm$   0.014 \\
MCG$+$12-02-001 & 11.5\tablenotemark{g} & 72.07 & 560.0752 & 3065.8076 & 0.52 & 4.606 & 9.8937 $\pm$   0.010 & 9.7051 $\pm$   0.012 & 10.3716 $\pm$   0.011 & 10.8149 $\pm$   0.010 & 11.253 $\pm$   0.014 \\
F03359$+$1523 & 11.53\tablenotemark{g} & 157.18 & 66.3665 & 415.0831 & 0.719 & 4.608 & 9.5629 $\pm$   0.010 & 9.454 $\pm$   0.012 & 10.1226 $\pm$   0.011 & 10.6238 $\pm$   0.011 & 10.9222 $\pm$   0.014 \\
NGC1614 & 11.66\tablenotemark{g} & 67.1 & 946.8895 & 5814.2104 & 0.474 & 4.728 & 10.0285 $\pm$   0.010 & 9.8215 $\pm$   0.012 & 10.5376 $\pm$   0.010 & 11.0308 $\pm$   0.014 & 11.3879 $\pm$   0.014 \\
UGC2369 & 11.66\tablenotemark{g} & 138.75 & 199.2178 & 1247.5616 & 0.294 & 4.189 & 10.2759 $\pm$   0.010 & 9.997 $\pm$   0.012 & 10.4916 $\pm$   0.011 & 10.9934 $\pm$   0.010 & 11.6352 $\pm$   0.014 \\
Arp236 & 11.71\tablenotemark{g} & 85.76 & 569.1014 & 2746.3096 & 1.044 & 3.959 & 10.1017 $\pm$   0.010 & 10.1226 $\pm$   0.012 & 10.5296 $\pm$   0.010 & 10.9181 $\pm$   0.010 & 11.461 $\pm$   0.014 \\
IC883 & 11.73\tablenotemark{g} & 107.12 & 201.8004 & 1022.3195 & 0.595 & 4.295 & 9.8904 $\pm$   0.010 & 9.732 $\pm$   0.012 & 10.2725 $\pm$   0.010 & 10.6822 $\pm$   0.010 & 11.2497 $\pm$   0.014 \\
UGC4881 & 11.75\tablenotemark{g} & 176.17 & 98.5039 & 418.2176 & 0.321 & 4.036 & 10.2297 $\pm$   0.010 & 9.9615 $\pm$   0.012 & 10.3932 $\pm$   0.010 & 10.7262 $\pm$   0.011 & 11.5891 $\pm$   0.014 \\
NGC3690 & 11.94\tablenotemark{g} & 51.22 & 3593.5107 & 21073.629 & 1.146 & 3.821 & 10.4689 $\pm$   0.010 & 10.5307 $\pm$   0.012 & 10.8823 $\pm$   0.010 & 11.3555 $\pm$   0.010 & 11.8282 $\pm$   0.014 \\
F17132$+$5313 & 11.95\tablenotemark{g} & 231.0 & 115.9448 & 428.5789 & 0.58 & 4.574 & 10.2095 $\pm$   0.010 & 10.0451 $\pm$   0.012 & 10.6993 $\pm$   0.012 & 10.9721 $\pm$   0.010 & 11.5688 $\pm$   0.014 \\
Mrk848 & 11.95\tablenotemark{g} & 180.69 & 167.5365 & 1321.332 & 0.622 & 4.667 & 10.1015 $\pm$   0.010 & 9.9537 $\pm$   0.012 & 10.6458 $\pm$   0.010 & 11.2477 $\pm$   0.010 & 11.4608 $\pm$   0.014 \\
IRAS10565$+$2448W & 11.99\tablenotemark{g} & 188.91 & 151.349 & 1045.468 & 0.725 & 4.46 & 10.1385 $\pm$   0.010 & 10.0318 $\pm$   0.012 & 10.6404 $\pm$   0.010 & 11.1847 $\pm$   0.011 & 11.4978 $\pm$   0.014 \\
VIIZw31 & 12.0\tablenotemark{g} & 230.22 & 124.3413 & 458.6414 & 0.518 & 4.464 & 10.307 $\pm$   0.010 & 10.1178 $\pm$   0.012 & 10.7267 $\pm$   0.011 & 10.9986 $\pm$   0.010 & 11.6663 $\pm$   0.014 \\
IRAS23365$+$3604 & 12.19\tablenotemark{g} & 269.83 & 70.1698 & 676.6337 & 0.707 & 4.6 & 10.0647 $\pm$   0.010 & 9.951 $\pm$   0.012 & 10.6162 $\pm$   0.011 & 11.3054 $\pm$   0.010 & 11.424 $\pm$   0.014 \\
\enddata

\tablenotetext{a}{\citet{Fisch10}}
\tablenotetext{b}{\citet{Gold95}}
\tablenotetext{c}{\citet{DH02}}
\tablenotetext{d}{\citet{DeL11}}
\tablenotetext{e}{\citet{Ar09}}
\tablenotetext{f}{\citet{Gao01}}
\tablenotetext{g}{\citet{San03}}




\end{deluxetable}
\end{longrotatetable}

\begin{longrotatetable}
\begin{deluxetable}{lccccccccccc}

\tablecaption{Measured \wise properties of Dwarf Sample}

\tablenum{5}

\tablehead{\colhead{Galaxy} & \colhead{W1}  & \colhead{W1f} & \colhead{W2}  & \colhead{W2f} & \colhead{W3}  & \colhead{W3f} & \colhead{W4}  & \colhead{W4f} & \colhead{Radius} & \colhead{b/a} & \colhead{P.A.} \\ 
\colhead{} & \colhead{(mag)}  & \colhead{(--)} & \colhead{(mag)}  & \colhead{--} & \colhead{(mag)}  & \colhead{--} & \colhead{(mag)} & \colhead{--} & \colhead{} & \colhead{} & \colhead{} } 

\startdata
NGC0625 & 8.536 $\pm$ 0.012 & 0 & 8.46 $\pm$ 0.02 & 0 & 5.423 $\pm$ 0.033 & 10 & 2.357 $\pm$ 0.024 & 10 & 227.06 & 0.406 & 93.7 \\
NGC1569 & 7.484 $\pm$ 0.011 & 0 & 7.258 $\pm$ 0.019 & 0 & 3.916 $\pm$ 0.018 & 10 & -0.016 $\pm$ 0.011 & 10 & 208.06 & 0.532 & 119.9 \\
\enddata




\end{deluxetable}
\end{longrotatetable}

\begin{longrotatetable}
\begin{deluxetable}{lccccccccccc}

\tablecaption{Derived \wise properties of Derived Sample}

\tablenum{6}

\tablehead{\colhead{Galaxy} & \colhead{$L_{\rm TIR}$} & \colhead{D$_{\rm lum}$} & \colhead{W3$_{\rm PAH}$} & \colhead{W4$_{\rm dust}$} & \colhead{W1$-$W2} & \colhead{W2$-$W3} & \colhead{log\,$L_{\rm W1}$}  & \colhead{log\,$L_{\rm W2}$}  & \colhead{log\,$L_{\rm W3}$} &  \colhead{log\,$L_{\rm W4}$}  & \colhead{log\,$L_{\rm W1_{\rm Sun}}$}  \\ 
\colhead{} & \colhead{(Wm$^{-2}$)} & \colhead{(Mpc)} & \colhead{(mJy)} & \colhead{(mJy)} & \colhead{(mag)} & \colhead{(mag)} & \colhead{}  & \colhead{}  & \colhead{} & \colhead{} & \colhead{} }

\startdata
NGC0625 & 8.50\tablenotemark{a} & 4.68 & 173.3133 & 870.4055 & 0.076 &  3.038 & 7.8515  $\pm$    0.010 &   7.4854 $\pm$     0.012  &  7.4878  $\pm$    0.018 &   7.8937 $\pm$     0.013  & 9.2108 $\pm$      0.014  \\
NGC1569  & 8.57\tablenotemark{a} & 1.59 & 738.7435 &  7969.3071 &   0.226 &   3.342  & 7.3458  $\pm$    0.010 &   7.0397 $\pm$     0.012  &  7.1791  $\pm$    0.013 &   7.9171 $\pm$     0.010  & 8.7052 $\pm$      0.014 \\
\enddata

\tablenotetext{a}{\citet{DeL11}}




\end{deluxetable}
\end{longrotatetable}

\end{document}